\documentclass[%
 reprint,
nofootinbib,
 amsmath,amssymb,
 aps,
]{revtex4-2}

\usepackage{graphicx}% Include figure files
\usepackage{dcolumn}% Align table columns on decimal point
\usepackage{bm}% bold math
\usepackage{adjustbox}
\usepackage{graphicx}% Include figure files
\usepackage{dcolumn}% Align table columns on decimal point
\usepackage{bm}% bold math
\usepackage{hyperref}% add hypertext capabilities
\usepackage{multirow}
\usepackage{cleveref}
\usepackage[caption=false]{subfig}
\usepackage[usenames]{color}
\usepackage{slashbox}
\usepackage{adjustbox}

\begin{document}

\preprint{APS/123-QED}

\title{Relic Neutrino Helicity Evolution in the Galactic Magnetic Field \\ and Its Implications}% Force line breaks with \\

\title{Relic Neutrino Helicity Evolution in the Galactic Magnetic Field }% Force line breaks with \\

\author{Kuo(i:to).k.Liao}
\email{kl4180@nyu.edu}
 \altaffiliation{%
 Center for Cosmology and Particle Physics, Department of Physics, New York University}%Lines break automatically or can be forced with \\
\author{Glennys R. Farrar}%
 \email{gf25@nyu.edu}
\affiliation{%
 Center for Cosmology and Particle Physics, Department of Physics, New York University  
}%

\date{\today}% It is always \today, today,
             %  but any date may be explicitly specified

\begin{abstract}
We simulate the evolution of the helicity of relic neutrinos as they propagate to Earth through a realistic model of the Galactic magnetic field, improving upon the rough estimates in the literature. 
%Following ideas from previous research, we apply Jansson and Farrar(JF-12) \cite{Jansson:2012pc} coherent galactic magnetic field model and updated model with random components to have a comprehensive understanding of relic neutrino helicity evolution in astrophysical magnetic field for the first time.
For magnetic moments consistent with experimental bounds and several orders of magnitude smaller, we confirm that the helicity of relic neutrinos and anti-neutrinos rotates so much that the spin projection changes by $\mathcal{O}$(1).  However, as we show, the total event rate in an inverse tritium beta decay (ITBD) experiment changes by less than a few percent, unless the lightest neutrino has mass of order 0.001 eV or less.  Such a tiny reduction in the absolute rate relative to the standard model value would be very difficult to establish, even if detecting relic neutrinos were routine.  However as we show, the \emph{directional anisotropy} of the rate in a \emph{polarized} ITBD detector is $\gtrsim \mathcal{O}$(10\%) as long as the lightest neutrino mass is $\gtrsim \mathcal{O}$(0.01 eV).  Thus with percent-level error bars on the absolute neutrino flux and its directional anisotropy, both the mass and magnetic moment of the relic neutrinos can in principle be probed if they are within a few orders of magnitude of current bounds.  
%In favor of Python package $Healpix$, relic neutrino's helicity-flip probability and its distribution are presented.%
%e report simulation results of relic neutrino helicity evolution in the astrophysical magnetic field. 
%Following ideas from previous research, we apply Jansson and Farrar(JF-12) \cite{Jansson:2012pc} coherent galactic magnetic field model and updated model with random components to have a comprehensive understanding of relic neutrino helicity evolution in astrophysical magnetic field for the first time.

\end{abstract}
\keywords{Suggested keywords}%Use showkeys class option if keyword
%display desired
\maketitle

%\tableofcontents

\section{Introduction}

Relic neutrinos carry information from the early Universe. They decoupled from the hot plasma about 1 second after the big bang, much earlier than the cosmic microwave background. Neutrinos and anti-neutrinos are decoupled in chirality eigenstates at temperatures $\simeq$ MeV.
When decoupled, they were highly relativistic, and their helicity eigenstates coincided with chiral eigenstates. 

The idea that the spin of a massive Dirac neutrino with a magnetic moment precesses in a magnetic field was originally studied by Fujikawa and Shrock \cite{PhysRevLett.45.963}. 
More recently, Baym and Peng studied the effects of astrophysical magnetic fields and gravitational inhomogeneities on the helicities of present-day relic neutrinos \cite{Baym:2020riw} \cite{Baym:2021ksj}. Gravitational effects can modify the helicity eigenstates of both Dirac and Majorana neutrinos having finite mass, but helicity modifications due to magnetic moments arise only for Dirac and not Majorana neutrinos \cite{Schechter:1981hw}. 
See \cite{Baym:2020riw,Baym:2021ksj} and citations therein for a recent survey of the field.

Inverse tritium beta decay \cite{PhysRev.128.1457} (ITBD) is a proposed experimental means \cite{PTOLEMY:2022ldz} to observe relic neutrinos. Since the ITBD cross-section depends on helicity, ITBD measurements can, in principle, provide a tool to detect helicity modification.  Moreover, directional anisotropies in the neutrino event rate are in principle accessible with a polarized ITBD target~\cite{Lisanti:2014pqa}.  Thus it may one day be possible to detect an angular anisotropy in the neutrino event rate due to helicity modification, which depends on the neutrino arrival direction with respect to the Galactic magnetic field.  

The aim of the present paper is twofold.  First, we improve on the very rough estimate of the amount of spin rotation in the Galactic magnetic field provided by Baym and Peng~\cite{Baym:2020riw,Baym:2021ksj}, by simulating the propagation of relic neutrinos and anti-neutrinos through a realistic model of the Galactic magnetic field (GMF).  Second, we determine the observational impact of such spin rotation, which has not previously been discussed.  We find that large spin rotation does not \emph{per se} imply a significant change in the interaction rate, unless the lightest neutrino remains relativistic today.  However counting rate and anisotropy measurements can together constrain, at least in principle, several decades of mass-magnetic moment parameter space.

%The understanding of the Galactic magnetic field developed over the decades. Today 
As of now, the Jansson and Farrar model (JF12) \cite{Jansson:2012pc} fits the largest range of observations and we adopt it as a sufficiently realistic model to study neutrino helicity evolution in the Galaxy.  At the time the JF12 models of the coherent and random field were developed, the polarized synchrotron emission of the Galaxy -- a key observable used by JF12 to constrain the GMF -- was quite uncertain, so we report results on neutrino spin rotation in both the coherent and combined coherent+random JF12 models as a measure of the range of uncertainty in the predictions.  Near-future GMF models based on the latest observations and modeling will make the predictions for neutrino spin rotations more precise, without however altering the basic conclusions of the work presented here.  
%Here we focus on Dirac neutrinos with non-zero diagonal magnetic moments. 

Propagation through the intense field which may exist in the radiative zone of the sun, could also potentially produce $\nu_{L}$ to $\nu_{R}$ transitions. The Magnetic fields could impact
electron neutrino resonant spin-flavor conversion in the solar environment \cite{Semikoz:1996up}\cite{Joshi:2019dcj}\cite{Giunti:2014ixa} and impact solar neutrino flux observations.  However we do not address those topics here.  We note recent papers addressing the impact of flavor oscillations and magnetic fields on the detection of VHE or UHE neutrinos, e.g.\cite{lichkunov2022three};  this is a totally different phenomenon than what is being discussed here.

\section{Theoretical Background}
\subsection{Relic Neutrino Spin Rotation in Astrophysical Magnetic Field}
Neutrinos are produced in flavor eigenstates, but they arrive at Earth in well-separated mass packets.
The three flavors are superpositions of mass eigenstates via the Pontecorvo-Maki-Nakagawa-Sakata (PMNS) mixing matrix.
While coupled to ambient plasma, relic neutrinos had a relativistic thermal distribution at temperature $T$. As momenta and temperature redshift identically with the expansion of the Universe, the Fermi-Dirac distribution is preserved even after the decoupling process, so the present momentum $\vec{p}_{0}$ and temperature $T_{0}$, for flavor $\alpha$ follows the distribution:
\begin{equation}
    f_{\alpha}(\vec{p_{0}},T_{0})= \frac{1}{e^{|\vec{p_{0}}|/T_{0}}+1},
    \label{fermi}
\end{equation}
where the present cosmic neutrino background temperature $T_{0}$ $\simeq$ 1.67 $\times$ $10^{-4}$ eV = $(\frac{4}{11})^{1/3}$ $T_{CMB}$.

Propagating in a magnetic field, the neutrino spin vector rotates in its rest frame \cite{PhysRevLett.45.963,Baym:2020riw,Baym:2021ksj}:
\begin{equation} 
\label{eq:basic}
     \frac{d\vec{S}}{d\tau} =2{\mathbf{\mu}}_{\nu}\index{\n}(\vec{S}\times{\vec{B}}),
\end{equation}
where $\vec{B} $ represents the magnetic field vector in its rest frame, $\tau$ is the neutrino proper time, and ${\mathbf{\mu}}_{\nu}\index{\nu}$ is the diagonal magnetic moment of the Dirac neutrino, expressed in Bohr magnetons, $\mu_{B}=1.4$ MHz/Gauss.

 Reference~\cite{Baym:2020riw} showed that the $\nu$ momentum direction does not change during propagation through the Galaxy.  Then, rewriting Eq.~\eqref{eq:basic} in terms of the lab position and using that the lab frame time $dt = \gamma d\tau$ (where $\gamma \equiv 1/\sqrt{1 - \beta^2}$ and $\beta \equiv v/c$) gives
\begin{equation}
\label{eq:dSdr}
    \frac{\partial\vec{S}}{\partial \vec{r}}=\frac{1}{|\vec{v}|} \frac{\partial \vec{S}}{\partial {t}}=\frac{2}{\gamma}\frac{{\mathbf{\mu}}_{\nu}\index{\nu}}{|\vec{v}|}\,{\vec{S}\times \vec{B}}.
\end{equation} 
Given that neutrino mass eigenstates are a mixing-matrix-weighted sum of  flavor eigenstates, the symbol $\mu_{\nu}$ appearing in \eqref{eq:dSdr} is the corresponding weighted sum of the flavor-basis magnetic moments, since the coupling to the magnetic field is diagonal in flavor. $\vec{S}_{final}$ can be calculated by integrating over the total propagation path length. 
\begin{equation}     
  \Delta \vec{S} =2{\mathbf{\mu}}_{\nu}\index{\n}\int{\vec{S}\times \vec{B}} \, \frac{dt}{\gamma}.
\end{equation}

The probability of helicity flipping can be expressed as a function of the rotation angle $\theta$: $P_{f}=sin^2(\theta/2)$. %For infinitesimal rotation ($\theta \ll 1$), $P_{f}$ $\approx$ $\theta^2$/4.
In the lab frame, with respect to the neutrino's lab frame velocity $\vec{B}_{\perp}^{rest frame}$=$\gamma$$\vec{B}_{\perp}^{lab frame}$ and $\vec{B}_{\parallel}^{lab frame}$=$\vec{B}_{\parallel}^{rest frame}$, so in terms of lab frame magnetic field and time \cite{Baym:2020riw}: 
\begin{equation}
\frac{d\vec{S}_{\perp}}{dt}=(\vec{S}_{\parallel}\times\vec{B}_{\perp}^{lab frame}
+\frac{1}{\gamma} \vec{S}_{\perp}\times \vec{B}_{\parallel}^{lab frame})2{\mu_{\nu}}.
\label{7}
\end{equation}
The cumulative rotation angle $\theta$ at a specific moment compared to neutrino's initial $\vec{S}$ state is $\theta=sin^{-1}({\vec{S}_{\perp}/\vec{S}} )$. When $\theta \ll 1$, $\vec{S}_{\perp}\approx0$, the second term can be neglected in the computation process; Baym-Peng make this approximation, but we do not.

We note that for non-relativistic neutrinos, $v/c << 1$, the spin rotation is linear in  $ m_\nu$. This is because the spin rotation scales linearly in the time spent traversing the field, hence inversely with the velocity.  But as noted above, the relic neutrino momentum distribution is fixed by cosmology so $v_\nu \sim p_0/m_\nu$.  When not otherwise specified, results given below refer to a lowest mass eigenstate value of 0.1 eV.  We also tested $\nu$ masses of $10^{-2}$eV and $10^{-3}$eV, corresponding to velocities approximately $10^{-3}$, $10^{-2}$ and $10^{-1}$ of the speed of light respectively.

\subsection{Neutrino magnetic moment}
In the (extended, to accommodate a Dirac neutrino mass) Standard Model~\cite{PhysRevLett.45.963}:
\begin{equation}
    \mu^{SM}_{\nu} \approx \frac{3G_{F}}{4\sqrt{2}\pi^{2}} m_{\nu}m_{e}\mu_{B} \approx 3\times10^{-21} \frac{m_{\nu}}{0.01eV}\mu_{B}.
    \label{SM}
\end{equation}
We also revisited the supersymmetric contribution to the magnetic moment of a massive neutrino. Based on the updated LHC limit and the supersymmetry pattern \cite{PhysRevD.28.671}, we found that the SUSY value cannot be larger than the contribution of the standard model result.

Current experimental bounds on the neutrino magnetic moment are  approximately $10^{9}$ higher than the $\mu_{\nu}$ predicted by extending the SM to include neutrino masses, with gauge-singlet right-handed neutrinos as given above in Eq.[\ref{SM}].
The experimental bound from the GEMMA experiment is among the best present bounds.  It is obtained based on analysis of the recoiling electron kinetic energy in $\nu$-electron scattering using reactor neutrinos.
With a nonzero magnetic moment, the $\nu$-electron scattering differential cross-section is a sum of the usual weak interaction cross-section ($d\sigma^{W}$/$dT_{e}$) and an electromagnetic cross-section ($d\sigma^{EM}$/$dT_{e}$) proportional to $\mu_\nu$.  These have different dependence on the recoiling electron energy $T_{e}$:  at low recoil energy($T_{e}$$\ll$ $E_{\nu}$), $\frac{d\sigma^{EM}}{dT_{e}}$ $\propto$ $T_{e}^{-1}$ while $d\sigma^{W}/dT_{e}$ remains almost constant. 
Having a low detector threshold thus increases sensitivity to a $\nu$ magnetic moment.  The GEMMA upper limit, $\mu_{\nu}^{\rm lim}$, in units of Bohr magnetons and energy threshold is given as\cite{PhysRevLett.95.151802}:
\begin{equation}
    \frac{|\mu_{\nu}^{\rm lim}|}{\mu_{B}} \approx \frac{G_{F} m_{e}}{\sqrt{2} \pi \alpha} \sqrt{m_{e} T_{e}},
\end{equation}
from which they find $\mu_{\nu}^{\rm lim} \lesssim$ 2.9 $\times$ $10^{-11} \mu_{B}$ \cite{Beda:2012zz}. In addition to reactor neutrino experiments,  
bounds have been derived by Borexino from their Phase-II solar neutrino data;
\cite{Borexino:2017fbd} shows $\mu_{\nu}$
$\lesssim$ 2.8$\times$$10^{-11}$ 
$\mu_{B}$.

\subsection{Simulation Method}

In this section, we summarize the essential elements of the simulation procedure.

Following the thermal distribution and its relationship with the neutrino total number density $n$= $\frac{1}{(2\pi)^{3}}$$\int$ f($\vec{p_0}$,$T_0$) $d^3\vec{p_0}$ $\approx$ 56.25 cm$^{-3}$ we generate velocity samples for non-relativistic cases based on:
\begin{equation}
    1=\frac{1}{n}\frac{1}{(2\pi)^{3}} \int \frac{1}{e^{m|\vec{v}|/T_{0}}+1}    4\pi m^{3} v^{2}dv.
\end{equation}
The $Astropy.units$ function is imported to track the conversion of units better.

Throughout the simulation of each neutrino trajectory, we assume that the $\nu$ momentum direction is fixed in the laboratory frame. The neutrino's helicity change depends purely on the $\vec{S}$ rotation during propagation through the astrophysical magnetic field.

The JF12 coherent Galactic Magnetic Field (GMF) model and the coherent model with random components are used in the simulation \cite{Jansson:2012pc}. The JF12 model contains large-scale regular fields, small-scale random fields, and striated random fields. The model itself was constrained by a simultaneous fit to the WMAP7 Galactic Polarized Synchrotron Emission and multiple extragalactic Faraday rotation measures.
The JF12 coherent model includes a striated component, poloidal component, toroidal halo, and spiral arm disk components. There are 8 spiral arms in the Galactic disk beyond 5 kpc. From 3-5 kpc is the purely azimuthal molecular ring.  Numerical grids of the JF12 and random field realizations used are available on the Cosmic Ray Tracking webpage~\cite{Krf}.
Field data are presented in Cartesian coordinates ranging [-20kpc, 20kpc], in units of $[\mu G]$, where the galaxy center coordinate is [0,0,0]. 
The JF12 random field realization we used has a coherence length of 30 pc. 

Given the magnetic field model, we apply the Riemann sum to compute the spin vector evolution
\begin{equation}
    \vec{S}_{N} - \vec{S}_{N-1}= 2\frac{\mu_{\nu}}{|\vec{v}|}\times(\vec{S}_{N-1}\times\vec{B}_{N-1}) \delta r,
\end{equation} 
where $\delta {r}$ = $R_{ total}$/$N_{\rm sample}$; $R_{total}$ is the propagation path length of a test neutrino coming from a known direction and $N_{\rm sample}$ is the number of steps. We ran the simulation with different step sizes $\delta$r, and the results for spin rotation and helicity-flip probability at given $\mu_{\nu}$ are well-converged. In this paper, we present our results based on the simulation with step size $\delta r$ = 10pc. 

The incoming directions of relic neutrinos are generated as pixelated data. Using the Hierarchical Equal Area isoLatitude Pixelization (HEALPix) library and the Python package $healpy$\cite{Zonca:2019vzt}, we plot values on a skymap with high resolution. The HEALPix resolution parameter (scalar integer) is set to 12288 pixels =$12\times N_{side}$; function $healpy.nside2npix$ from the package returns pixel number at $N_{side}$=32. Every pixel on the map represents a specific direction with latitude and longitude. For our purpose, each neutrino's incoming direction in the simulation is identical to each pixel on the mollweide view projection map. The area of each pixel is $\Omega_{pixel}$=$\pi/3(N_{side})^{2}$ and angular resolution is $(\Omega_{pixel})^{1/2}$ = 1.83 degrees.

We propagate $\nu$ at each pixel outward from Earth to a distance where the local field vanishes and far away from the galactic center.  Then, we reverse the neutrino momentum vector and propagate the neutrino
inward starting with relic helicity eigenstate -1. Along the propagation inward we track the spin rotation of $\nu$.
The Riemann sum process yields the final direction of $\vec{S}$; the angle difference between $\vec{S}_{final}$ and $\vec{S}_{initial}$ is calculated and the
expectation values of rotation for each pixel is computed. %based on simulation results at different velocity samples.
Magnetic moment values $\mu_{\nu}$ used are the GEMMA reactor experimental upper limit $\mu_{\nu}$$<$ 2.9 $\times$ $10^{-11}$ $\mu_{B}$ and $10^{-1}$, $10^{-2}$ and $10^{-3}$ times smaller cases. The $\nu$ mass is assigned to be 0.1eV during most of the simulations; for a given mass, refer to Eq.~\eqref{fermi} to have the $\nu$ velocity. 

\begin{figure*}[!htp]
\hspace*{-6.1em}
\subfloat{%
\includegraphics[height=37mm,width=56mm]{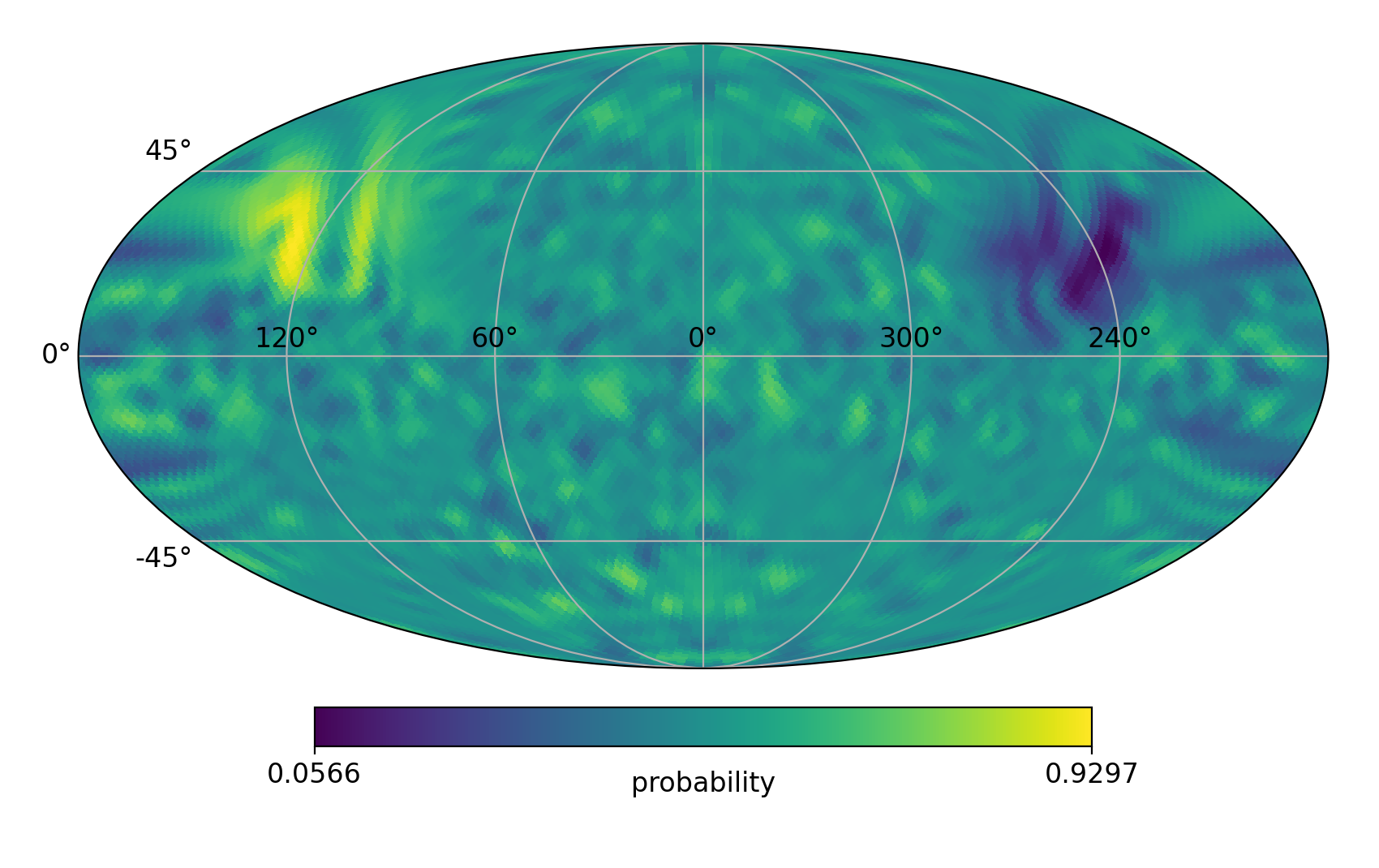}%
}\hspace*{-0.9em}
\subfloat{%
\includegraphics[height=37mm,width=56mm]{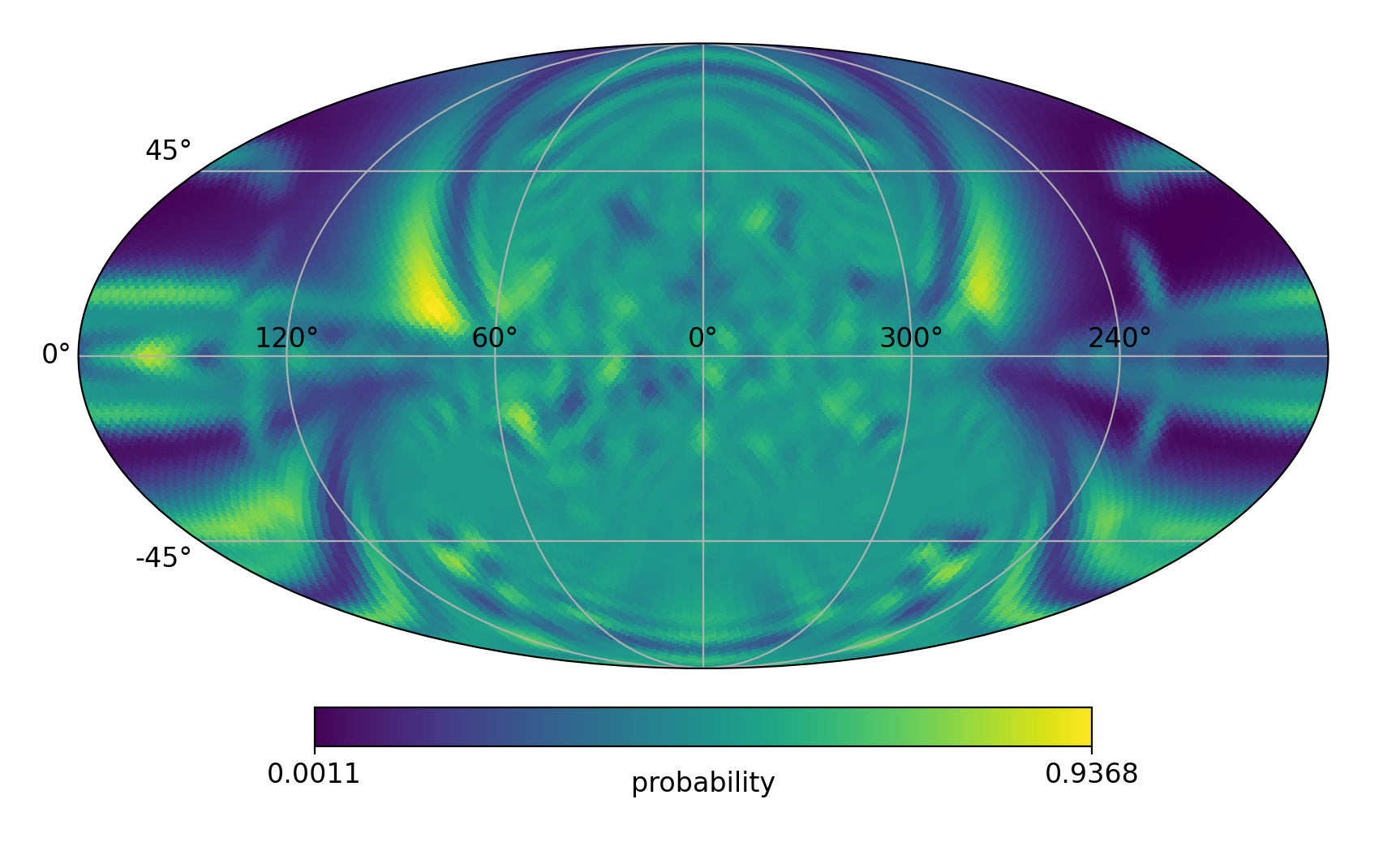}%
}\hspace*{-0.9em}
\subfloat{%
\includegraphics[height=37mm,width=56mm]{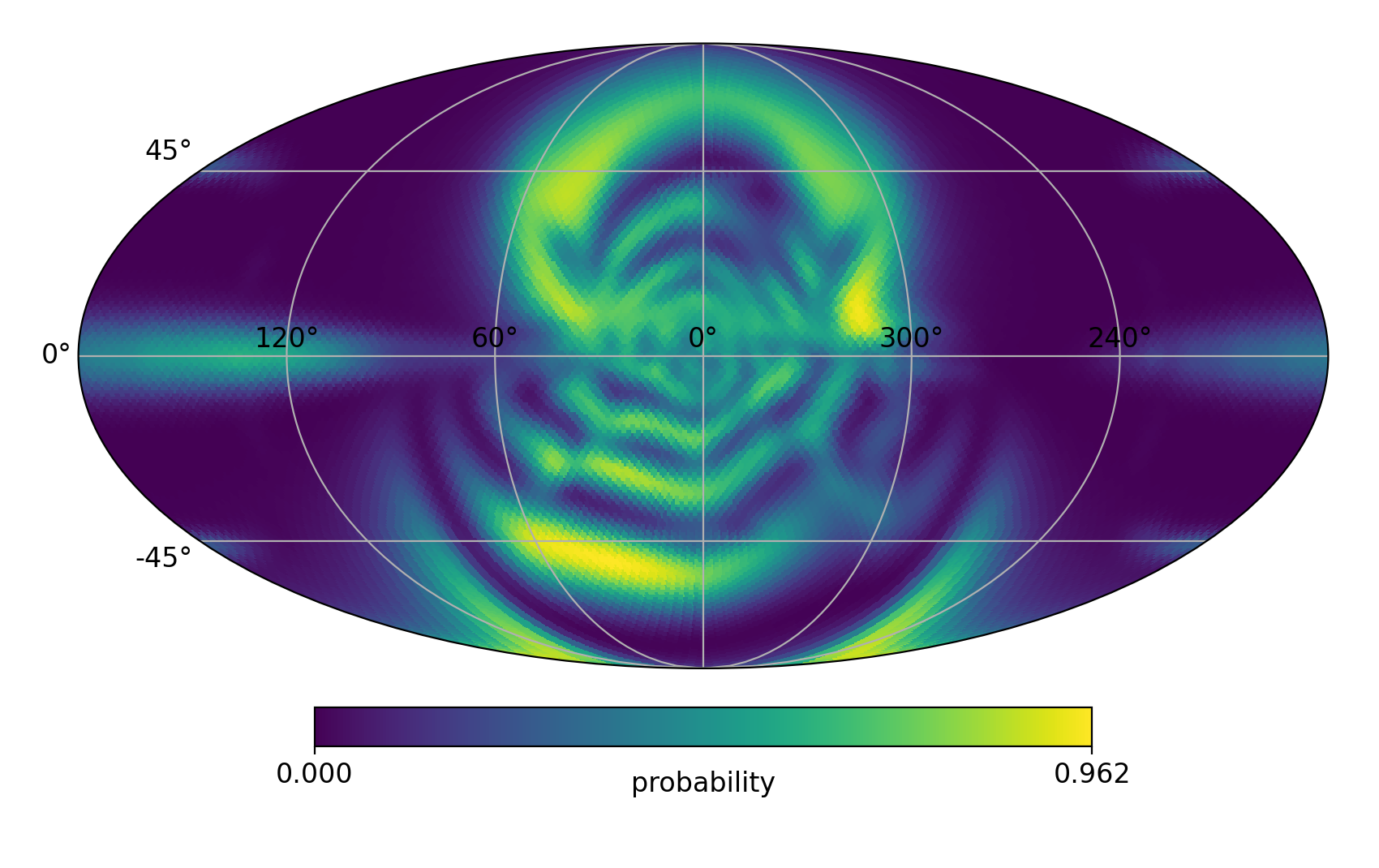}%
}\hspace*{-0.9em}
\subfloat{%
\includegraphics[height=37mm,width=56mm]{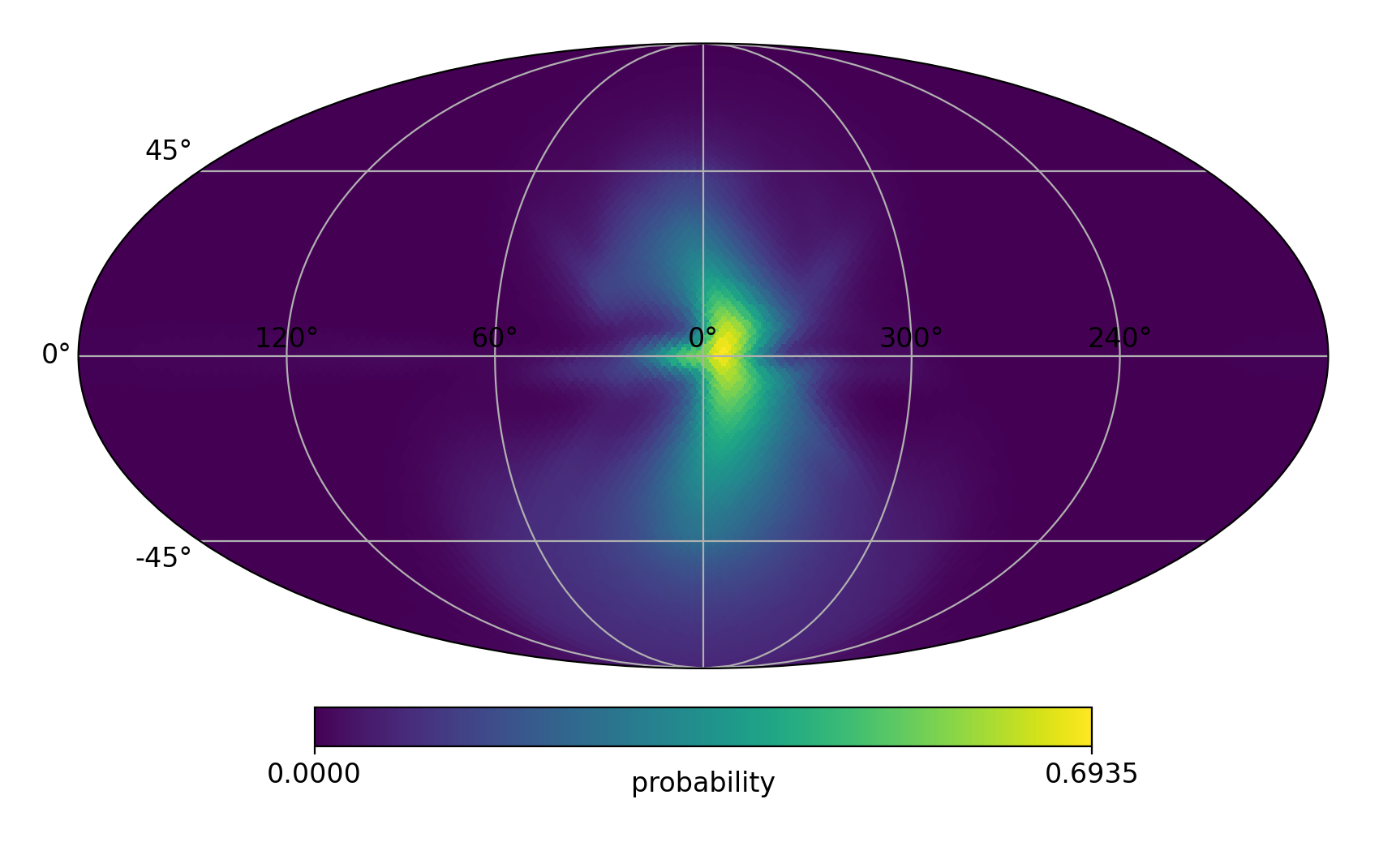}%
}\hfill
\hspace*{-6.1em}
\subfloat{%
\includegraphics[height=37mm,width=56mm]{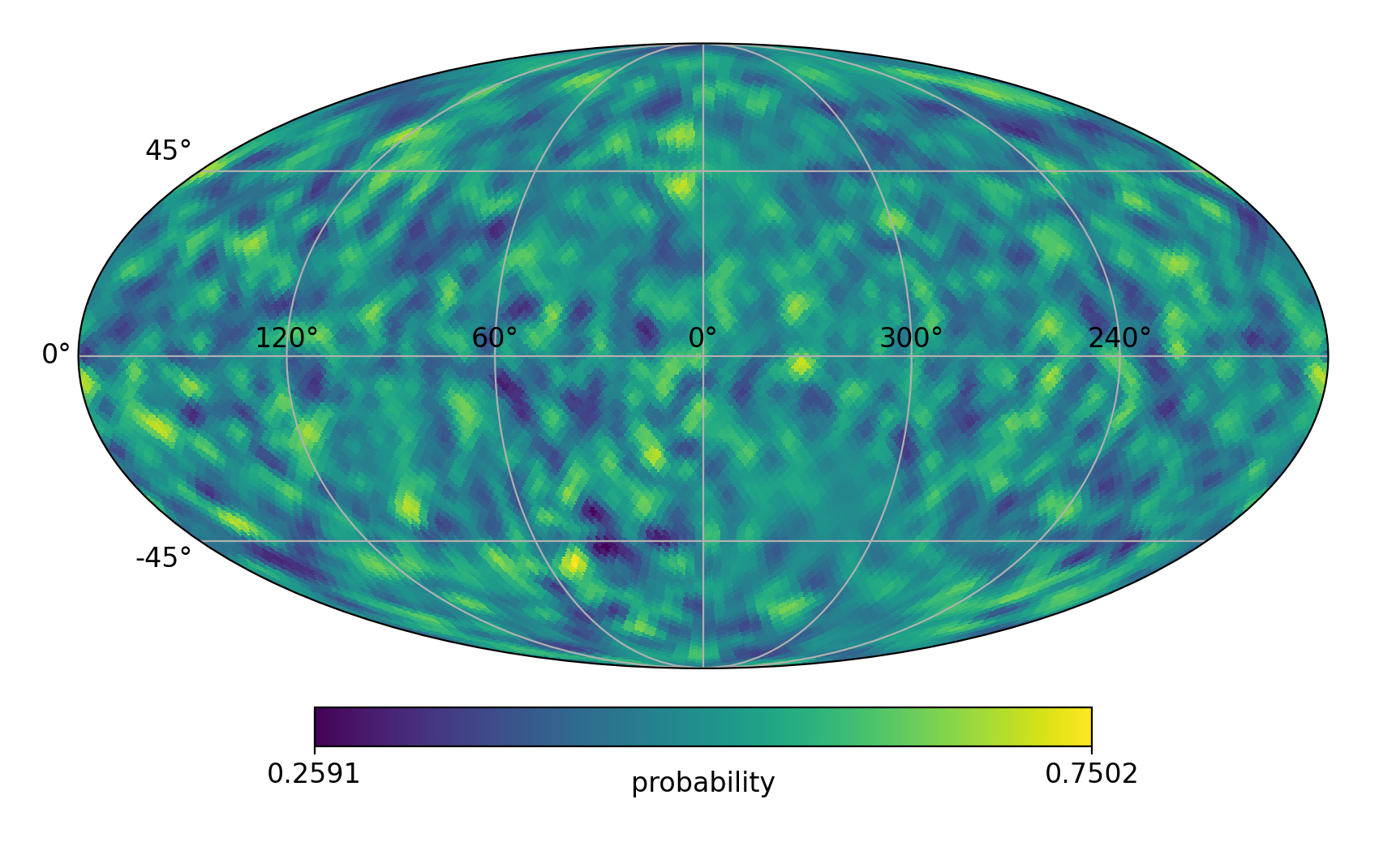}%
}\hspace*{-0.9em}
\subfloat{%
\includegraphics[height=37mm,width=56mm]{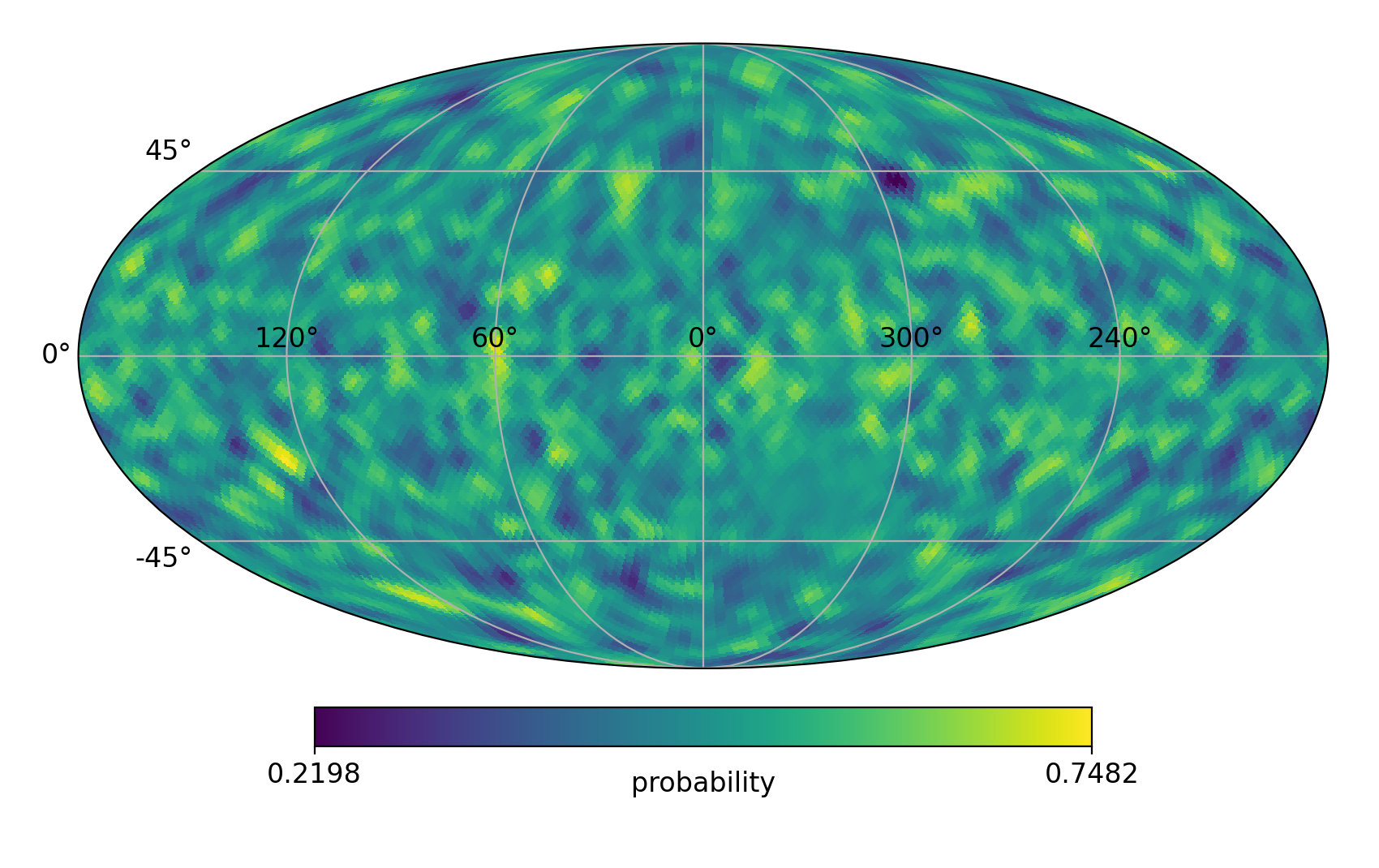}%
}\hspace*{-0.9em}
\subfloat{%
\includegraphics[height=37mm,width=56mm]{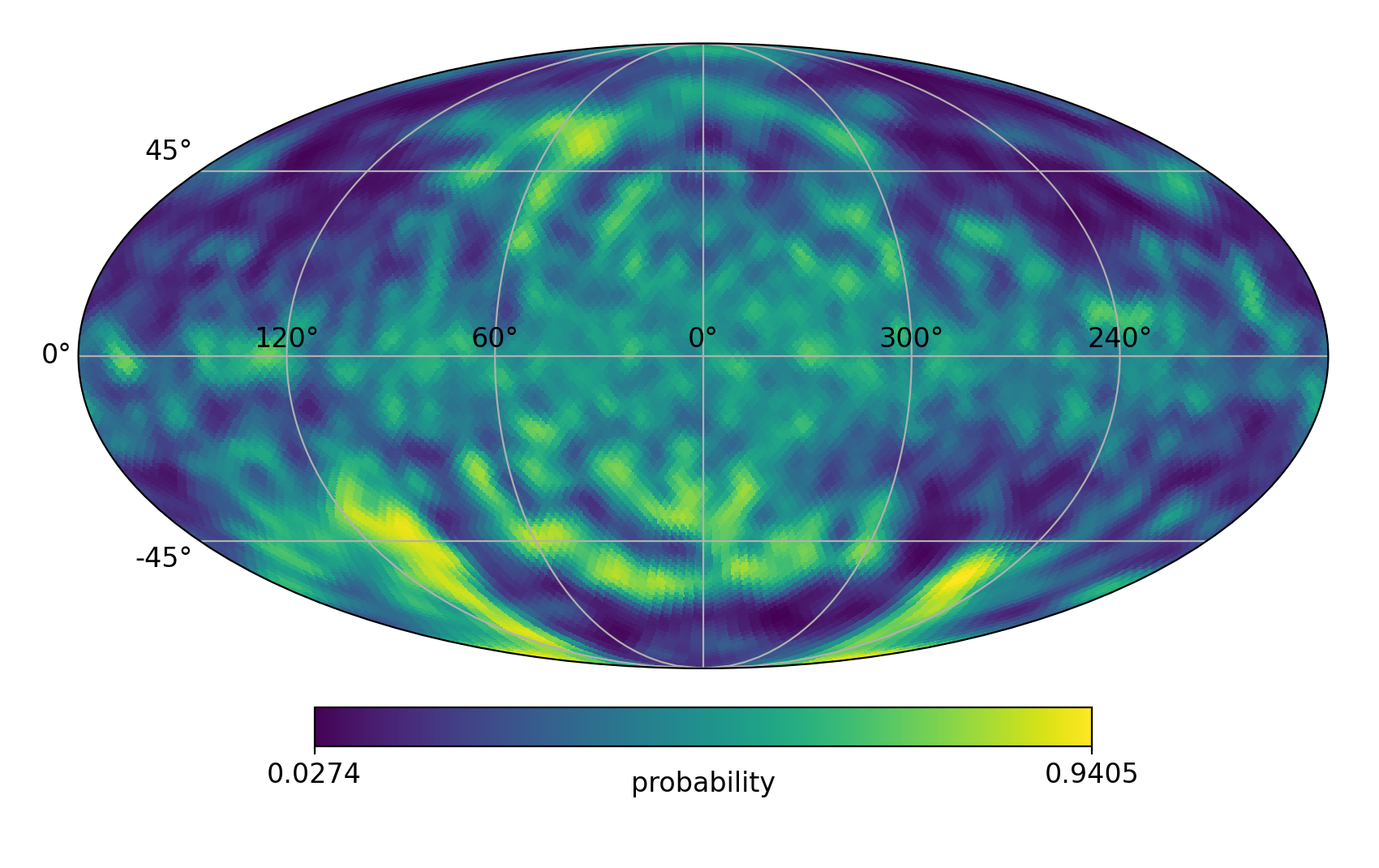}%
}\hspace*{-0.9em}
\subfloat{%
\includegraphics[height=37mm,width=56mm]{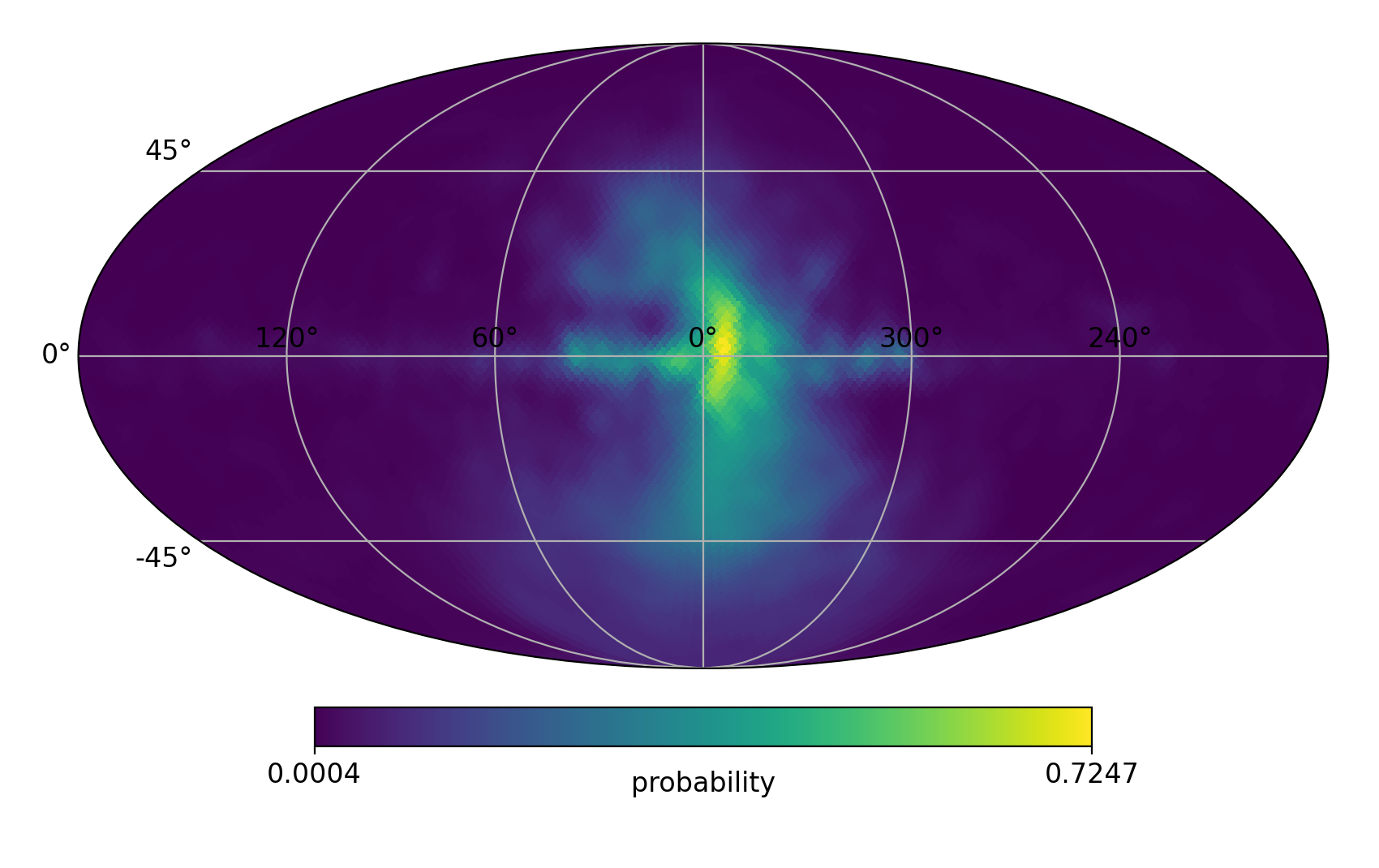}%
}
\caption{Relic Neutrino Helicity-Flip Probability in Molleweide view, for the same cases shown in Appendix.A Fig. \ref{rotation A}. 
(first row) the JF12 coherent Galactic magnetic field and (second row) the full coherent plus random field.
}
\label{probability flip}
\end{figure*}

\section{Discussion of the results}

We show in Fig. \ref{probability flip} and Appendix.A Fig. \ref{rotation A}, two different representations of the helicity change  of an $m=0.1$ eV neutrino as a function of arrival direction.  Appendix.A Fig. \ref{rotation A}, made with Healpy Projview, shows the net rotation angle in radians, in a sky projection centered on the North Galactic pole and with the South pole on the rim.  Fig. \ref{probability flip} displays the helicity flip probability using the Molleweide projection, in which the Galactic center (longitude L = 0 and colatitude B = 0) is in the center of the plot and the N and S Galactic poles are at the top and bottom, respectively.  From left to right, the plots show decreasing $\mu_\nu$ starting with the maximum value allowed by the GEMMA bound and decreasing progressively by factors of 10.

The upper row of both figures is for the coherent field only and the lower row is for the more realistic case with random field as well.  The strength of the JF12 random field is probably larger than the true random field by a factor-few, due to the JF12 analysis being based on the original WMAP synchrotron intensity maps which have since been revised to take better account of contamination by dust contributions, thus the real pattern of helicity rotations can be expected to be somewhere in between these two sets of plots.  

For an incoming neutrino at the upper bound value of magnetic moment, the total net deviation in spin angle between entering the Galaxy and arriving at Earth ranges from $\simeq$ 0.40 to $\simeq$ 2.64 radians in the coherent field model and $\simeq$  1.01 to $\simeq$ 2.18 radians in the field model including the random field.  In fact, the spin rotates multiple times for many arrival directions when the moment is within two orders of magnitude of the GEMMA limit. However, at $10^{-3}$ $\times$ $\mu_{\nu}^{\rm lim}$, the rotation angles are in general less than $\pi/2$. 

A large-scale directional anisotropy pattern is visible in the coherent field model for magnetic moments $\lesssim 0.1$ times the experimental magnetic moment upper bound, with the directions of high helicity-flip probability being toward the Galactic center.  However, for the GMF model including the random components, large-scale patterns in the flip probability start to be observable only at or below $10^{-2}$ $\times$ $\mu_{\nu}^{\rm lim}$. 
For larger magnetic moments the rotation is so large that the neutrino is in a nearly equal superposition of helicity eigenstates over most of the sky. The angular power spectrum is shown in Appendix.A Fig. \ref{power} to quantify how the angular scale of the flip-probability anisotropy depends on the magnetic moment.  We used the $healpy$ functions $anafast$ to perform these calculations.

Overall, the plots of  Fig.\ref{probability flip} and Appendix.A Fig. \ref{rotation A} show the expected feature that directions with longer propagation paths and those which traverse regions of stronger fields have a higher helicity-flip probability -- hence the noticeable difference in the Galactic and antigalactic center directions.  The other important general features are that for a large magnetic moment or relatively stronger random field, there is less large-scale structure in the helicity flip and effectively half the neutrinos have a flipped spin.  A forthcoming analysis of the GMF with the latest data will remove most of the uncertainty in these predictions due to GMF uncertainties~\cite{uf23}.  Then, if a large scale pattern in the helicity flip probability were observed, that would constrain the value of the magnetic moment to within an order-of-magnitude range. For magnetic moment below or above this special range, the overall average probability flip would be small or 1/2, respectively, thus providing an upper or lower bound on the magnetic moment.  In spite of such large differences in the flip probability, the actual impact on observables is very small, as discussed below in Sec.~\ref{ITBD}, so for now this is a futuristic aspiration.

\begin{figure}[h!]
\hspace{-5em}
    %\centering
    \includegraphics[width=10.5cm,height=8.2cm]{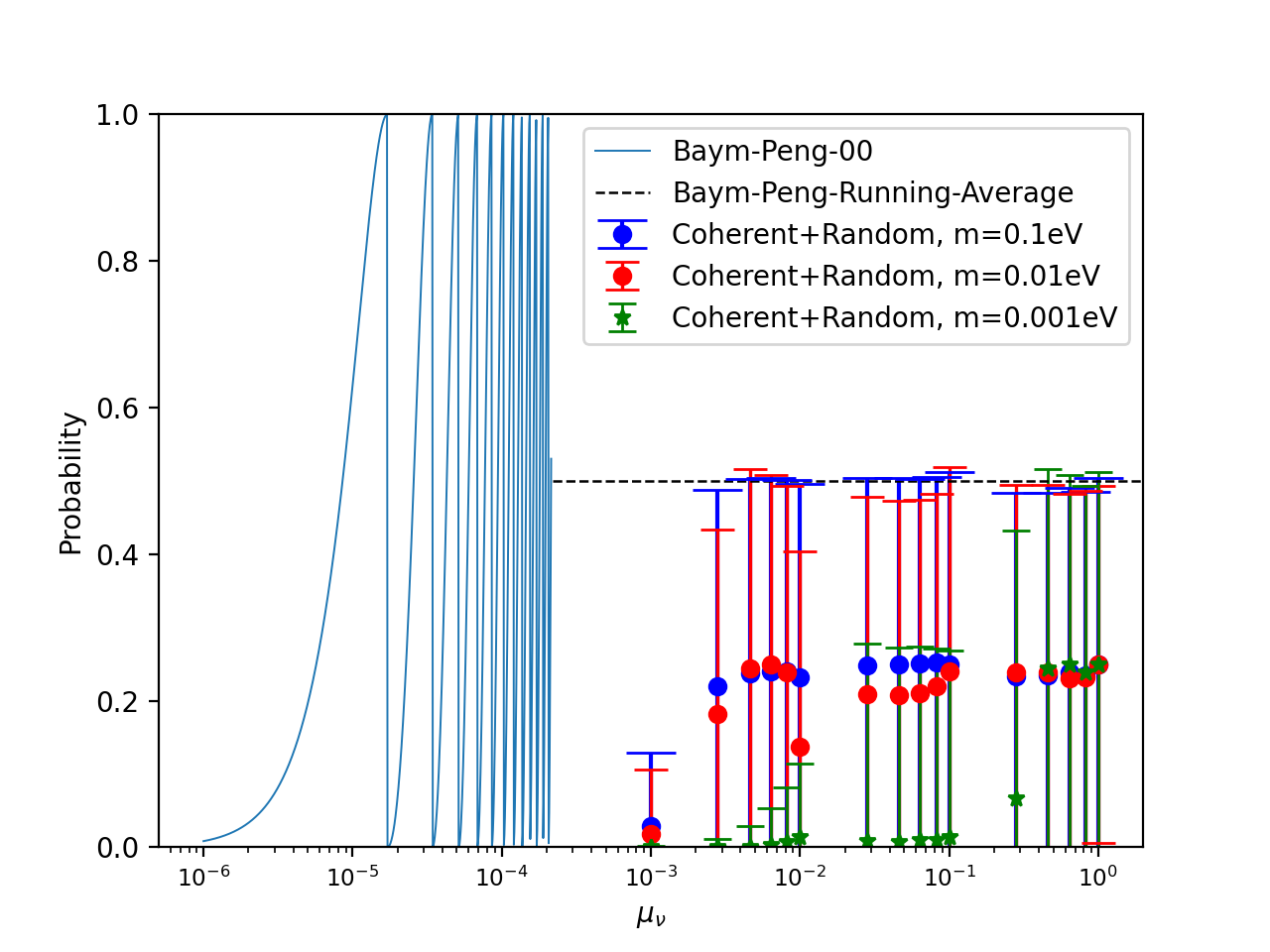}
    \caption{The sky-averaged helicity flip probability in the JF12 Coherent and Coherent + Random Magnetic Field models, for $m_{\nu}$=0.1 eV, 0.01eV and 0.001 eV
    (blue, red and green data points, respectively, with RMS spread on the sky shown as error bar), overlaid on the prediction using the Baym-Peng approximation. For $\mu_{\nu}$ $\gtrsim$ $10^{-4}$ $\mu_{\nu}^{\rm lim}$, the oscillation in the Baym-Peng model is so rapidly varying, we show only the average value $\frac{1}{2}$.}
    \label{Baym and I}
\end{figure}

\section{Comparison with earlier work}

The Baym-Peng treatment~\cite{Baym:2021ksj} took the
Galactic magnetic field to be entirely random with magnitude $B_{g}$ $\simeq$ 10 $\mu$Gauss, with a coherence length $\Lambda_{g}$ $\simeq$ kpc within a mean crossing distance of the galaxy of $l_{g}$$\simeq$ 16 kpc. In a uniform random field the spin vector of a neutrino undergoes a random walk in the field during propagation, generating -- in the limit of small rotations in each patch -- a mean-square spin rotation \cite{Baym:2021ksj}:
\begin{equation}
    <\theta^2>\simeq (2\mu_{\nu}B_{g}\frac{\Lambda_{g}}{v})^{2} \frac{l_{g}}{\Lambda_{g}}.
     \label{eq:BP}
\end{equation}
%For ($\mu_{\nu}$$B_{g}$$\frac{\Lambda_{g}}{v}$) of order unity and sufficiently large magnetic moment, this predicts cumulative rotation angles large enough for significant $\nu$ helicity-flip.  
For $m_\nu = 0.1$ eV,  the Baym-Peng estimate predicts spin rotation which is generically large so the flip probability is of order one, when $\mu_{\nu} \gtrsim 2 \times 10^{-6}$ $\mu_{\nu}^{\rm lim}$, as shown in FIG. \ref{Baym and I}.  By contrast, we find the maximum sky-averaged flip probability is about 20\%, and becomes very small for $\mu_{\nu} \leq 10^{-3}$ $\mu_{\nu}^{\rm lim}$.

\begin{figure*}[htp!]
\hspace*{-2.2em}
\subfloat[\label{sfig:testa}]{%
\includegraphics[height=70mm,width=95mm]{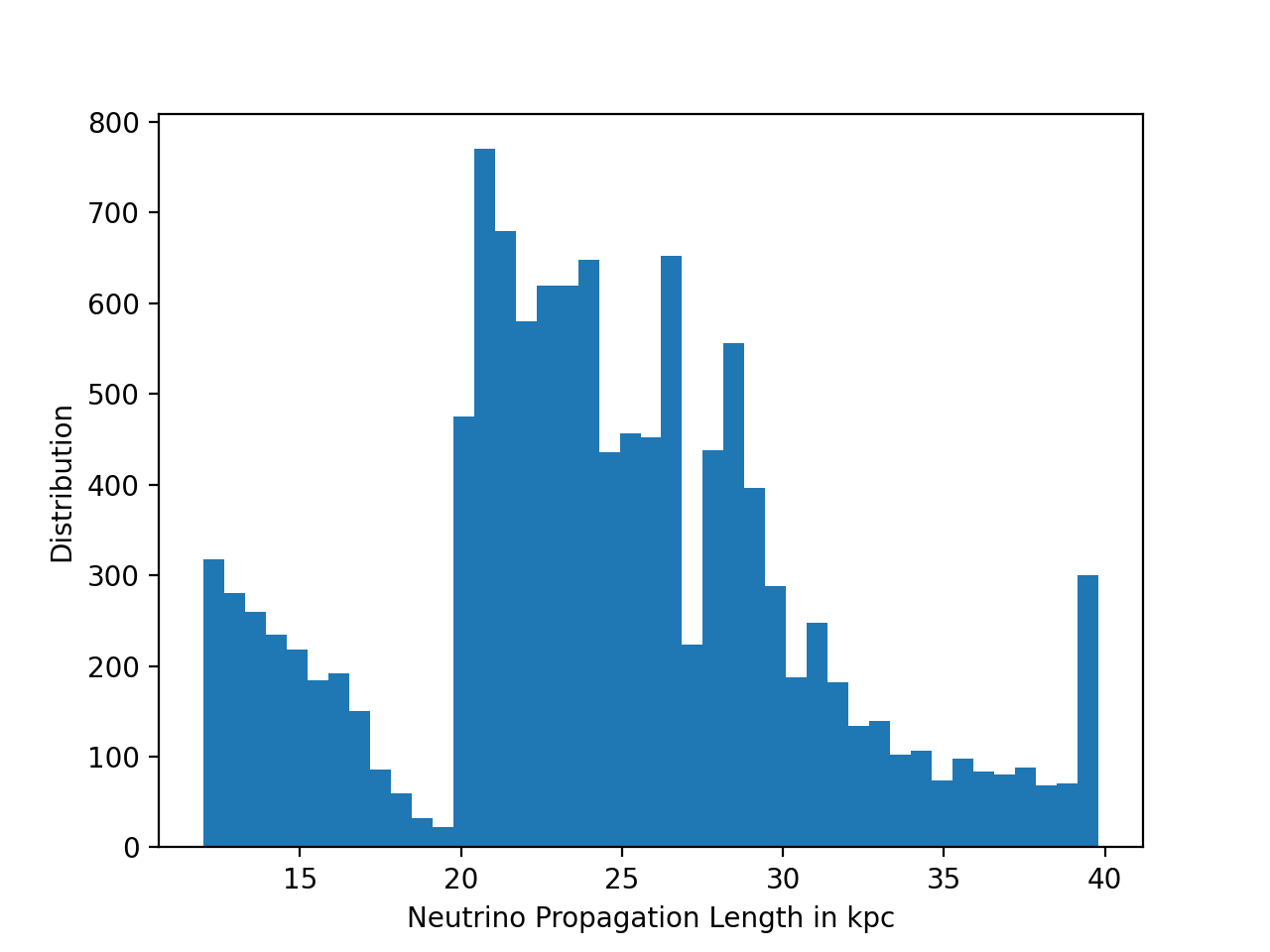}%
}\hspace*{-0.7em}
\subfloat[\label{sfig:testa}]{%
\includegraphics[height=70mm,width=95mm]{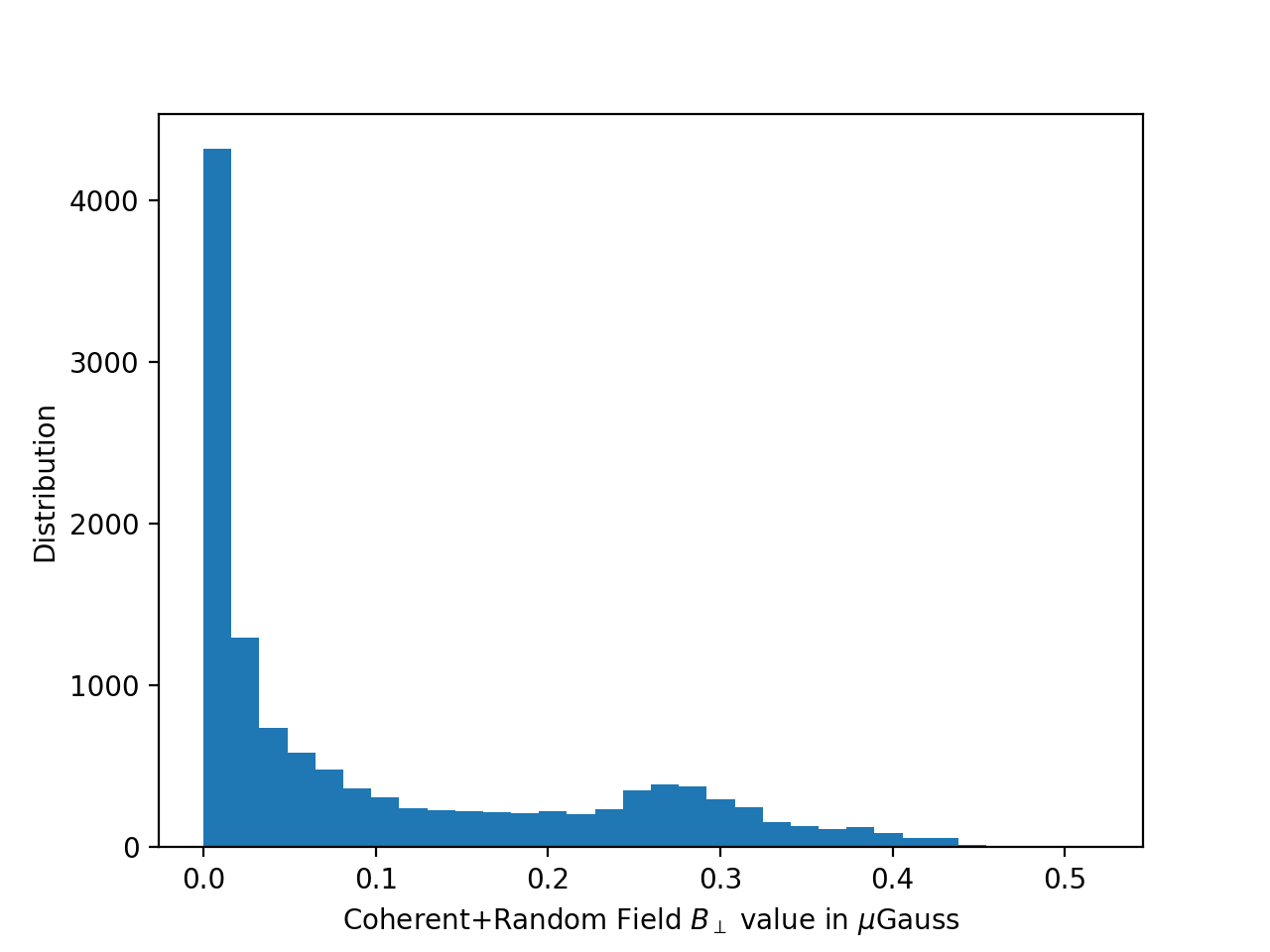}%
}
\caption{Histograms of (a) neutrino propagation length $l_g$  and (b) $|B_{\perp}|$ for all the lines of sight in JF 12 coherent plus random magnetic field model.}
\label{fig:GMFhists}
\end{figure*}

Much of the discrepancy between our results and those of~\cite{Baym:2021ksj} is due to the unrealistic field parameters assumed in~\cite{Baym:2021ksj}.  Averaging over the 20 kpc cube centered on the Galactic center in which the Jansson and Farrar (2012) model is given, the average path length to Earth is 24.4 kpc and the average value of $|B_\perp|$ is $ 0.104 \,\mu$G. (Histograms of $|B_\perp|$ and $l_g$ for all the lines of sight are shown in Fig.~\ref{fig:GMFhists}.) 

Moreover, the coherence length of the random component of the GMF is typically estimated to be  10 to 100 pc rather than the 1 kpc adopted by~\cite{Baym:2021ksj}; for our calculations we took it to be 30 pc.  Using our more realistic magnetic field properties in the Baym-Peng expression Eq.~\eqref{eq:BP}, decreases their estimated $<\theta^2>$ by a factor $ \sim 5 \times 10^{-6}$.

\section{Effects of helicity flip on the detection of Relic Neutrinos through Inverse Tritium Beta Decay}
\label{ITBD}
\subsection{Total ITBD event rate}
A proposed experimental method for detecting relic neutrinos is based on neutrino capture through the inverse tritium beta decay process 
$\nu_{e}$ +${}^{3}H$ $\rightarrow$ ${}^{3}He$ $+$ $e^{-}$~\cite{PhysRev.128.1457}\cite{Peng:2022nvi}. 
The cross-section times neutrino velocity for capturing $\nu_{e}$ of helicity $h$ is the mixing-matrix-weighted sum over the contribution of each mass eigenstate which is\cite{Peng:2022nvi}
\begin{equation}
\begin{aligned}
\sigma^{h}(E_{\nu})v= {} &
\frac{G_{F}^2}{2\pi}|V_{ud}|^{2} F(Z,E_{e})\frac{m_{{}^{3} He}}{m_{{}^{3}H}} E_{e}P_{e}\times \\
&(\langle f_{F} \rangle^{2} + (g_{A}/g_{V})^{2} \langle g_{GT} \rangle^{2} ) A^{h},
\end{aligned}
\end{equation}
where $v$ is the neutrino velocity, $V_{ud}$ is the up-down quark element of the CKM matrix and $F(Z,E_{e})$ is the Fermi
Coulomb correction.  From Eq~\eqref{fermi}, the magnitude of the neutrino momentum is of order $10^{-4}$ eV/c so unless $m_
\nu \lesssim 10^{-4}$ eV, even the least massive neutrino is non-relativistic.
The helicity-dependent factor $A^h \equiv (1 - \hat{s}\cdot \vec{v}/c)$ is: 
\begin{equation}
\label{A:factor}
A^{\pm}=1\mp\beta,
\end{equation}
$+$ for right-hand helicity eigenstate and $-$ for left-hand helicity eigenstate.  The helicity-averaged cross-section is
$\overline{\sigma} \simeq$ 3.834 $\times$$10^{-45}$ cm$^{2}$.  One sees that 
if all mass eigenstates are heavy enough such that $\beta$ $\ll$ 1, the cross-sections for different helicity states are practically indistinguishable so the impact of spin rotation by the GMF is difficult to observe in the rate. 

\begin{figure*}[htp!]
\hspace*{-2.9em}
\subfloat[\label{sfig:testa}]{%
\includegraphics[height=70mm,width=101mm]{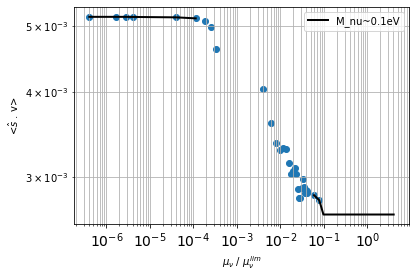}%
}\hspace*{-0.7em}
\subfloat[\label{sfig:testa}]{%
\includegraphics[height=69.5mm,width=94mm]{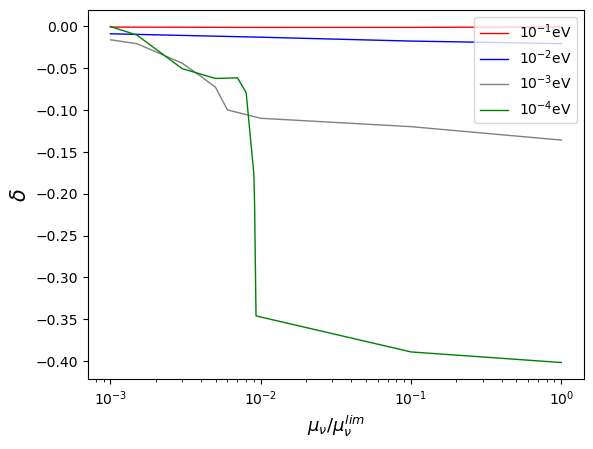}%
}
\caption{ (left) Arrival-direction-averaged value of  $<$$\hat{s}_{\nu}$$\cdot$$\vec{v}_{\nu}$$>$ and (right) $\delta$  versus $\mu_{\nu}/\mu_{\nu}^{\rm lim}$ ranging from the standard model magnetic moment $\mu_{\nu}^{SM}$ $\sim$ $10^{-18}$ $\mu_{B}$ (left) and $10^{-3}$ (right) to the experimental upper bound $\mu_{\nu}^{\rm lim}$$\approx$$2.9\times10^{-11}\mu_{B}$, in the JF12 coherent+random components GMF. As shown in the text, $<$$\hat{s}_{\nu}$ $\cdot$ $\vec{v}_{\nu}$$>$ is independent of neutrino mass for non-relativistic neutrinos.}
    \label{masses}
\end{figure*}

One might think since $v_\nu \sim p_0/m$ for non-relativistic neutrinos, that a lower neutrino mass and hence larger $<v>$, would imply larger sensitivity to the spin rotation via $<\hat{s}_{\nu}$$\cdot$$\vec{v}_{\nu}>$.  However, the spin rotation is proportional to the time spent traversing the GMF and is thus inversely proportional to the velocity when non-relativistic, so that $<\hat{s}_{\nu}$$\cdot$$\vec{v}_{\nu}>$ is independent of neutrino mass if the rotation angle does not wrap. We show in FIG. \ref{masses} the variation of $<\hat{s}_{\nu}$$\cdot$$\vec{v}_{\nu}>$ as $\mu_{\nu}$ ranges from standard model magnetic moment to experimental upper bound taking $m_{\nu}$= 0.1eV.

The event rate in an ITBD experiment is proportional to $\phi^+ \sigma^+ + \phi^- \sigma^-$, where $\phi^{-}$ and $\phi^{+}$ are the flux of negative and positive helicity neutrinos:
\begin{equation}
  \phi^- = \phi_{\rm relic} P_{\rm nf};~~\phi^+ = \phi_{\rm relic} P_{\rm f} 
\end{equation}
where $P_{\rm f}$ is the probability that the neutrino helicity flipped prior to reaching the detector and $P_{\rm nf} = 1 - P_{\rm f}$ is the probability that the neutrino arrives with its original negative helicity.  As for the computation of the spin rotation, we take the coherent mixing-weighted sum over the contribution of the mass eigenstates.  To simplify the discussion, focus on the contribution of the lightest mass eigenstate, $m_1$; it has a significant projection on $\nu_e$ so since $m_1$ may be much smaller than the others, its $\beta$ may be large. Substituting into the expressions, we have 
\begin{equation}
\label{veldep}
    {\rm Rate} \sim \beta \,\bar{\sigma} \, [1 + \beta - 2 \beta \,P_{\rm f}]~.
\end{equation}

As one would expect, helicity-flip decreases the ITBD rate. In the absence of helicity-rotation, the sky-averaged ITBD rate scales as $ \beta \bar{\sigma}(1 + \beta)$, while with fully-randomized helicities the rate scales as $\beta \bar{\sigma}$. With a nonzero helicity flip probability and a neutrino mass small enough that $\beta$ $\simeq$ 1, the rate is simply proportional to 1-$P_{f}$.

As emphasized above, the velocity-weighted spin rotation is independent of neutrino mass as long as the neutrino is non-relativistic.  However the detection probability depends on the velocity-weighted helicity flip probability, which is non-linear in the spin rotation.  Defining $\delta$ to be the fractional change in the total ITBD detection rate due to helicity rotation, from Eq.~\eqref{veldep} we have
\begin{equation}
    \delta \equiv \frac{ -2 \beta \, P_{\rm f}}{1 + \beta} ~. 
\end{equation}

%One might think that when $\beta$ is large, i.e., $m_1 \lesssim 10^{-3.5}$ eV, it would be easiest to see the impact of a neutrino magnetic moment on the event rate in an ITBD experiment, but that is not the case. If $\mu_\nu$ has a value such that the spin rotation angle is significant but small compared to $2 \pi$, then $\beta \,P_{\rm f}$ is independent of the mass of the neutrino because $\beta \sim 1/m $ while $P_{\rm f}\sim m$.  Thus to reiterate, unless the mass of the lightest neutrino eigenstate is $\lesssim 10^{-4}$ eV and relativistic corrections must be taken into account, the event rate as a function of neutrino mass is -- for fixed magnetic moment -- essentially independent of mass because the neutrino momentum spectrum is fixed by cosmology.
\begin{figure*}[t!]
\hspace*{-6.2em}
\subfloat{%
\includegraphics[height=39mm,width=55mm]{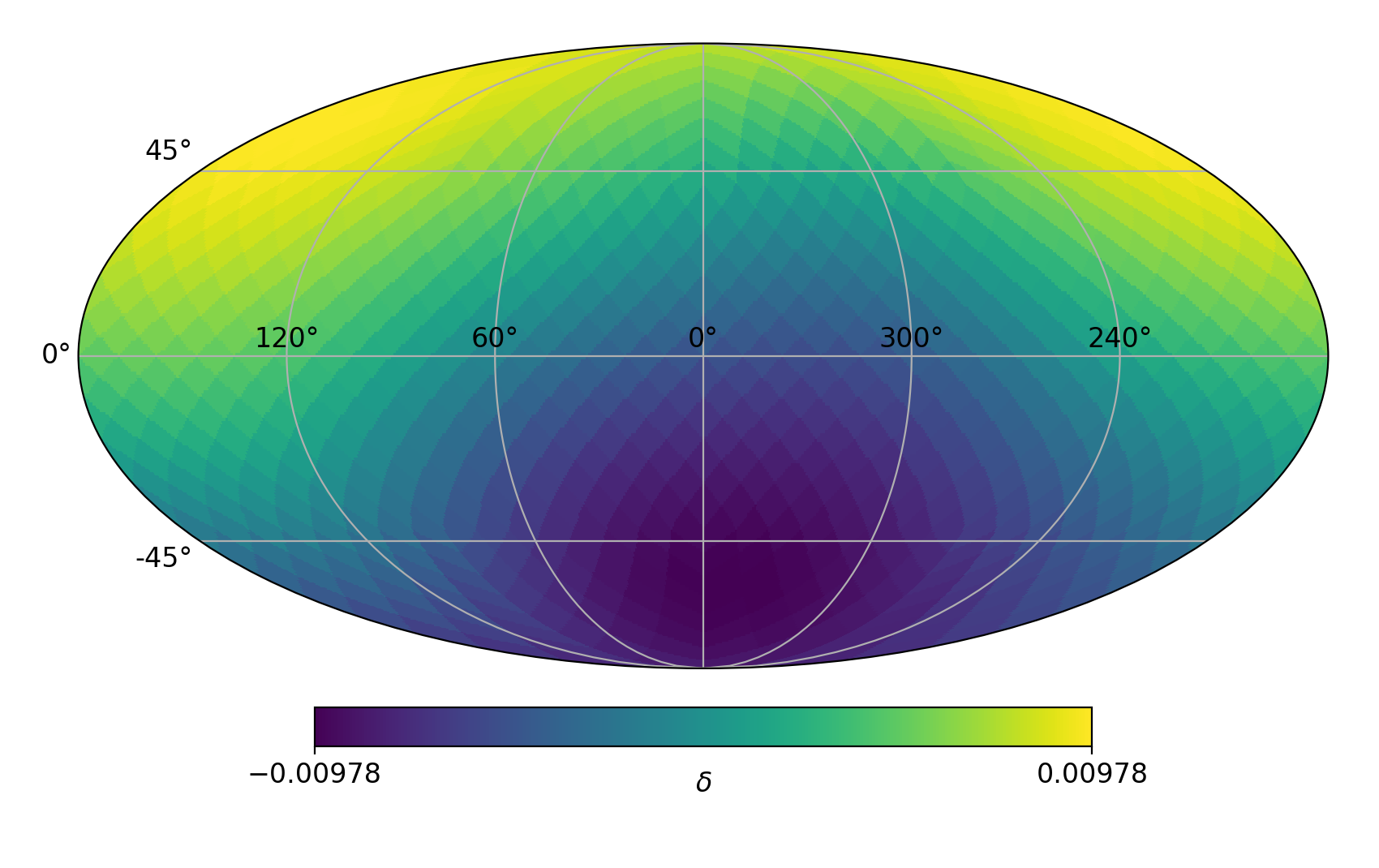}%
}\hspace*{-0.7em}
\subfloat{%
\includegraphics[height=39mm,width=55mm]{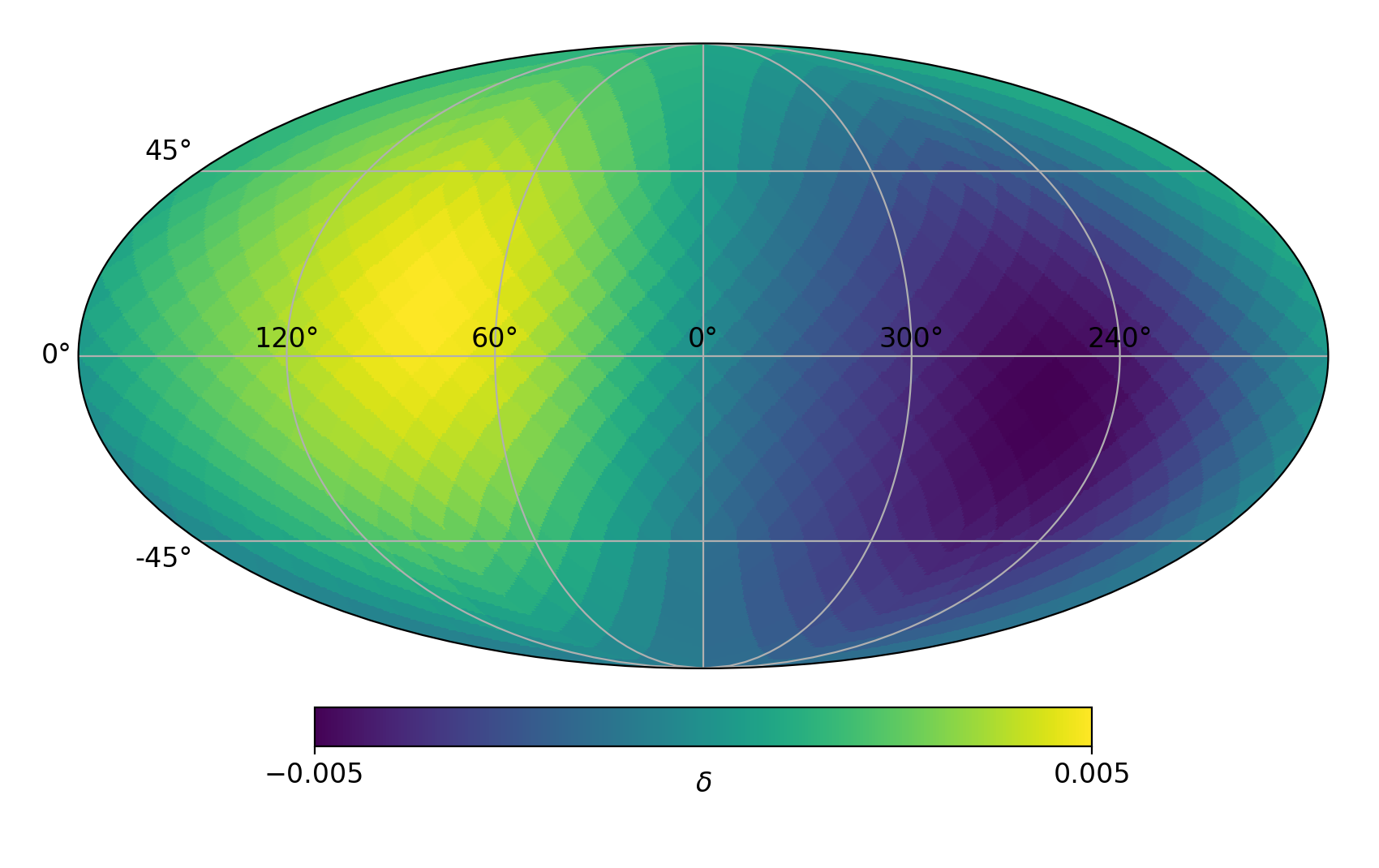}%
}\hspace*{-0.4em}
\subfloat{%
\includegraphics[height=39mm,width=55mm]{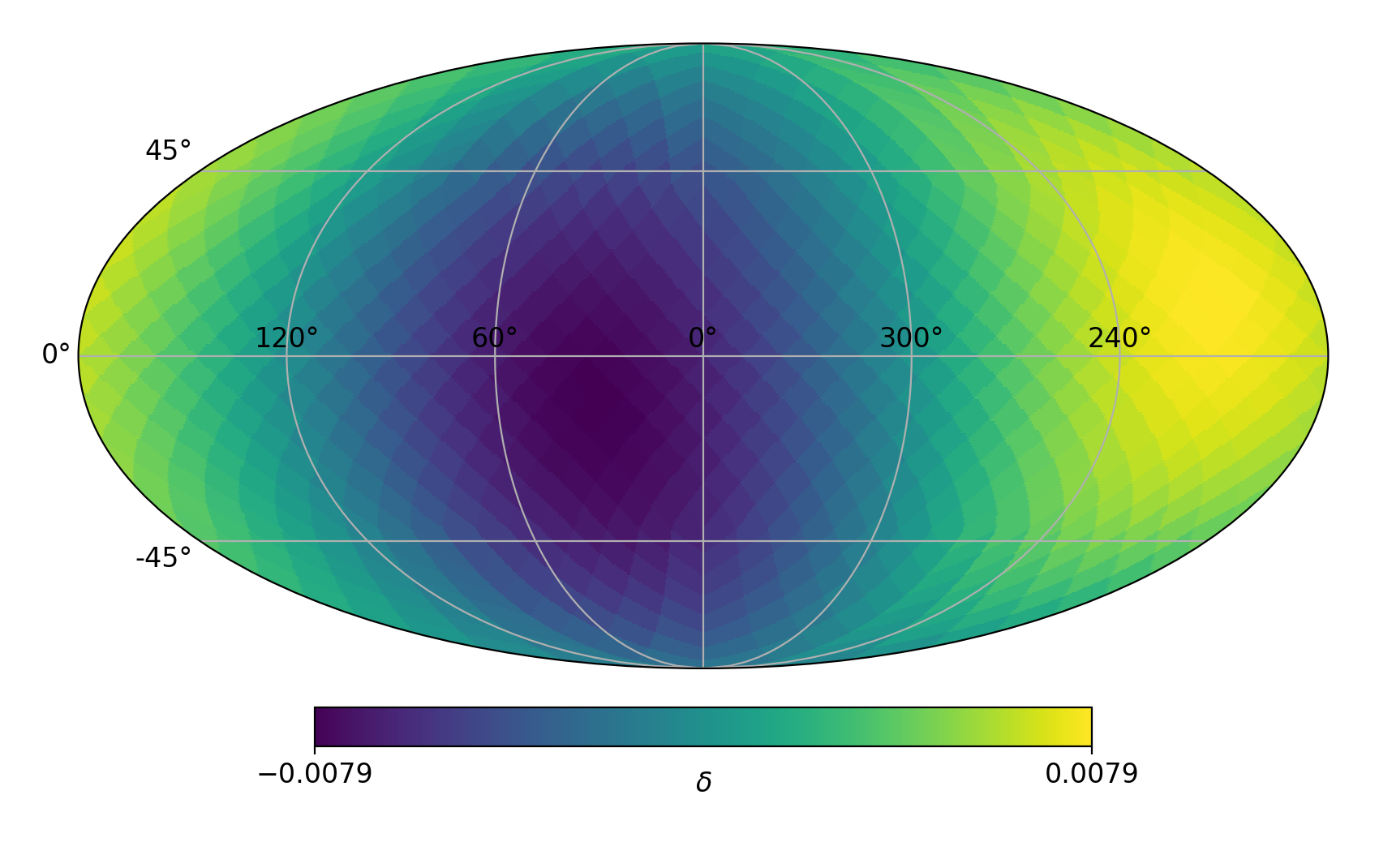}%
}\hspace*{-0.4em}
\subfloat{%
\includegraphics[height=39mm,width=55mm]{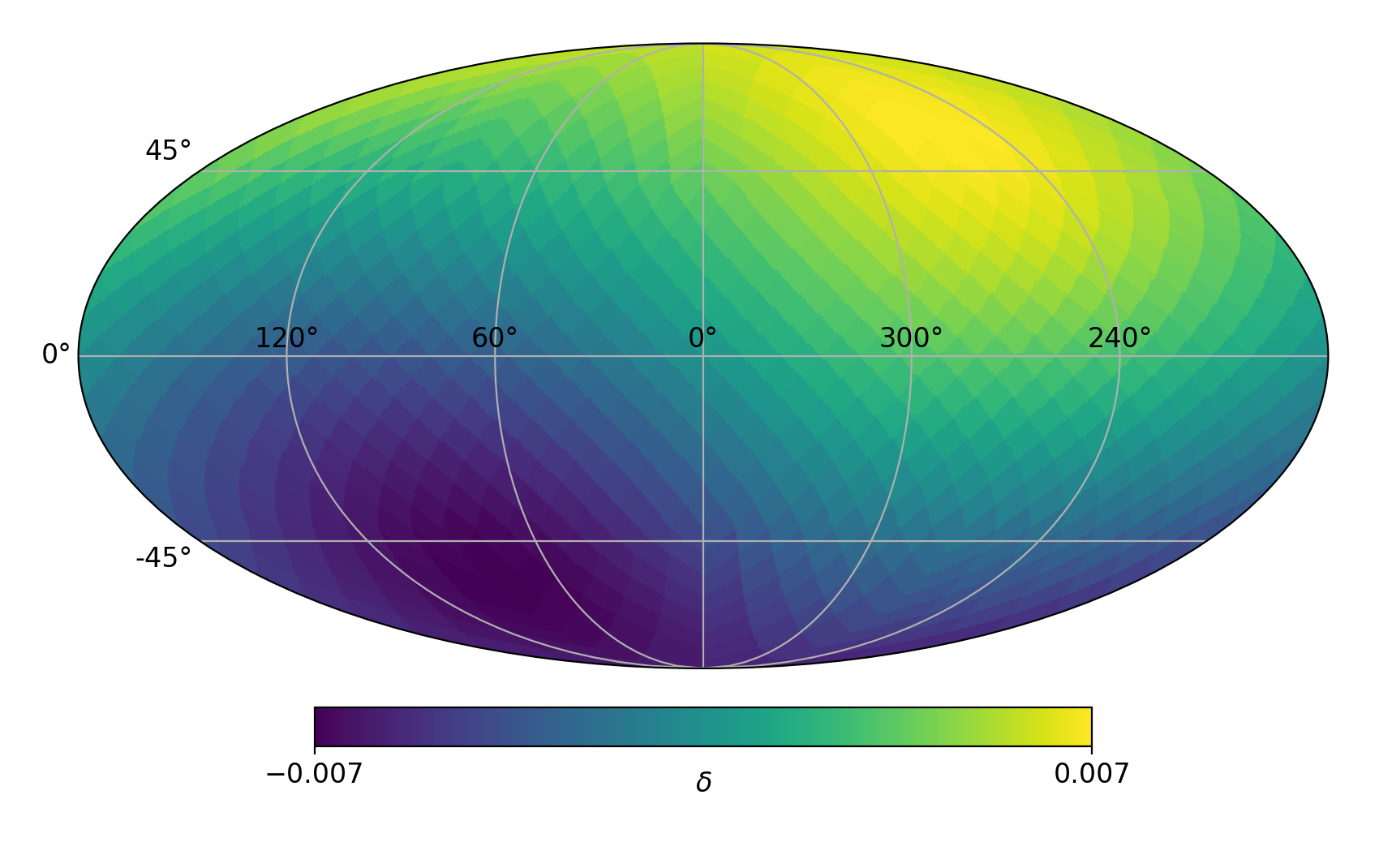}%
}\hfill
\hspace*{-6.2em}
\subfloat{%
\includegraphics[height=39mm,width=55mm]{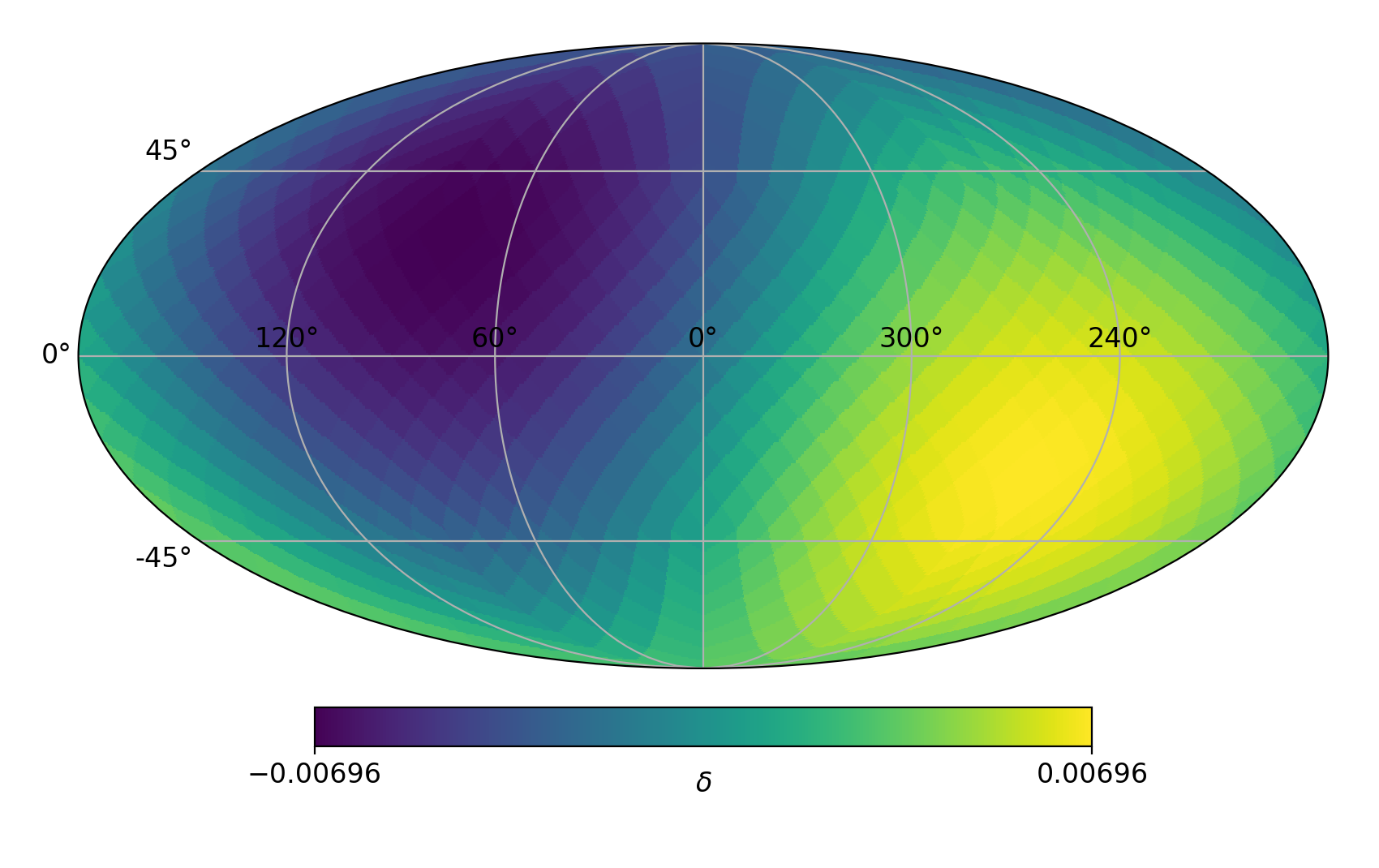}%
}\hspace*{-0.7em}
\subfloat{%
\includegraphics[height=39mm,width=55mm]{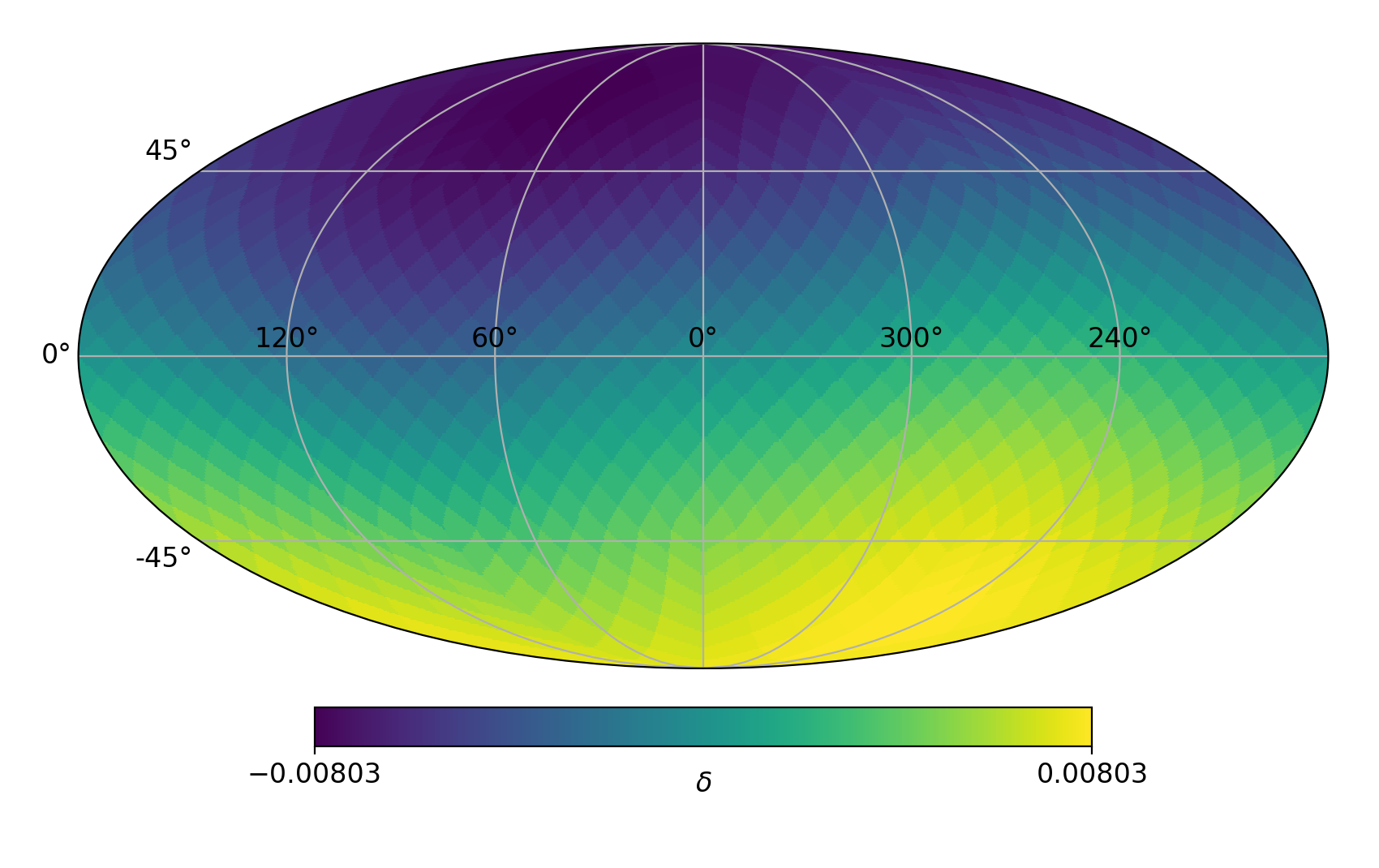}%
}\hspace*{-0.4em}
\subfloat{%
\includegraphics[height=39mm,width=55mm]{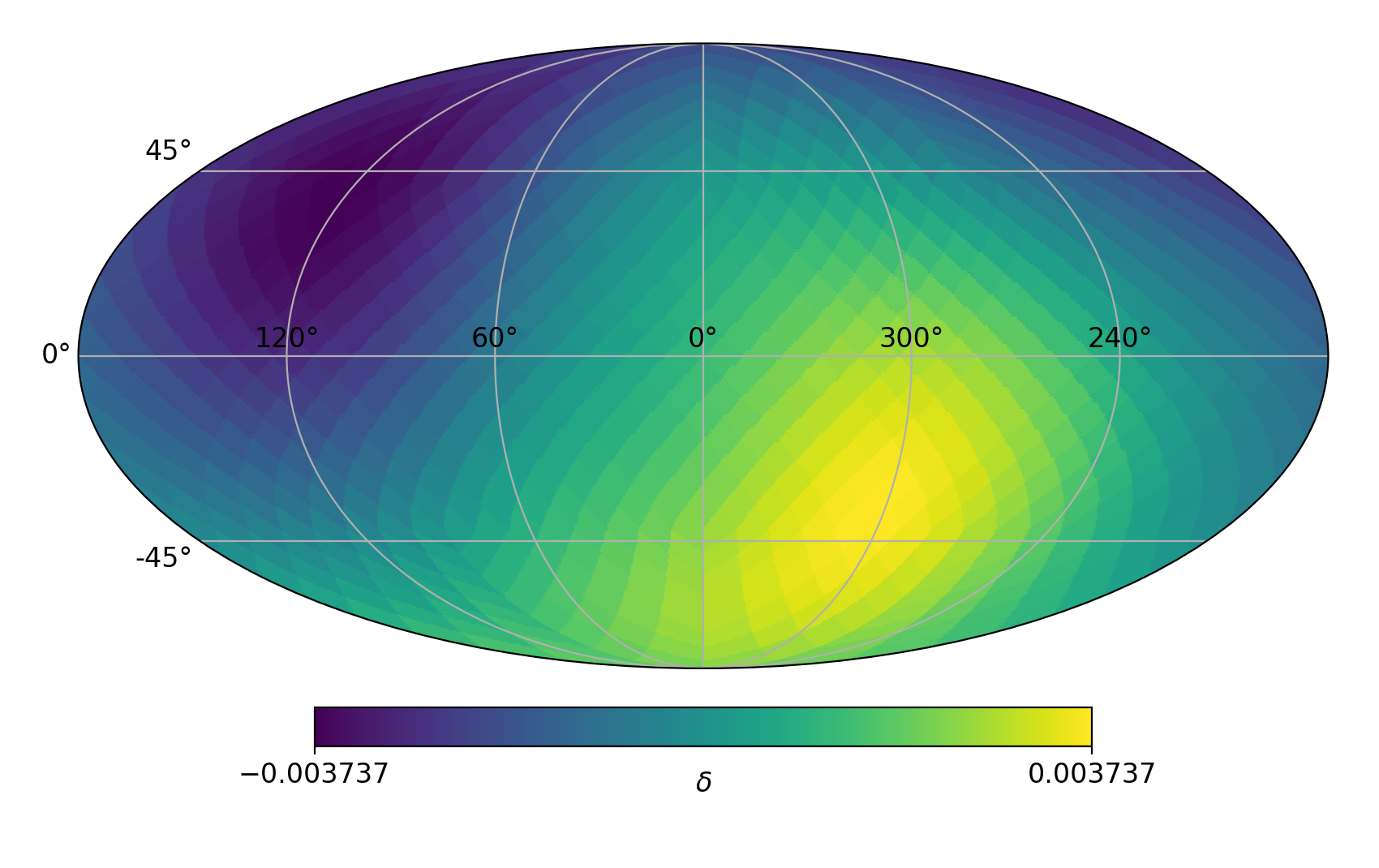}%
}\hspace*{-0.4em}
\subfloat{%
\includegraphics[height=39mm,width=55mm]{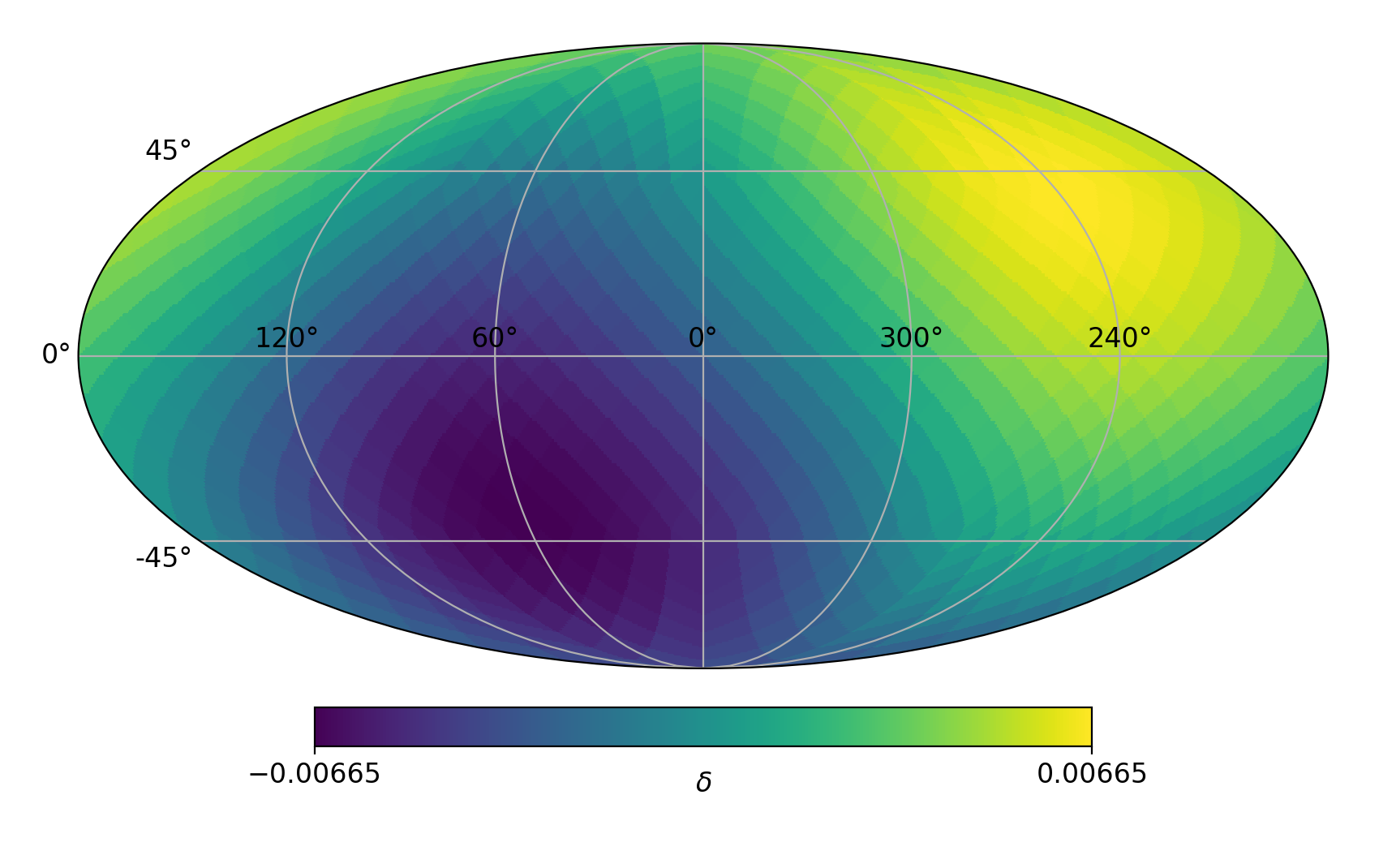}%
}

\caption{
Polarized ITBD event rate relative to the unpolarized rate, as a function of the tritium polarization direction, in Galactic coordinates.  Here $m_{\nu} = 0.1$eV and $\nu$ spin rotation in JF12 Coherent Galactic Magnetic Field Model (1st Row) JF12 coherent+Random field components (2nd Row) , for $\mu_{\nu}$= $\mu_{\nu}^{\rm lim}$, $10^{-1}$ $\times$ $\mu_{\nu}^{\rm lim}$, $10^{-2}$ $\times$ $\mu_{\nu}^{\rm lim}$ and $10^{-3}$ $\times$ $\mu_{\nu}^{\rm lim}$ in sequence from left to right}
\label{anisotropy}
\end{figure*}  

\subsection{Spin-polarized ITBD event rate}
References \cite{Lisanti:2014pqa}\cite{Tully:2021key}\cite{PhysRevD.90.073006} discuss the detection of a cosmic neutrino background arrival direction anisotropy. Such an anisotropy can appear due to density perturbations as for the CMB, or in our case, due to anisotropic impacts of helicity rotation in the Galactic magnetic field.  Equation (4.3) of Ref.\cite{Tully:2021key} gives the differential cross section's dependence on the direction of ${}^{3}H$ polarization, $\hat{s}_{H}$, the direction of the outgoing electron, the $\nu$ velocity, and the $\nu$ spin direction. Neglecting terms giving small contributions:
\begin{equation}
\label{eq:diffXcn}
\frac{d\sigma(\hat{s}_{H},\hat{v}_e)}{d\Omega_{e}}\approx\frac{\overline{\sigma}}{4\pi} [(1-\hat{s}_{\nu}\cdot \vec{v}_{\nu})+ B\, \hat{s}_{H}\cdot(\vec{v}_{\nu}-\hat{s}_{\nu})],
\end{equation}
in units with $c=1$.  Here $B$ $=$ $\frac{2g_{A}(1+g_{A})}{1+3g^{2}_{A}}$ $\simeq$ 0.99 is an asymmetry parameter from the amplitude calculation of $\nu$ to be captured on the neutron, and the term (1-$\hat{s}_{\nu}$$\cdot$$\vec{v}_{\nu})$ is identical to the $A^{\pm}$ in Eq.(\ref{A:factor}), the expression of helicity here is in terms of spin operators. This expression must be summed over all neutrinos contributing to the measurement, i.e., summed over all arrival directions in the sky since the direction of the incident neutrino is unknown. 
For unpolarized ${}^{3}H$, only the $<\hat{s}_{\nu}$$\cdot$$\vec{v}_{\nu}>$ term is of interest because the second term averages to zero. 

Neutrino spin rotation not only impacts the total event rate but, as can be seen in Eq.~\eqref{eq:diffXcn}, it generates a sensitivity of the event rate to the direction of polarization of the tritium target.
Using the results from our simulations,
we calculate the fractional difference of ITBD event rate for $\hat{s}_{H}$ pointing to a given Galactic coordinate, summing over the neutrino arrival directions, due to the term 
$\hat{s}_{H}\cdot(\vec{v}_{\nu}-\hat{s}_{\nu})$. Repeating this for each direction $\hat{s}_{H}$$ $ on the sky, and dividing by the average at the given $\mu_{\nu}$, yields a ``dipole anisotropy" map for the ITBD signal. Figures \ref{anisotropy},  Appendix.A Fig. \ref{anisotropy2} Fig. \ref{anisotropy3} and \ref{anisotropy4} show the fractional difference in ITBD rate as a function of the tritium polarization direction, taking $m_{\nu}$ = 0.1 eV, 0.01 eV, 0.001 eV and 0.0001 eV, for the suite of values of $\mu_\nu$ and assuming that the flux of cosmic background neutrinos on the Galaxy is isotropic.  

Table~\ref{table2} shows the dipole anisotropy of the polarized ITBD $\alpha$, and the fractional change in the total ITBD detection rate $\delta$, for four masses and our four standard $\mu_\nu$ values. 
\begin{table}[ht!]
\resizebox{.51\textwidth}{!}{% Resize the table to fit text width
\hspace{-1.5em}
  \begin{tabular}{|l|l|l|l|l|l|l|l|l|} 
    \hline
    \multirow{2}{*}{\backslashbox [3mm]{$m_{\nu}$(eV)}{$\mu_{\nu}$}}&
      \multicolumn{2}{c|}{$\mu_{\nu}^{\rm lim}$} &
      \multicolumn{2}{c|}{$0.1$ $\mu_{\nu}^{\rm lim}$} &
      \multicolumn{2}{c|}{$0.01$ $\mu_{\nu}^{\rm lim}$}&
      \multicolumn{2}{c|}{$0.001$ $\mu_{\nu}^{\rm lim}$}\\
      \cline{2-9}
    & $\delta$ & $\alpha$ & $\delta$ & $\alpha$& $\delta$ & $\alpha$& $\delta$ & $\alpha$ \\
    \hline
    $10^{-1}$ (C) & -0.0023 & 0.009 & -0.0022 & 0.005 & -0.0019  & 0.007 & -0.0007& 0.006  \\
    \hline
    $10^{-1}$(CR) & -0.0024 & 0.010& -0.0023& 0.005 & -0.0019 & 0.008 &-0.0007 &0.007 \\
    \hline\hline
    $10^{-2}$ (C) & -0.0222 & 0.006 & -0.0151  & 0.004 & -0.0143 & 0.004 & -0.0050& 0.009 \\
    \hline
    $10^{-2}$ (CR) & -0.0229 & 0.006 & -0.0179 & 0.005 & -0.0145  & 0.005 & -0.0057 & 0.009 \\
    \hline\hline
    $10^{-3}$ (C) & -0.146 & 0.002& -0.0489 & 0.008 & -0.0252 & 0.014 & -0.0162 & 0.018\\
    \hline
    $10^{-3}$ (CR) & -0.136 & 0.002 & -0.0542 & 0.008 & -0.0271& 0.014 & -0.0163 & 0.019\\
    \hline\hline
    $10^{-4}$ (C) & -0.381 & 0.011& -0.206 & 0.017 & -0.0090 & 0.065 & -9.7 $\times$ $10^{-5}$& 0.065\\
    \hline
    $10^{-4}$ (CR) & -0.402 & 0.014& -0.208 & 0.017 & -0.0093 & 0.070 & -9.7 $\times$ $10^{-5}$& 0.068\\
    \hline
  \end{tabular}
  }
  \caption{Modification factor $\delta$ of the ITBD event rate and dipole amplitude $\alpha$ of the polarized ITBD rate, for neutrino masses from $10^{-1}$eV to $10^{-4}$eV and four neutrino magnetic moment values from $1$ to $10^{-3}$ times $\mu_{\nu}^{\rm lim}$; the current experimental upper bound from GEMMA. (C) represents JF12 coherent field, (CR) represents JF12 coherent field plus random components.}
    \label{table2}

\end{table}

\section{\label{sec:level6}Summary}

We have explored the effects of a neutrino magnetic moment on the evolution of the helicity of relic neutrinos propagating through the Galaxy to a detector on Earth, by tracking through a realistic model of the Galactic magnetic field with numerical simulations. 
We find that for magnetic moments consistent with experimental bounds and even several orders of magnitude smaller, a relic neutrinos' helicity has a significant rotation, which depends on the arrival direction.

Using these results, we studied the effects of a neutrino magnetic moment on relic neutrino detection through inverse tritium beta decay and we calculated the dipole anisotropy in the neutrino detection rate as a function of tritium polarization direction.  The results are summarized in Table~\ref{table2}.  We find that the absolute modification in the detection rate due to spin rotation is very small, $\lesssim \mathcal{O}$(1\%),
%on account of the low velocity of cosmic neutrinos. 
in spite of significant helicity rotation, except for relativistic neutrinos.  The fractional directional anisotropy in a polarized ITBD experiment is generally $\lesssim$ 1\% for most values of magnetic moment and mass within 2 orders of magnitude of their current bounds, but for particular combinations could approach $\mathcal{O}$(10\%).  %Many difficulties still remain, however, in reaching the required levels of sensitivity or indeed even detecting a relic neutrino signal.

Although neutrino helicity modification due to propagation through the Milky Way's magnetic field is not measurable by current experiments, the phenomenon may eventually be observable if the neutrino magnetic moment is within a few orders of magnitude of the current experimental limit. In that case, observations of a directional anisotropy in the event rate with a polarized ITBD detector could complement a precise measurement of the absolute event rate and contribute to probing both the magnetic moment and mass of the lightest neutrino mass eigenstate.  The most important consequence of establishing the existence of a neutrino flux anisotropy correlated to the Galactic magnetic field, would be that that it would unambiguously imply the lightest neutrino mass eigenstate is Dirac and not Majorana (since only Dirac neutrinos experience spin rotation in a magnetic field)\cite{RevModPhys.87.531}.

\begin{acknowledgments}
We appreciate helpful discussions with G. Baym, C. Tully, A. Gruzinov, and X. Xu.
The research of GRF is supported by National Science Foundation Grant NSF-PHY-2013199 and by the Simons Foundation. 

\end{acknowledgments}

\appendix
\section{Additional Plots}

\begin{figure*}[!htp]
\hspace*{-6.0em}
\subfloat{%
\includegraphics[height=45mm,width=53.0mm]{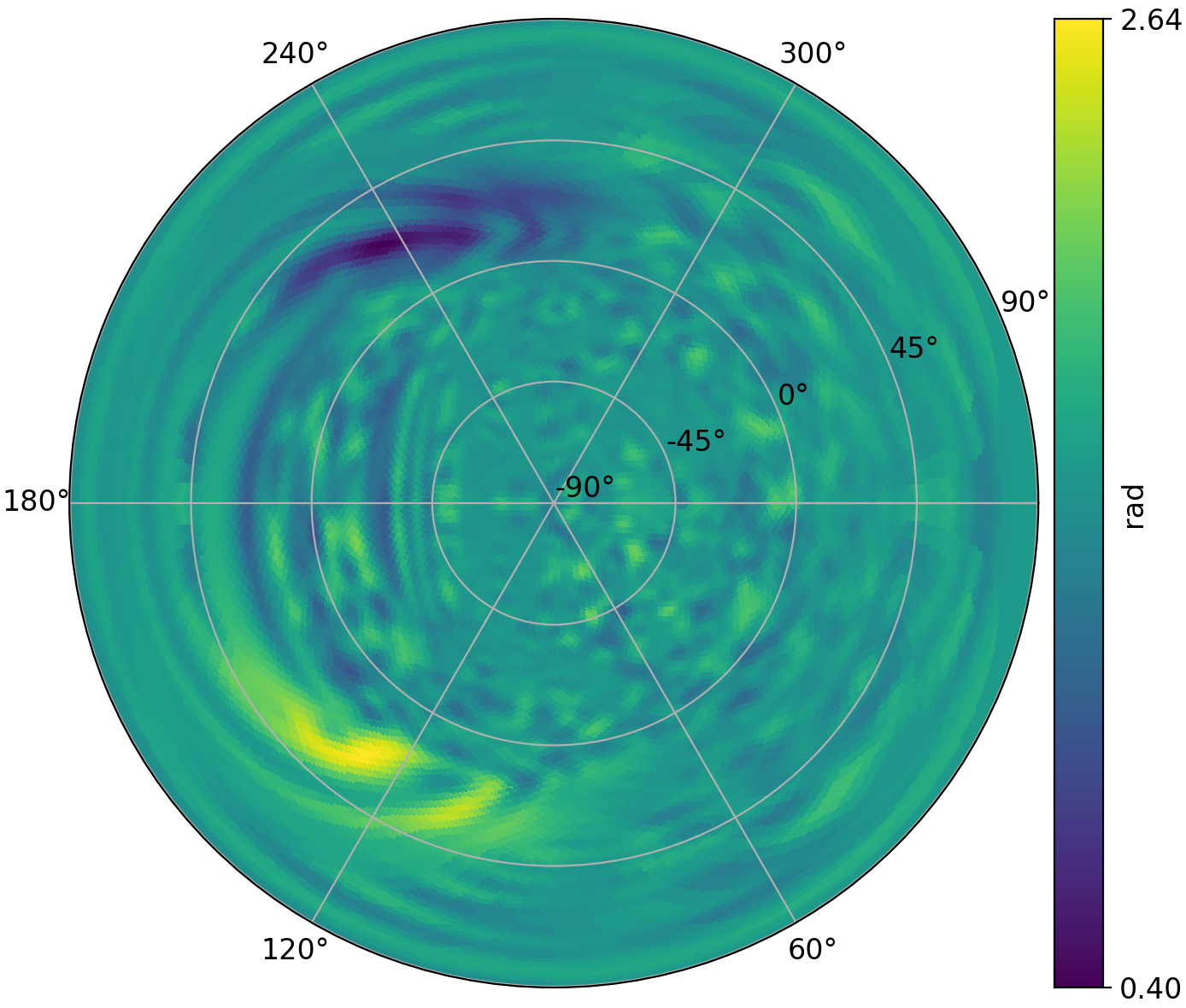}%
}\hspace*{-0.311em}
\subfloat{%
\includegraphics[height=45mm,width=53.0mm]{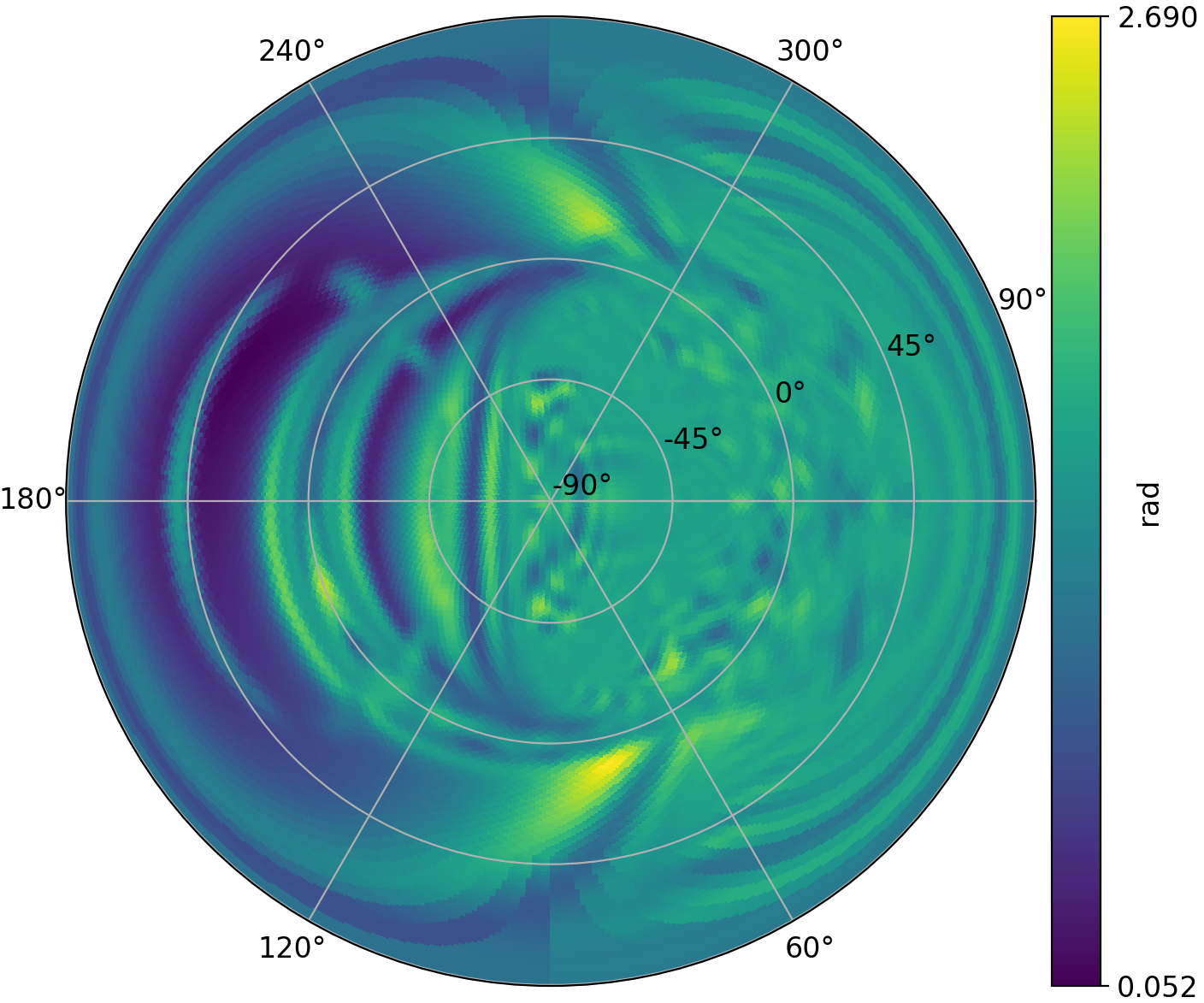}%
}\hspace*{-0.32em}
\subfloat{%
\includegraphics[height=45mm,width=53.0mm]{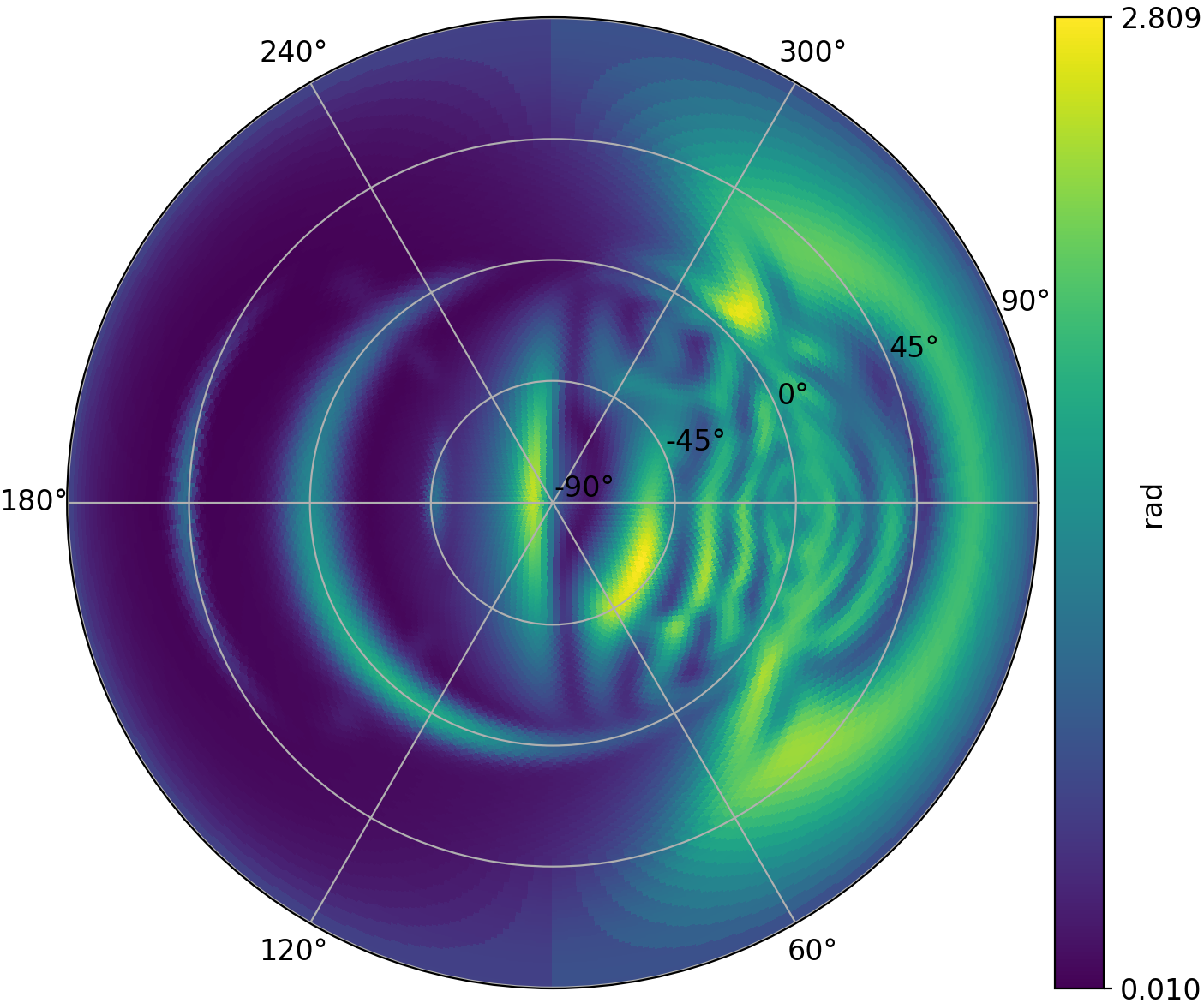}%
}\hspace*{-0.30em}
\subfloat{%
\includegraphics[height=45mm,width=53.0mm]{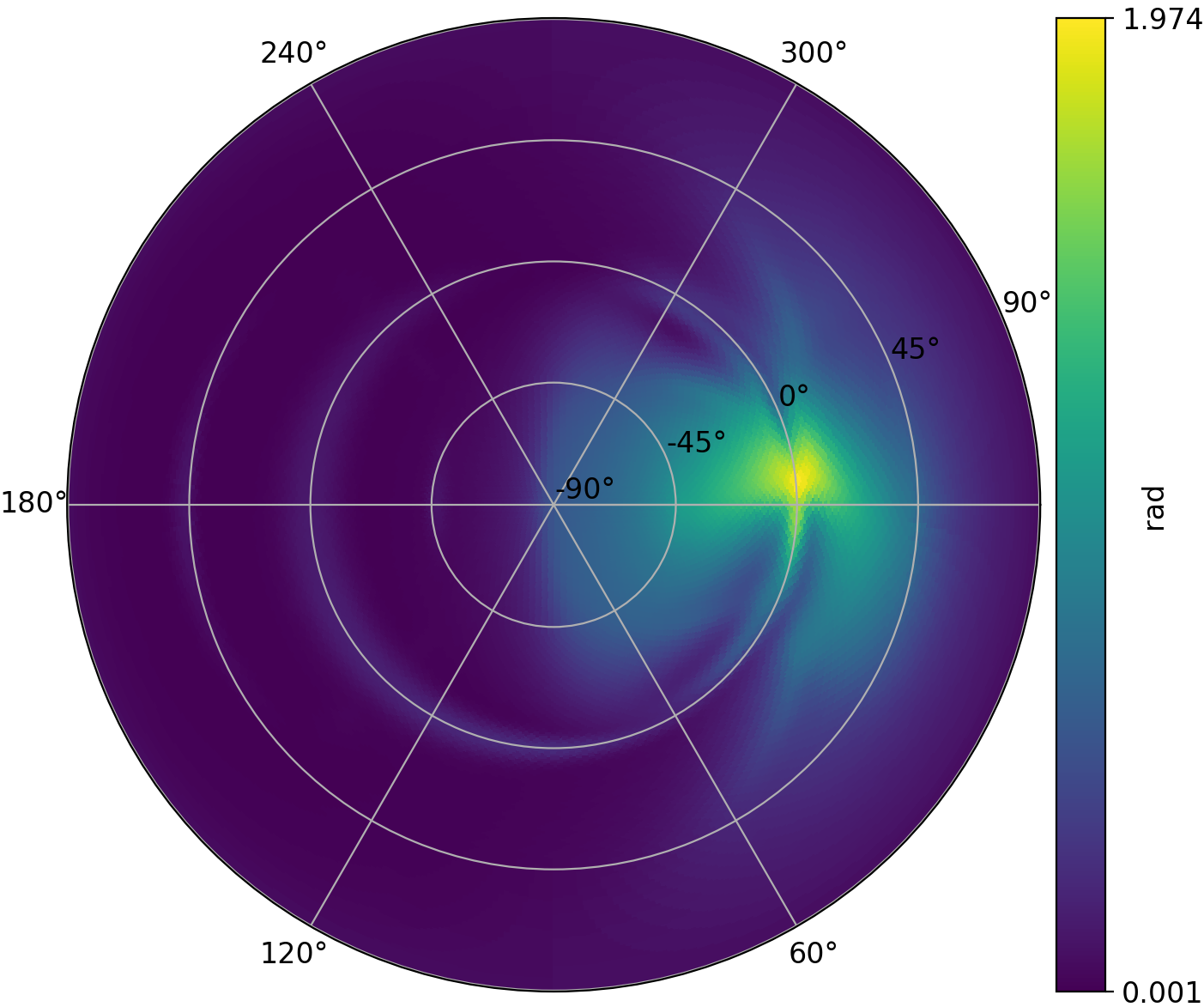}%
}\hfill
\hspace*{-6.0em}
\subfloat{%
\includegraphics[height=45mm,width=53.0mm]{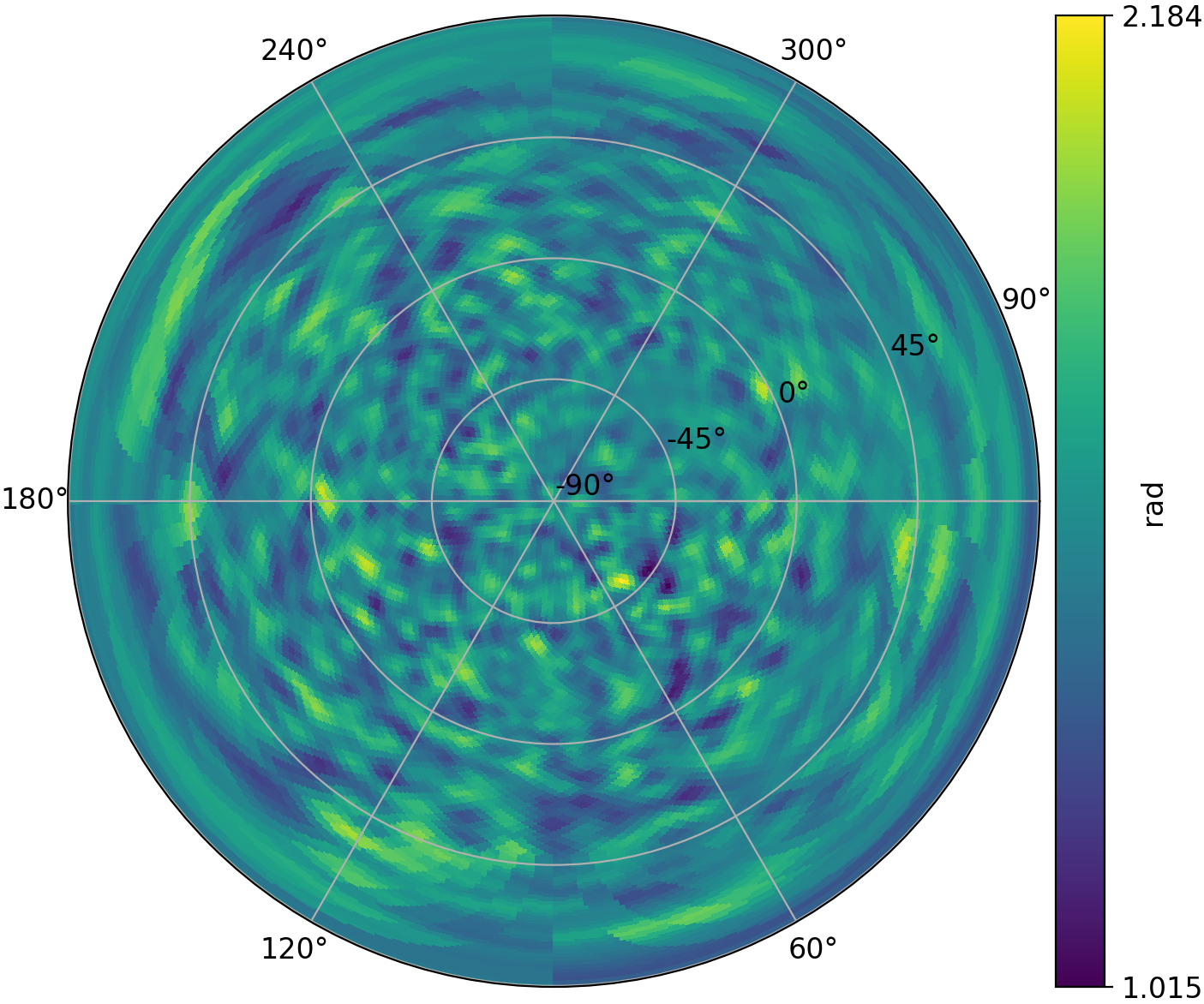}%
}\hspace*{-0.29em}
\subfloat{%
\includegraphics[height=45mm,width=53.0mm]{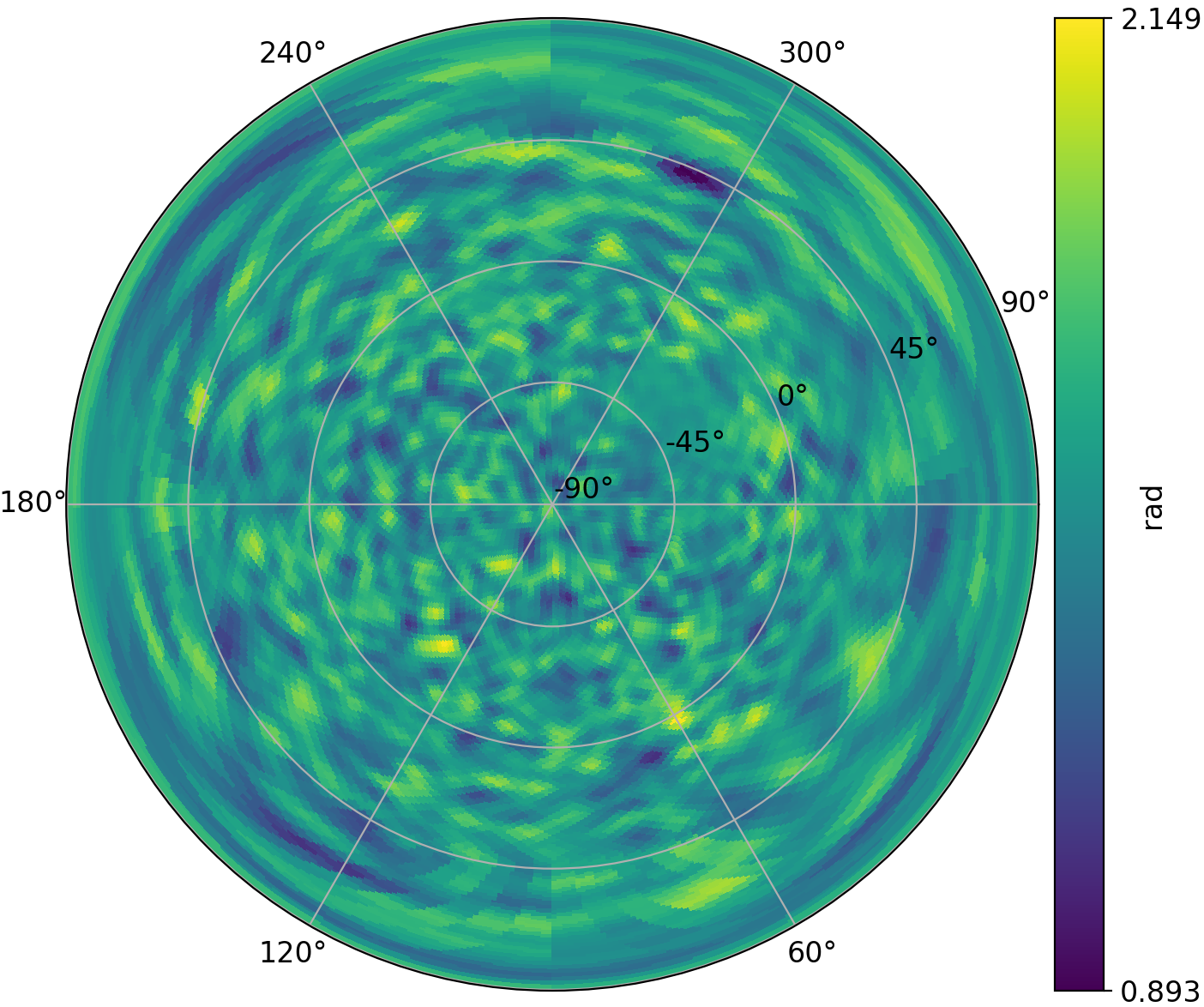}%
}\hspace*{-0.29em}
\subfloat{%
\includegraphics[height=45mm,width=53.0mm]{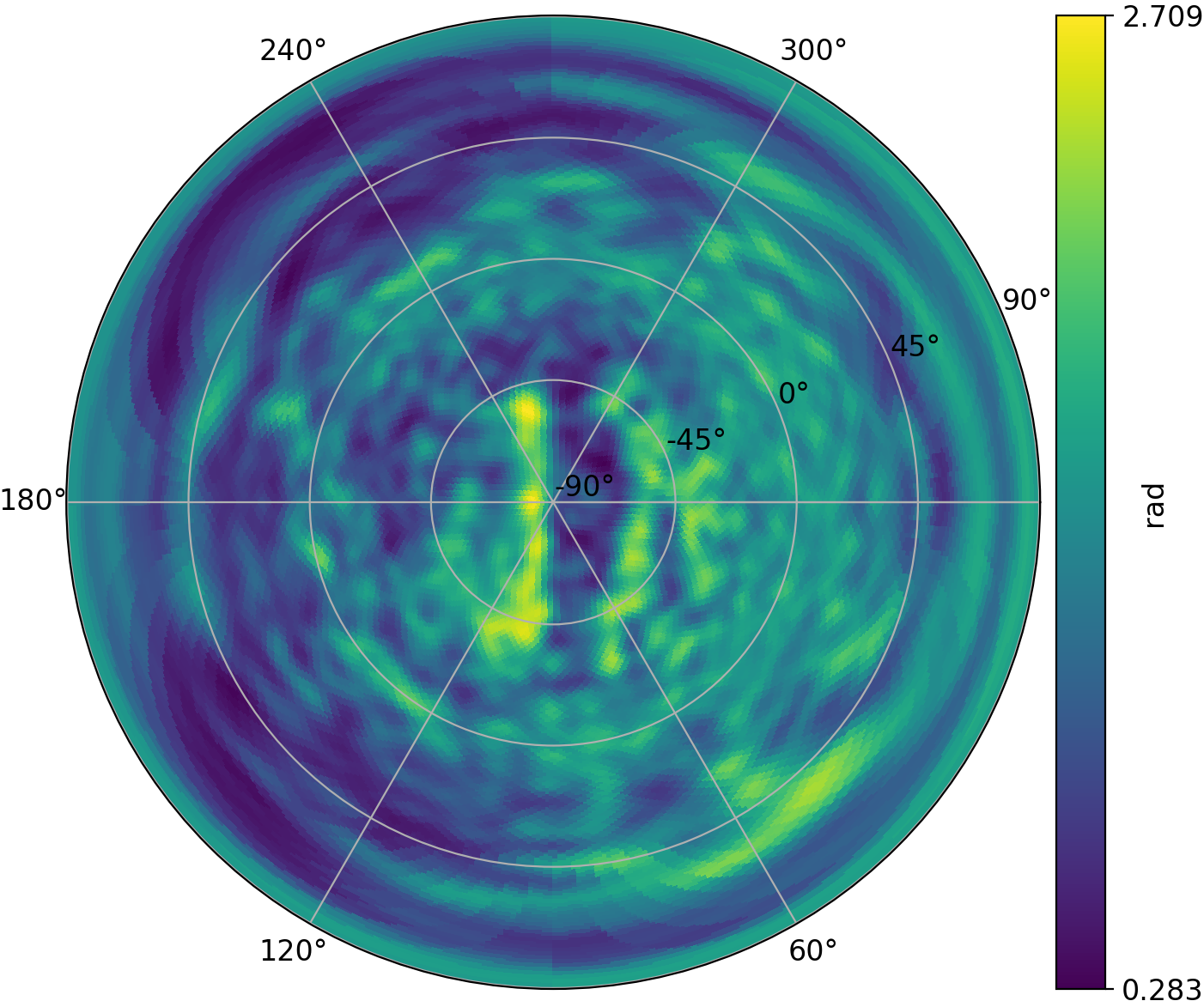}%
}\hspace*{-0.265em}
\subfloat{%
\includegraphics[height=45mm,width=53.0mm]{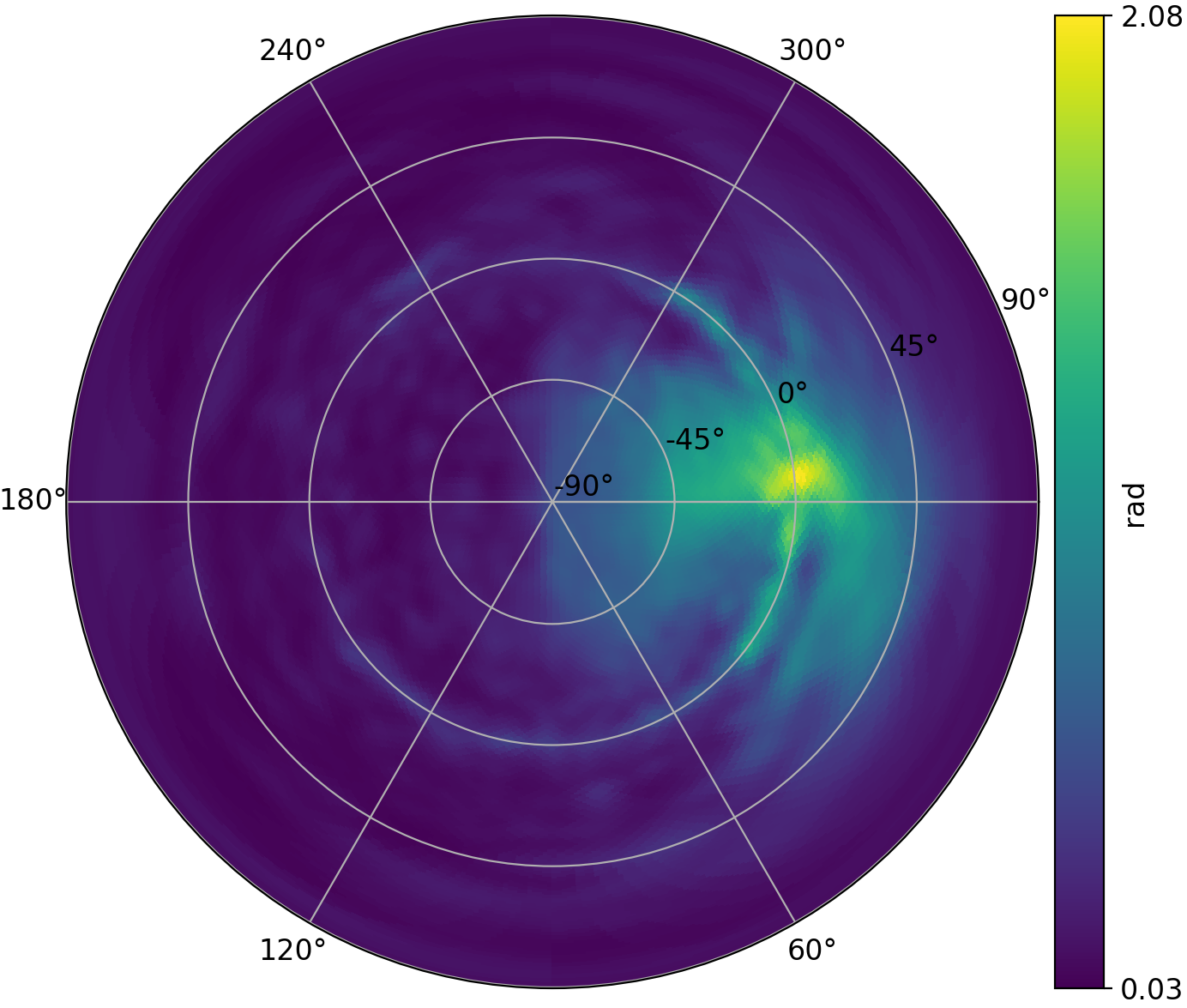}%
}
\caption{Spin vector rotation angle of a 0.1 eV relic neutrino traversing (first row) the JF12 coherent Galactic magnetic field and (second row) the full coherent plus random field.  From left to right the columns are for 
a magnetic moment equal to the GEMMA Reactor experiment bound $\mu_{\nu}^{\rm lim}\approx2.9\times10^{-11} \mu_{B}$ and factor-10 smaller values: 0.1$\mu_{\nu}^{\rm lim}$, 0.01$\mu_{\nu}^{\rm lim}$, and 0.001$\mu_{\nu}^{\rm lim}$.  $Healpy$ $Projview$ polar view has been used for sky-view mapping, with the coordinates being longitude from 0 to 2$\pi$ and colatitude from -$\pi/2$ to $\pi$/2. }
\label{rotation A}
\end{figure*}  

\hspace{-2cm}
\begin{figure*}[!htp]
\hspace*{-5.87em}
\subfloat[\label{sfig:testa}]{%
\includegraphics[height=35mm,width=59mm]{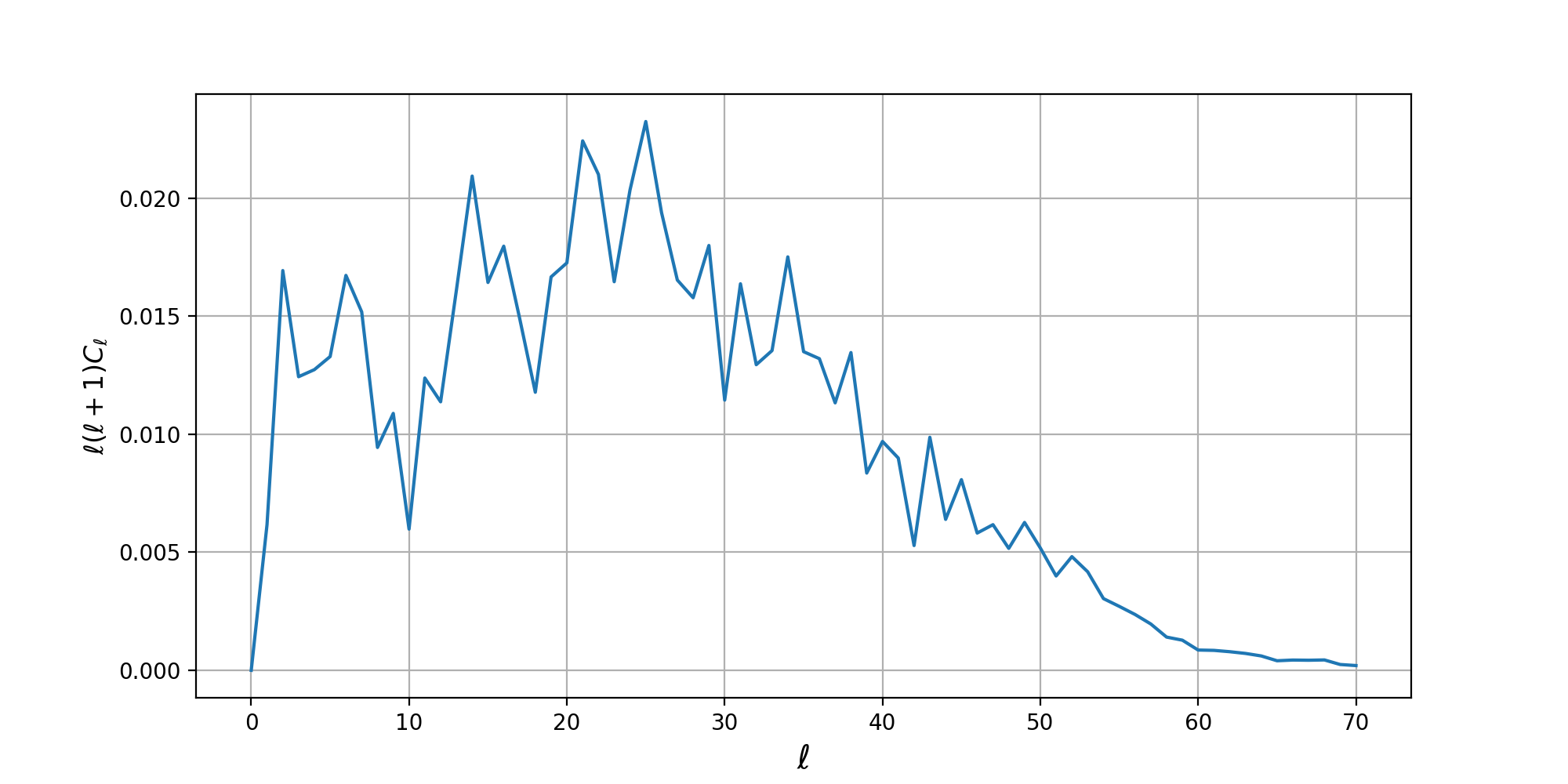}%
}\hspace*{-2.0em}
\subfloat[\label{sfig:testa}]{%
\includegraphics[height=35mm,width=59mm]{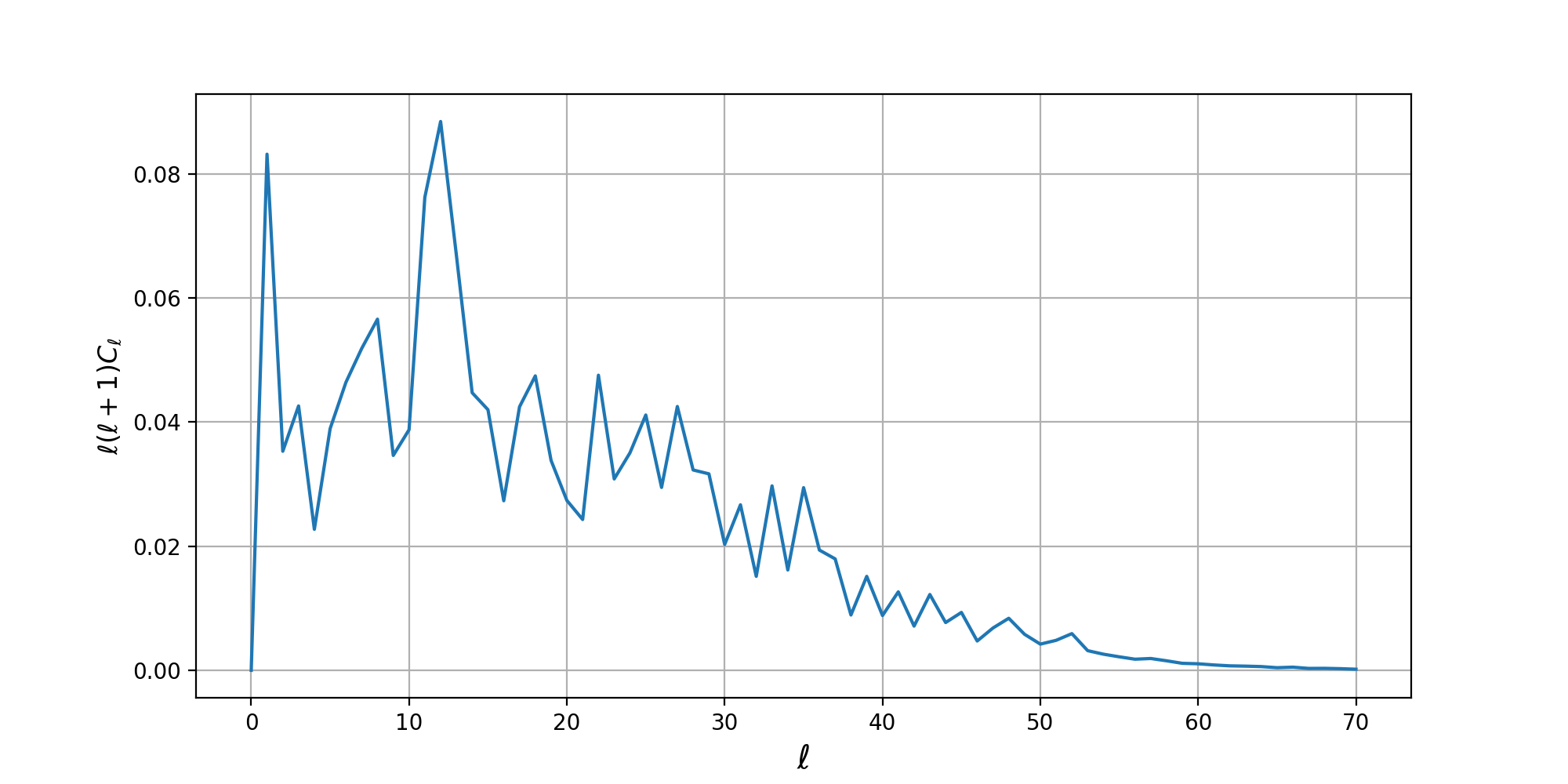}%
}\hspace*{-2.0em}
\subfloat[\label{sfig:testa}]{%
\includegraphics[height=35mm,width=59mm]{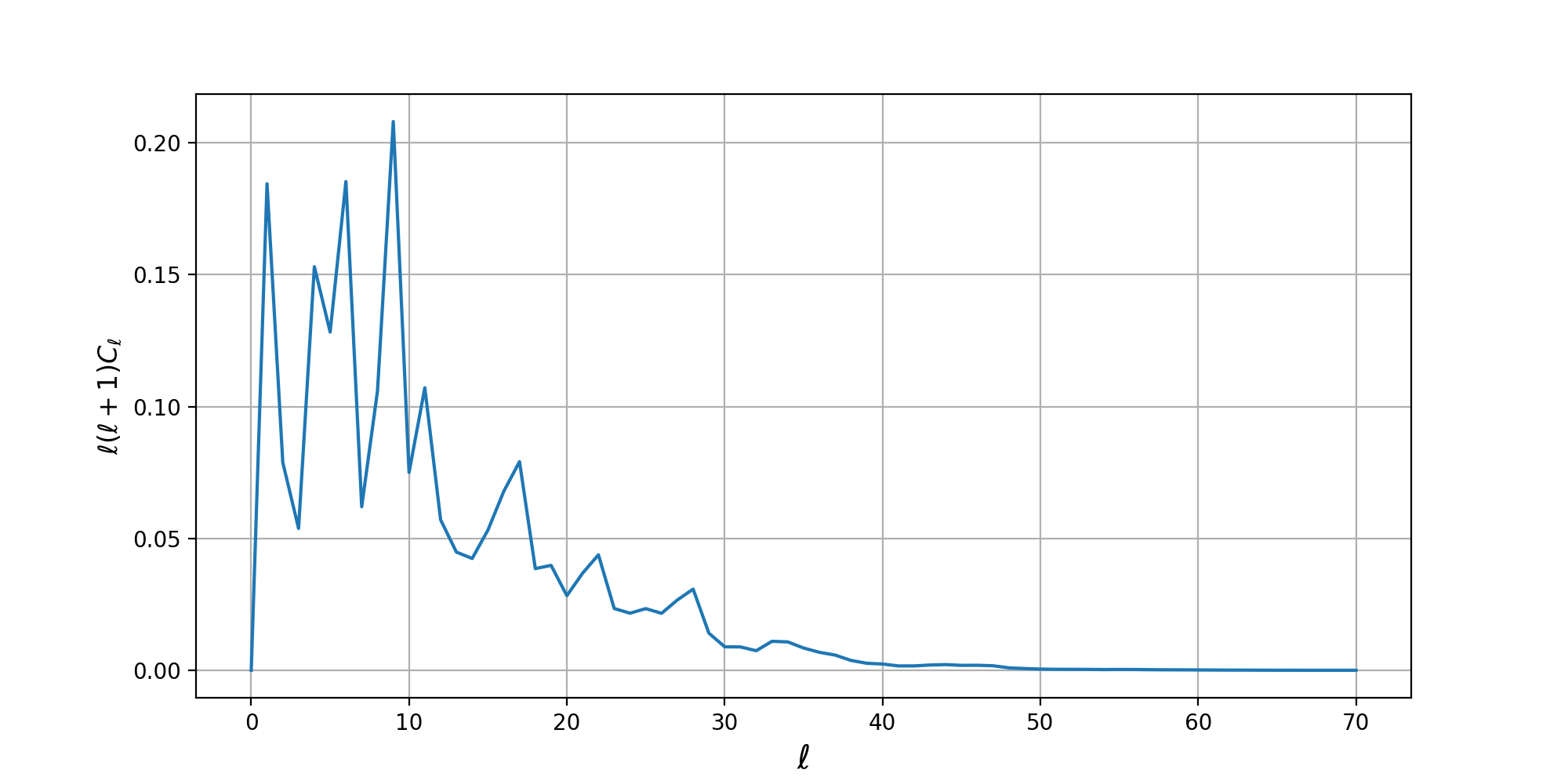}%
}\hspace*{-2.0em}
\subfloat[\label{sfig:testa}]{%
\includegraphics[height=35mm,width=59mm]{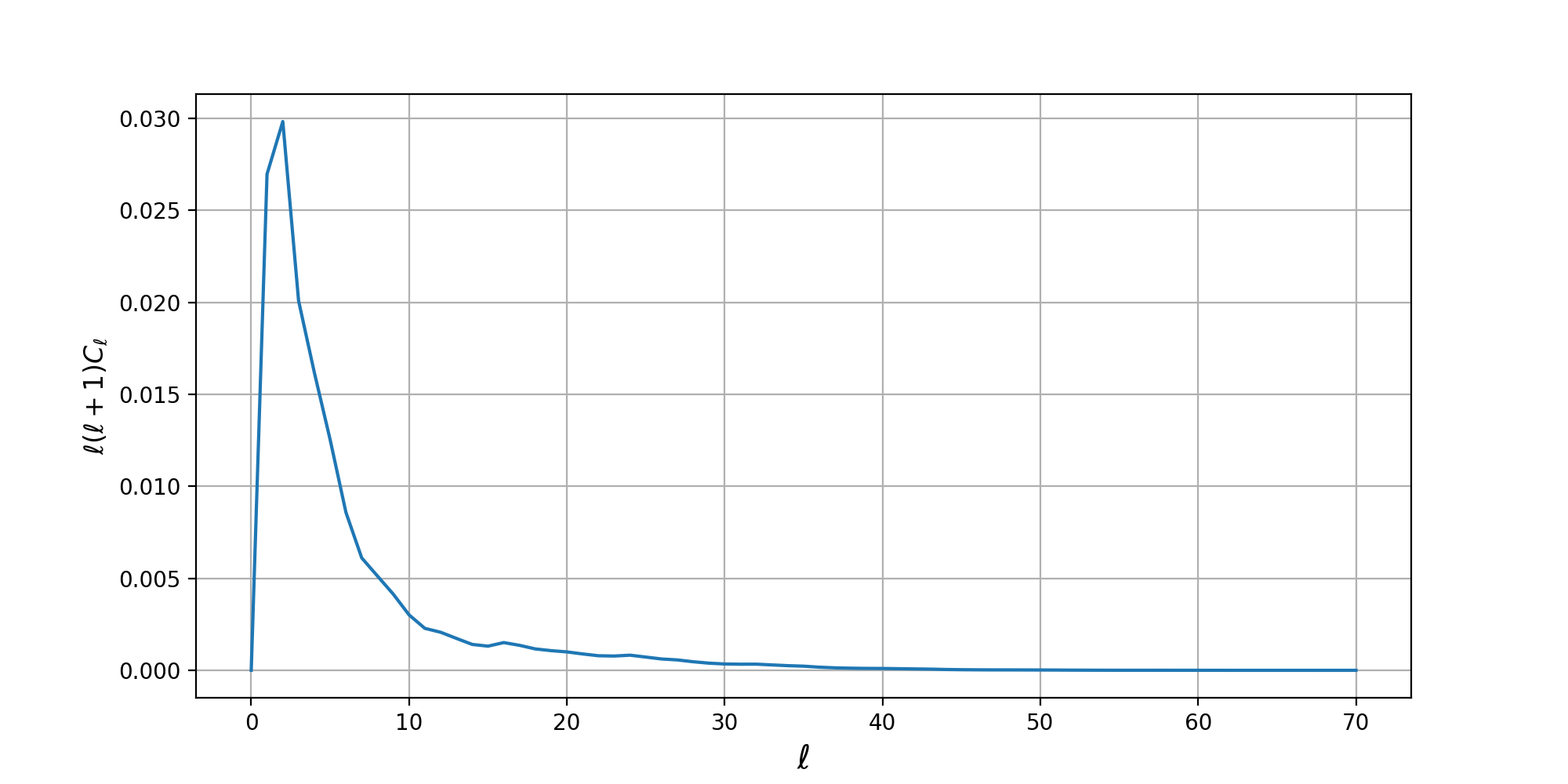}%
}\hfill
\hspace*{-5.87em}
\subfloat[\label{sfig:testa}]{%
\includegraphics[height=35mm,width=59mm]{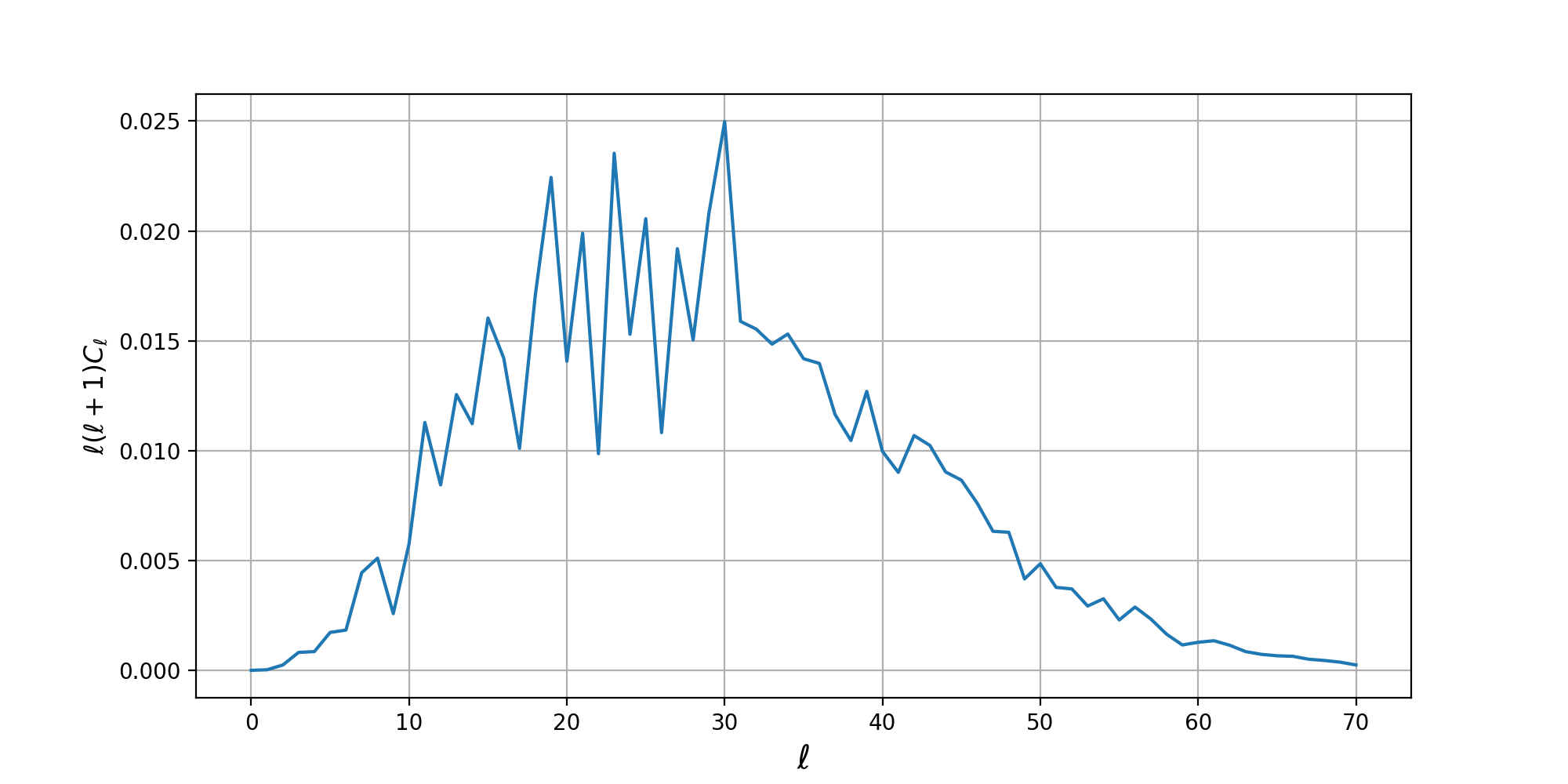}%
}\hspace*{-2.0em}
\subfloat[\label{sfig:testa}]{%
\includegraphics[height=35mm,width=59mm]{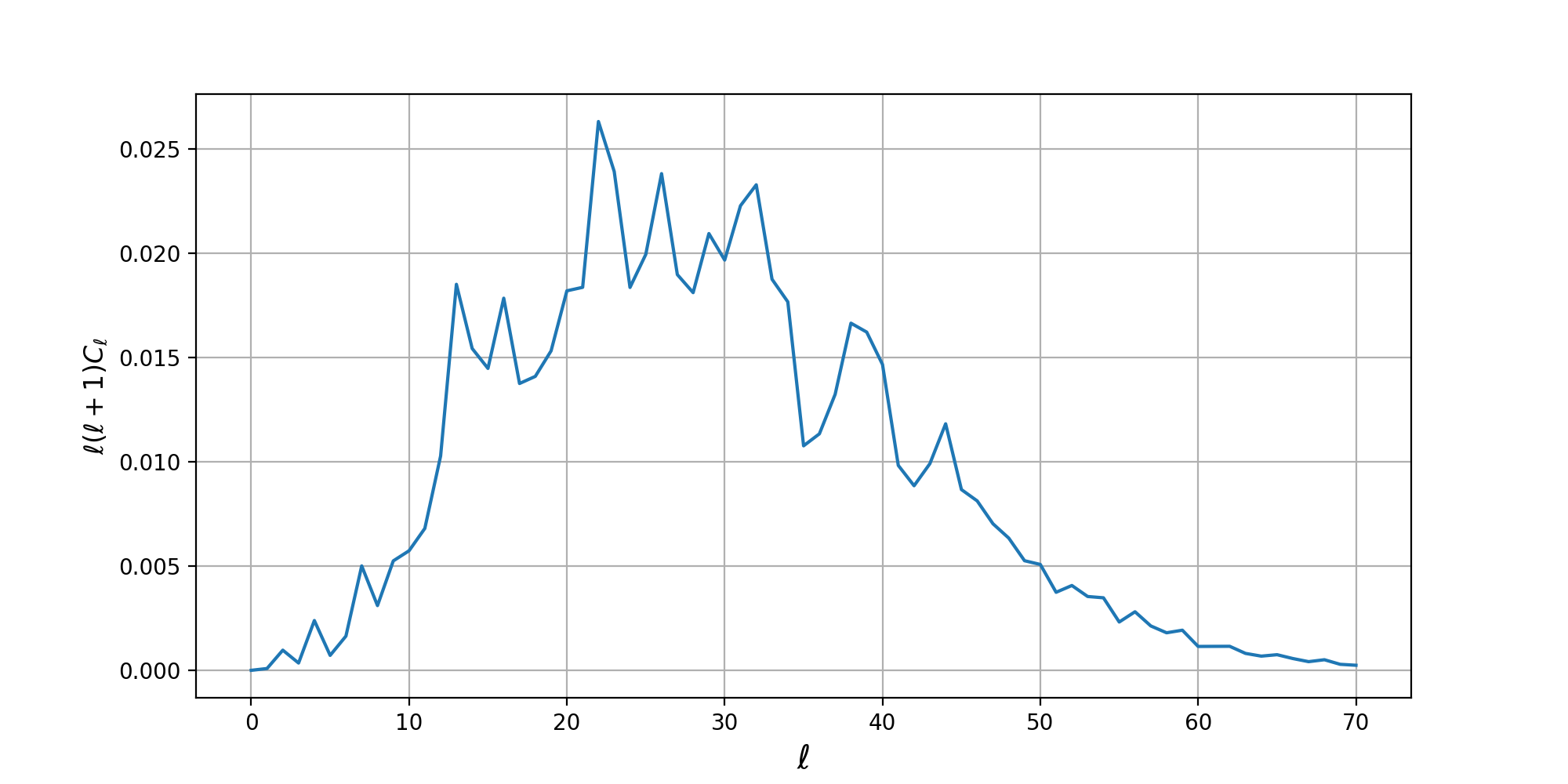}%
}\hspace*{-2.0em}
\subfloat[\label{sfig:testa}]{%
\includegraphics[height=35mm,width=59mm]{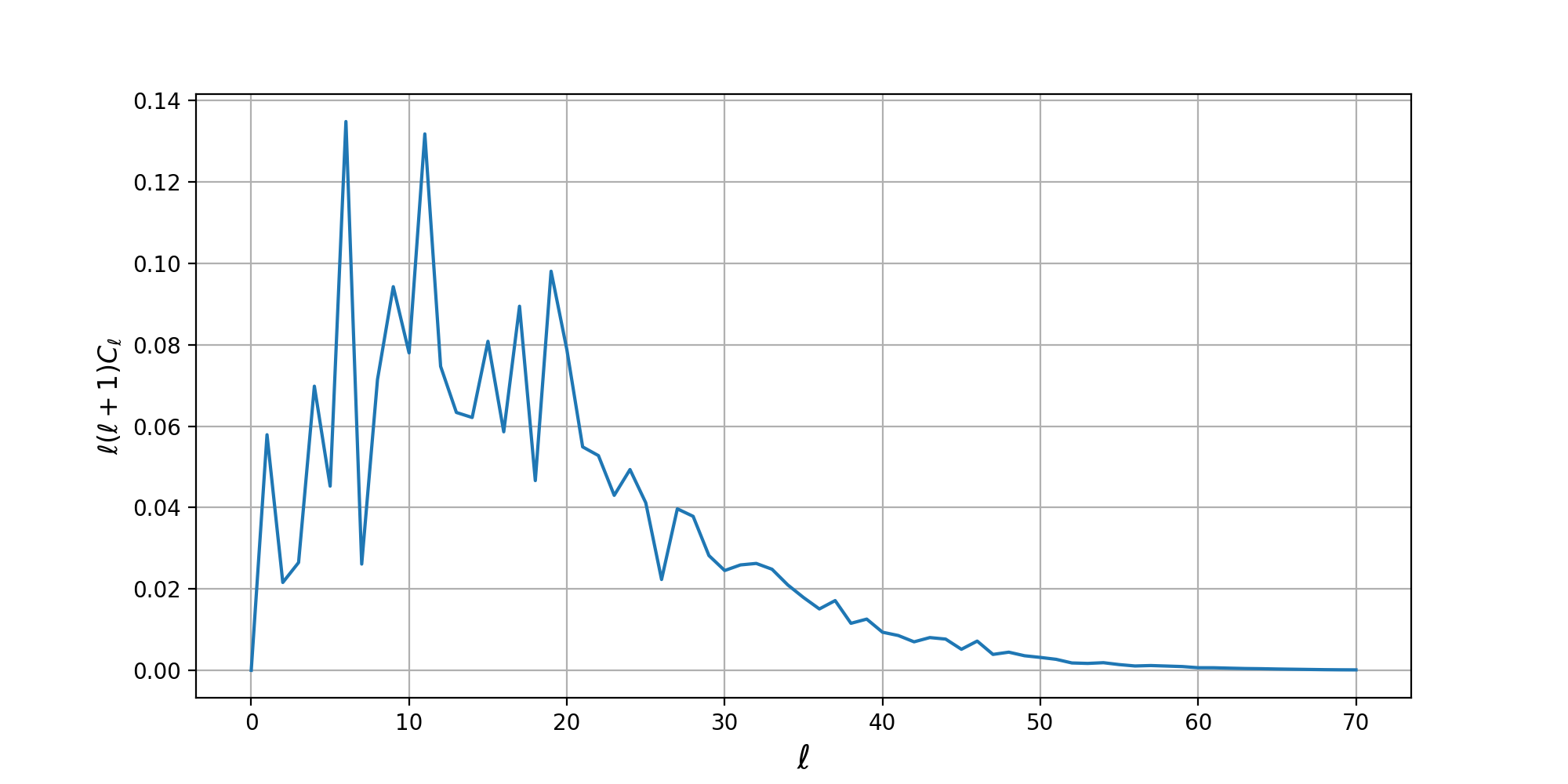}%
}\hspace*{-2.0em}
\subfloat[\label{sfig:testa}]{%
\includegraphics[height=35mm,width=59mm]{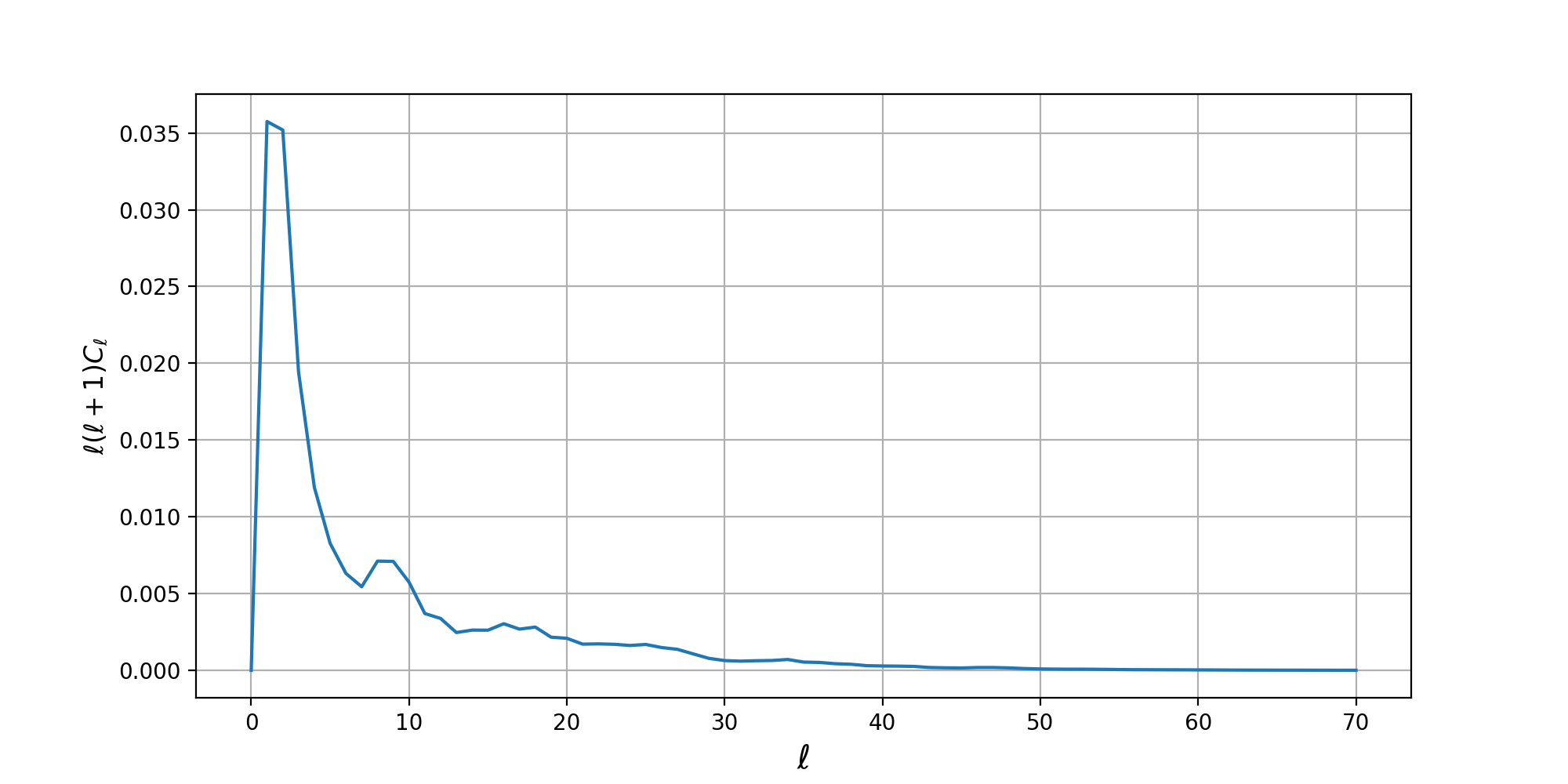}%
}
\caption{The angular power spectrum of the helicity-flip probability of relic neutrinos, for (left to right) $\mu_{\nu}^{\rm lim}\approx2.9\times10^{-11}\mu_{B}$, 0.1$\mu_{\nu}^{\rm lim}$, 0.01$\mu_{\nu}^{\rm lim}$, and 0.001$\mu_{\nu}^{\rm lim}$.  The upper row (a)-(d) are the simulation results in the JF-12 coherent GMF model, while (e)-(f) are for the coherent+random field model.}
\label{power}
\end{figure*}  

\hspace{-7cm}
\begin{figure*}[htp!]
\hspace*{-6.2em}
\subfloat{%
\includegraphics[height=39mm,width=55mm]{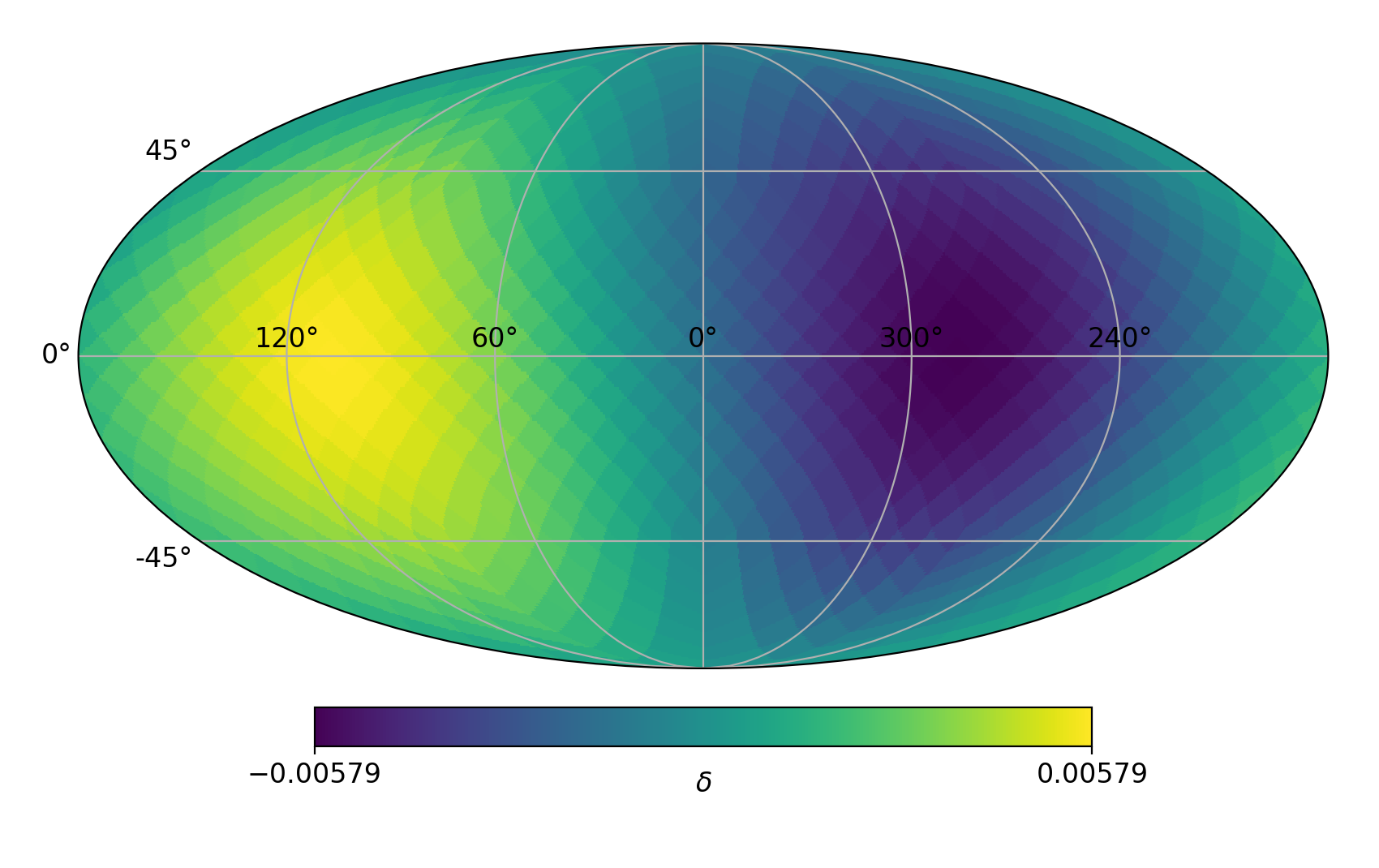}%
}\hspace*{-0.7em}
\subfloat{%
\includegraphics[height=39mm,width=55mm]{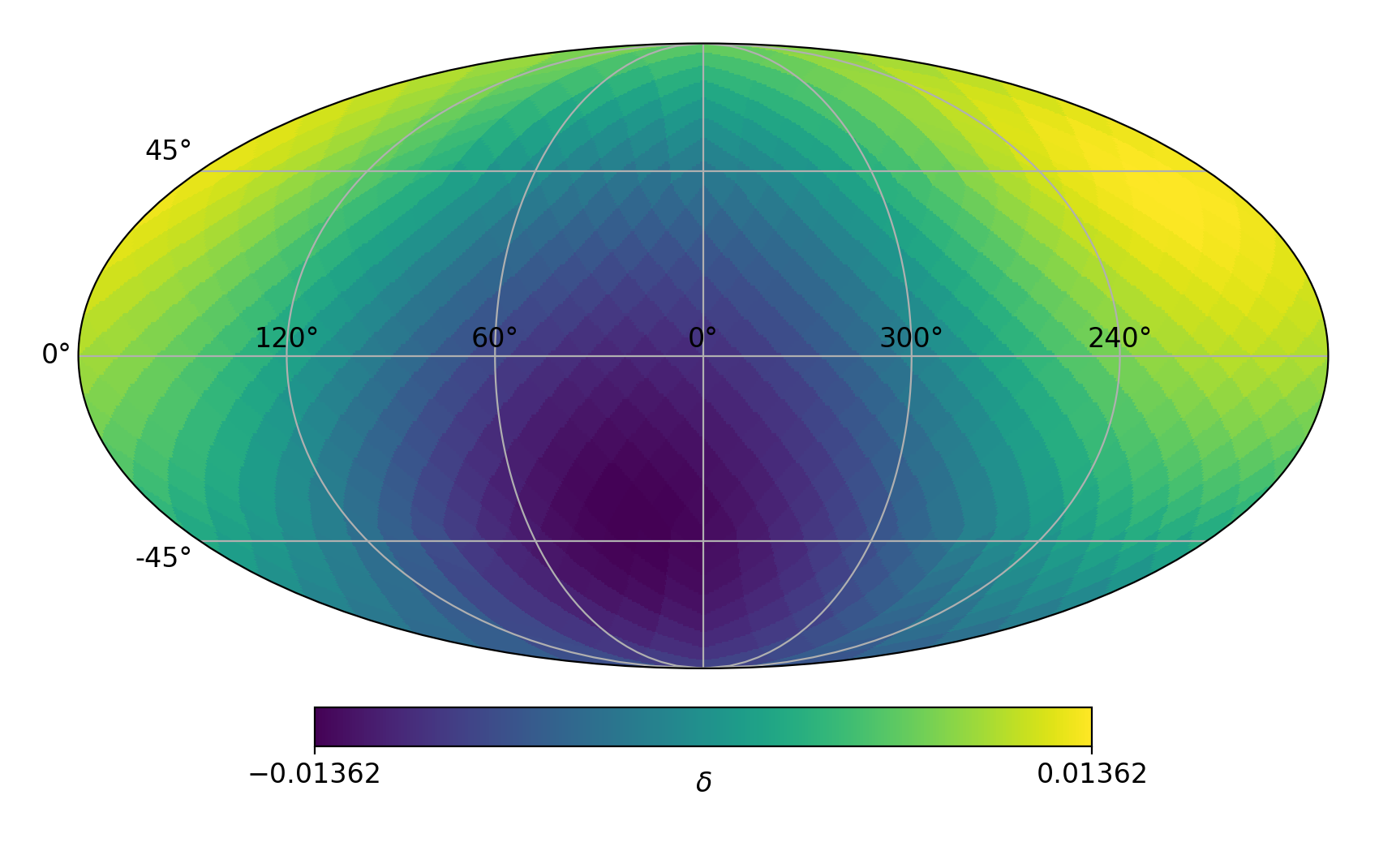}%
}\hspace*{-0.4em}
\subfloat{%
\includegraphics[height=39mm,width=55mm]{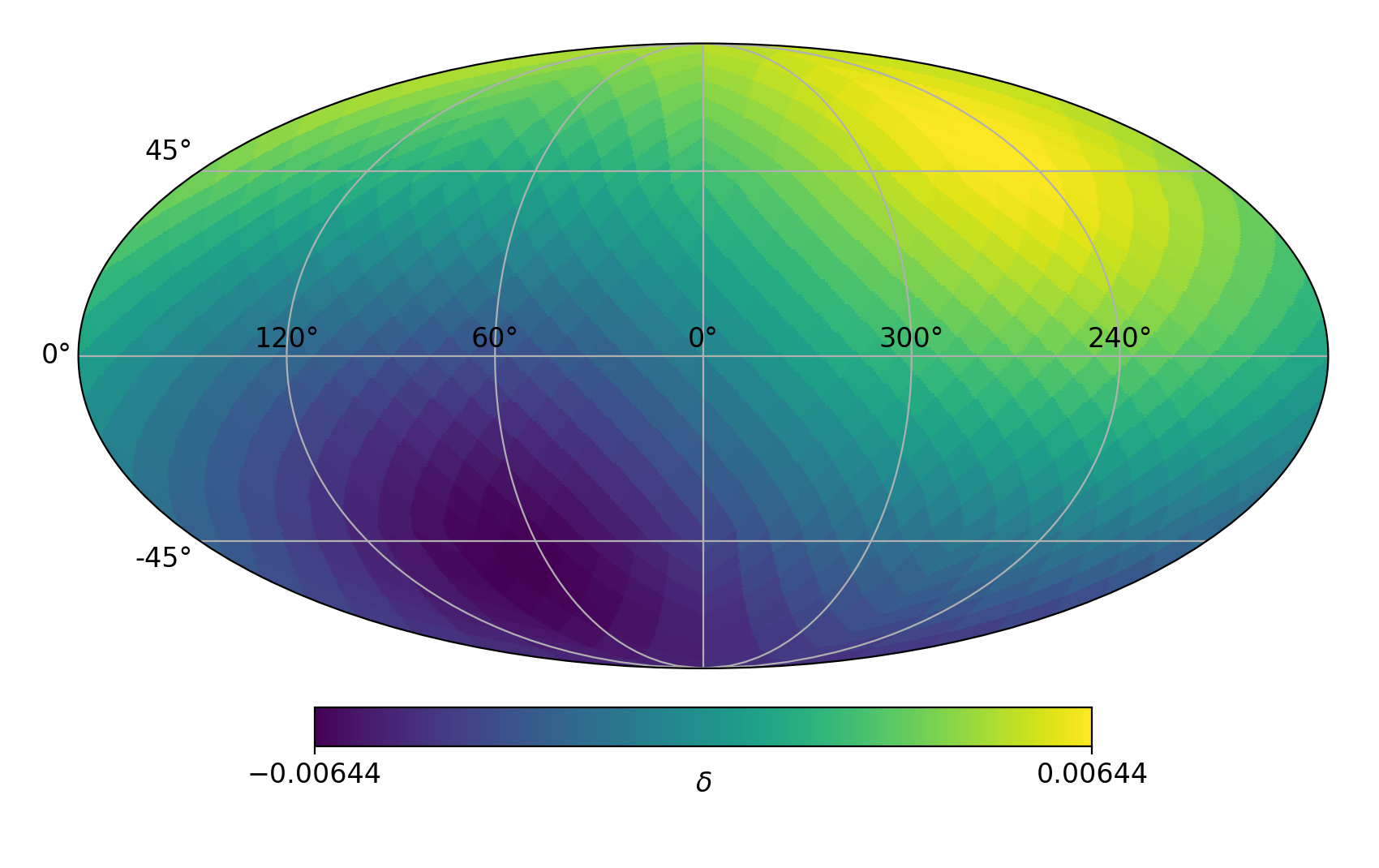}%
}\hspace*{-0.4em}
\subfloat{%
\includegraphics[height=39mm,width=55mm]{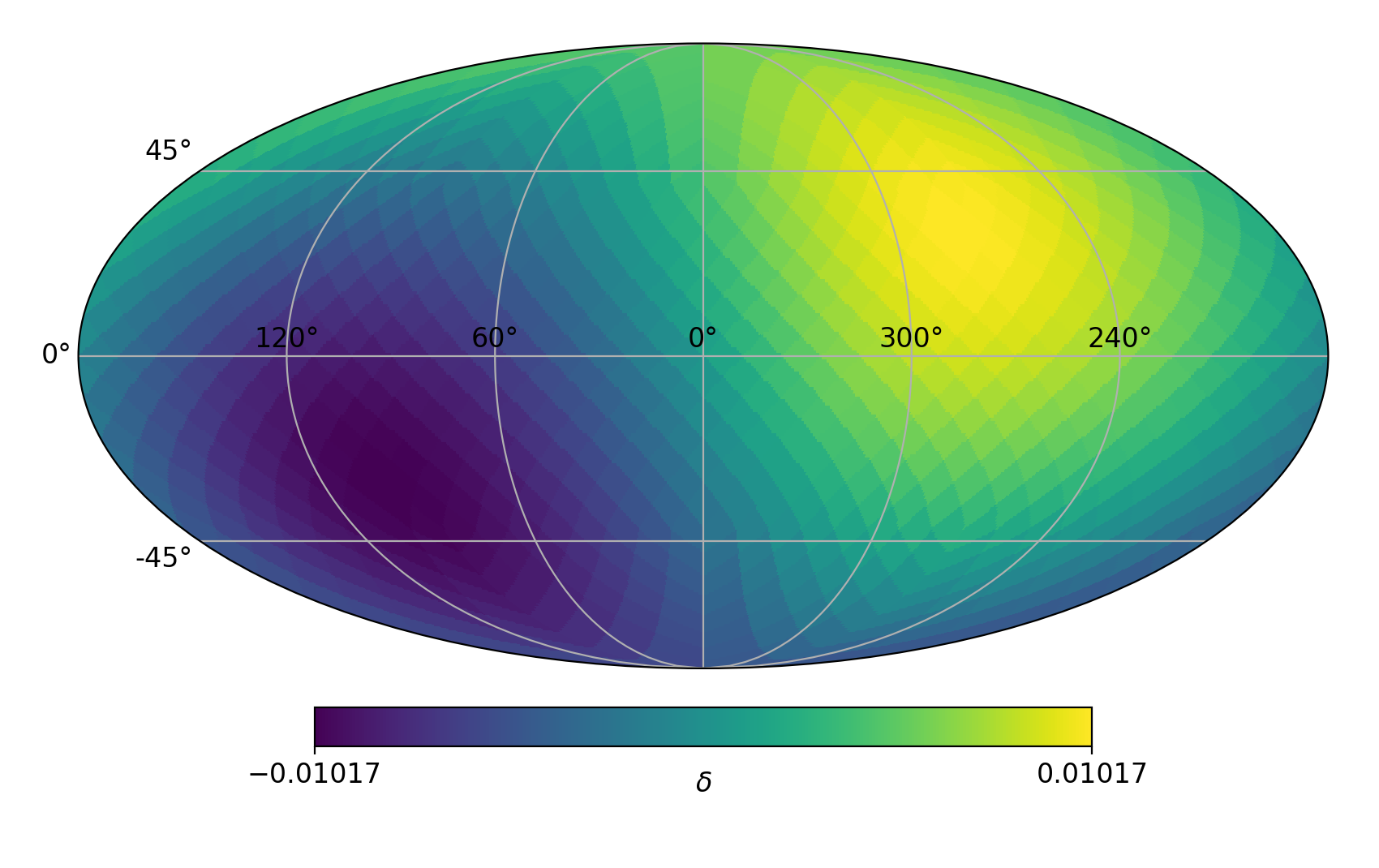}%
}\hfill
\hspace*{-6.2em}
\subfloat{%
\includegraphics[height=39mm,width=55mm]{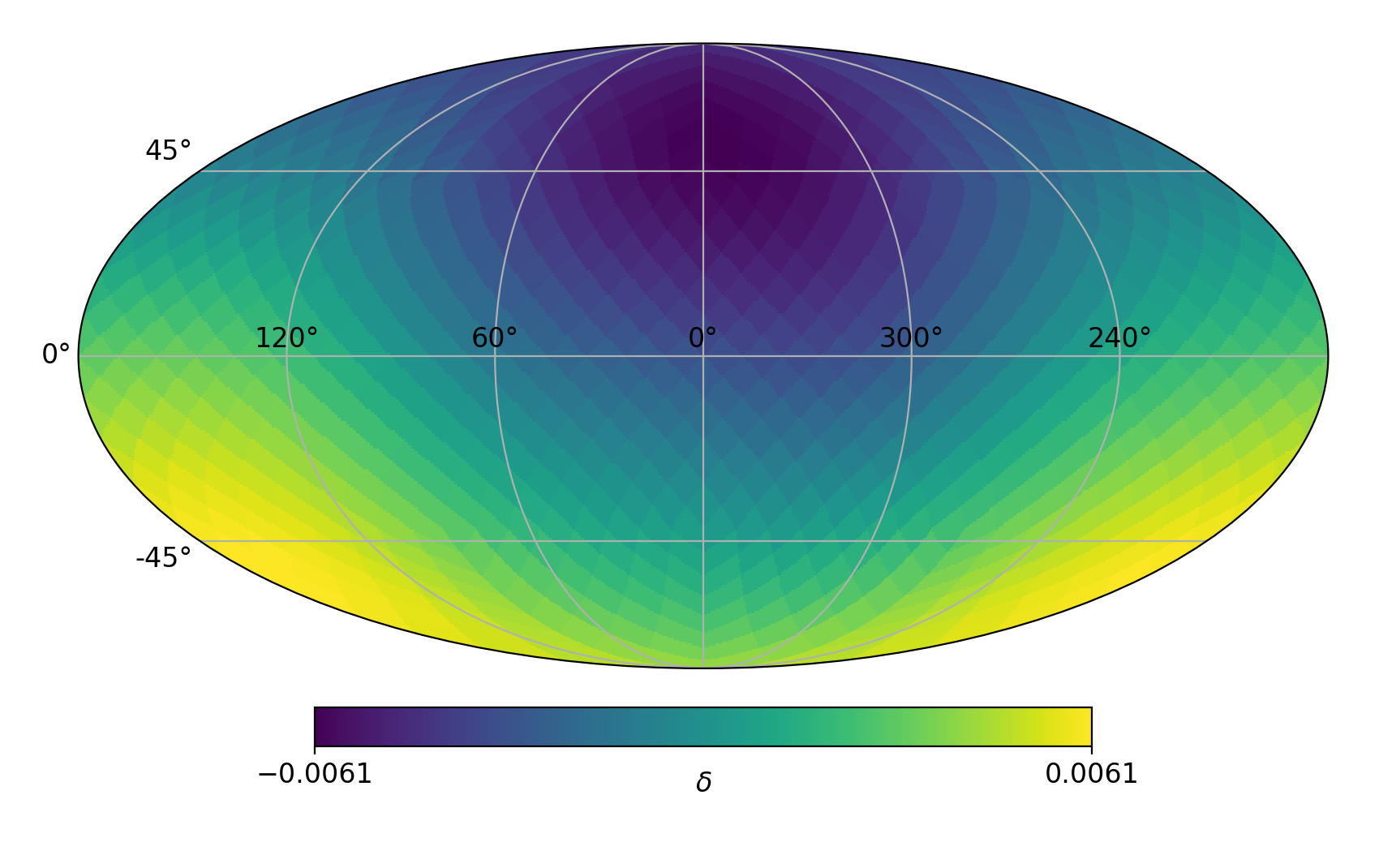}%
}\hspace*{-0.7em}
\subfloat{%
\includegraphics[height=39mm,width=55mm]{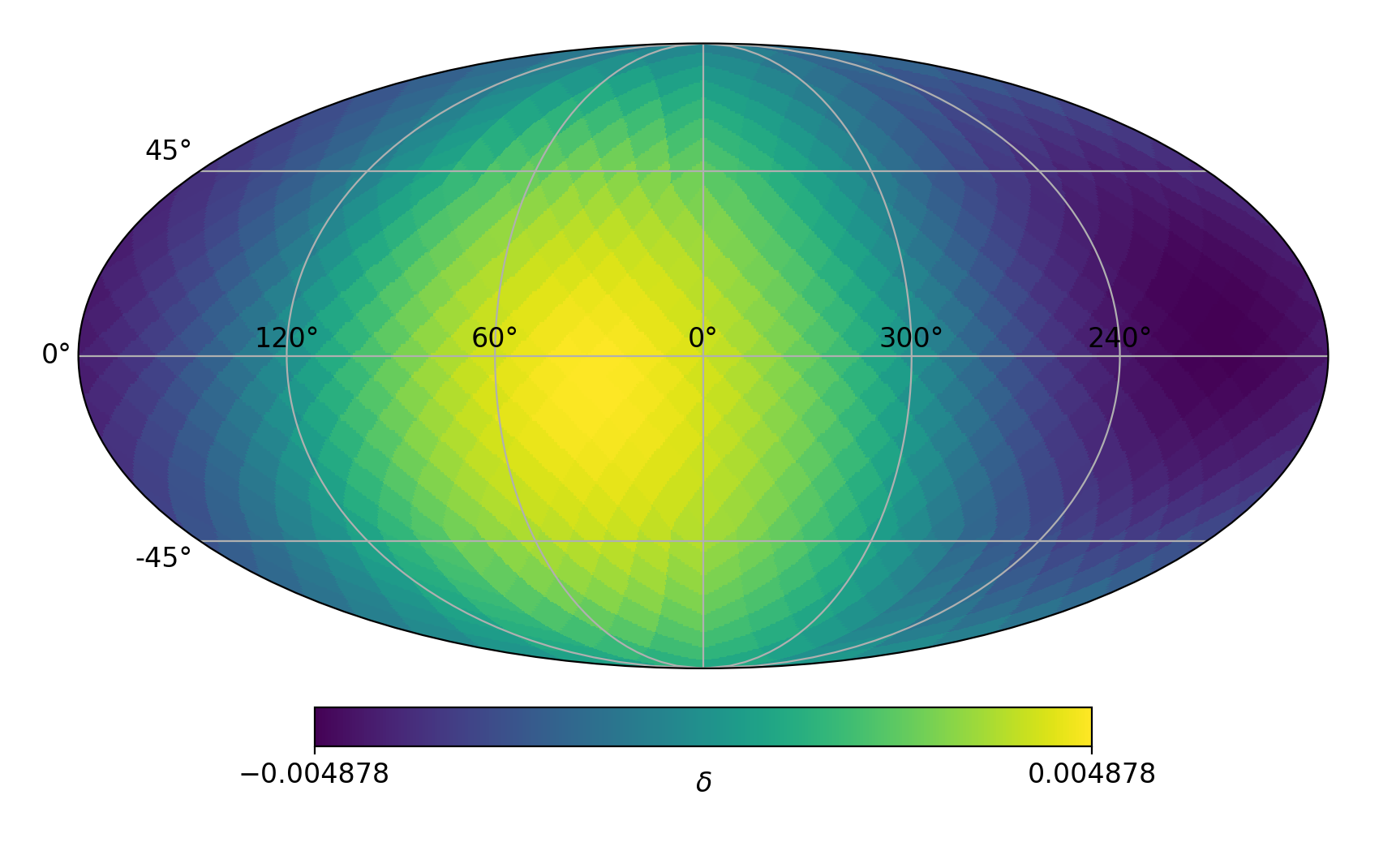}%
}\hspace*{-0.4em}
\subfloat{%
\includegraphics[height=39mm,width=55mm]{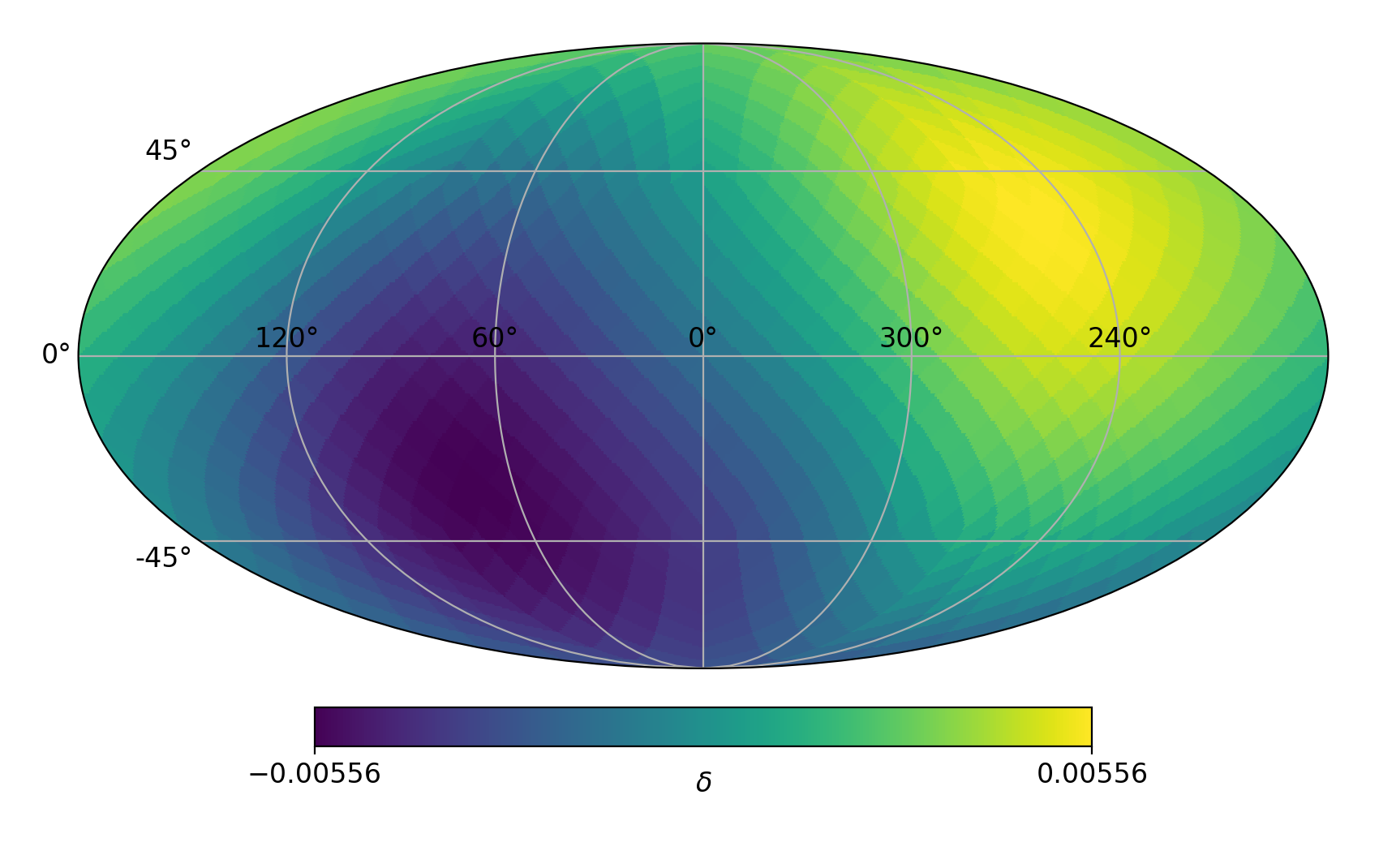}%
}\hspace*{-0.4em}
\subfloat{%
\includegraphics[height=39mm,width=55mm]{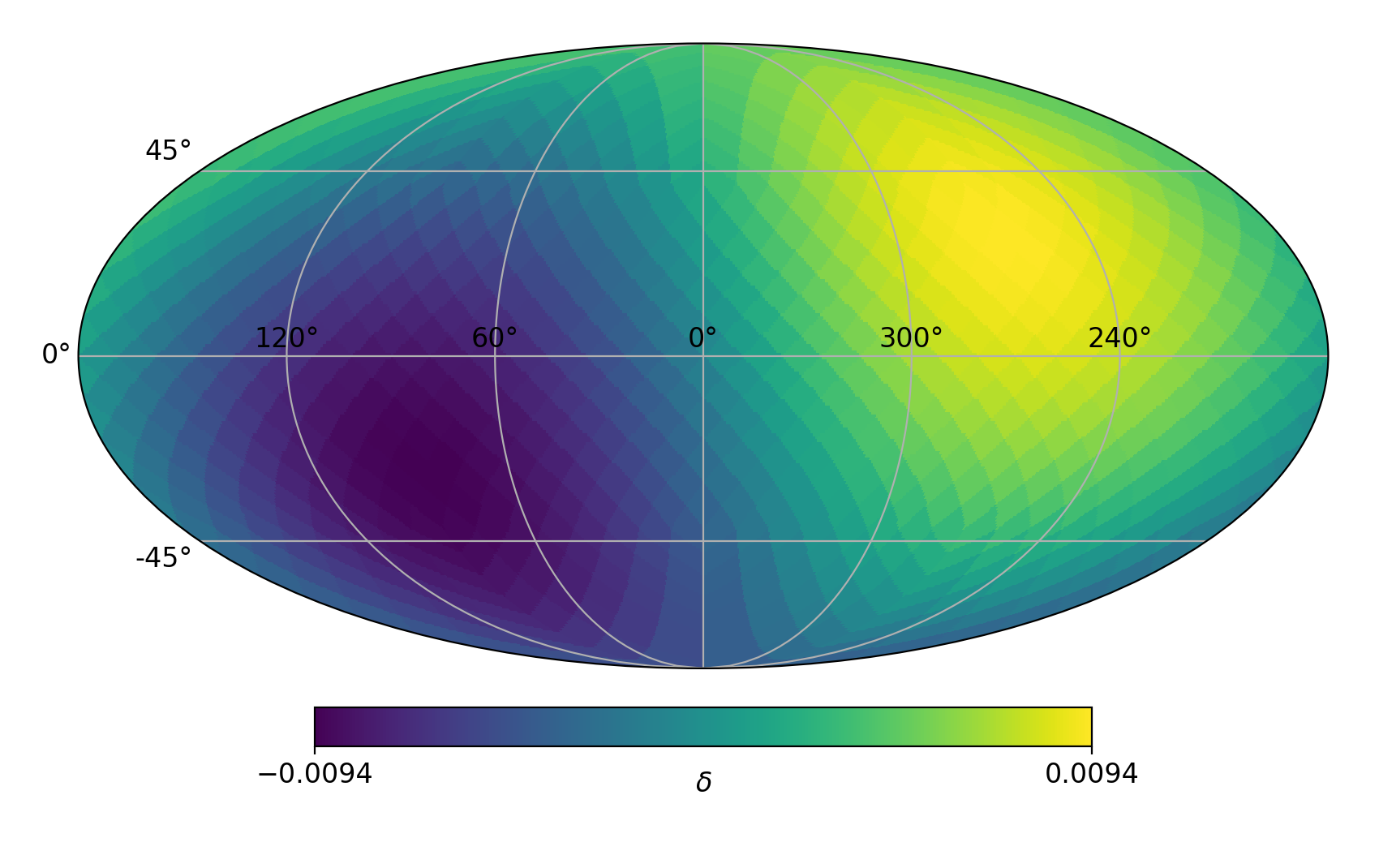}%
}

\caption{ Polarized ITBD event rate relative to the unpolarized rate at $m_{\nu} = 0.01$ eV. The sequence of plots follows FIG.5. 
}
\label{anisotropy2}
\end{figure*}  

\hspace{-3cm}
\begin{figure*}[htp!]
\hspace*{-6.2em}
\subfloat{%
\includegraphics[height=39mm,width=55mm]{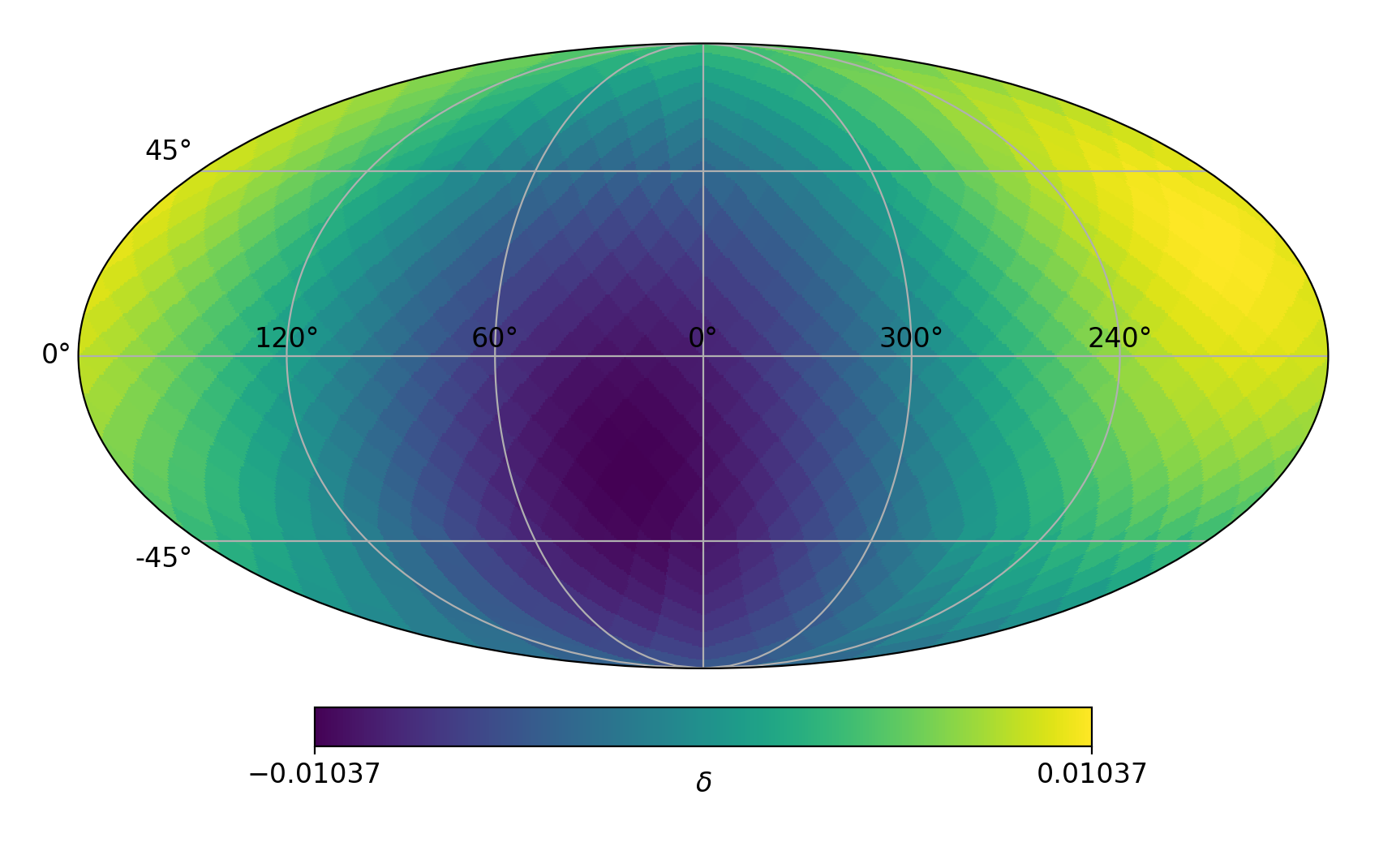}%
}\hspace*{-0.7em}
\subfloat{%
\includegraphics[height=39mm,width=55mm]{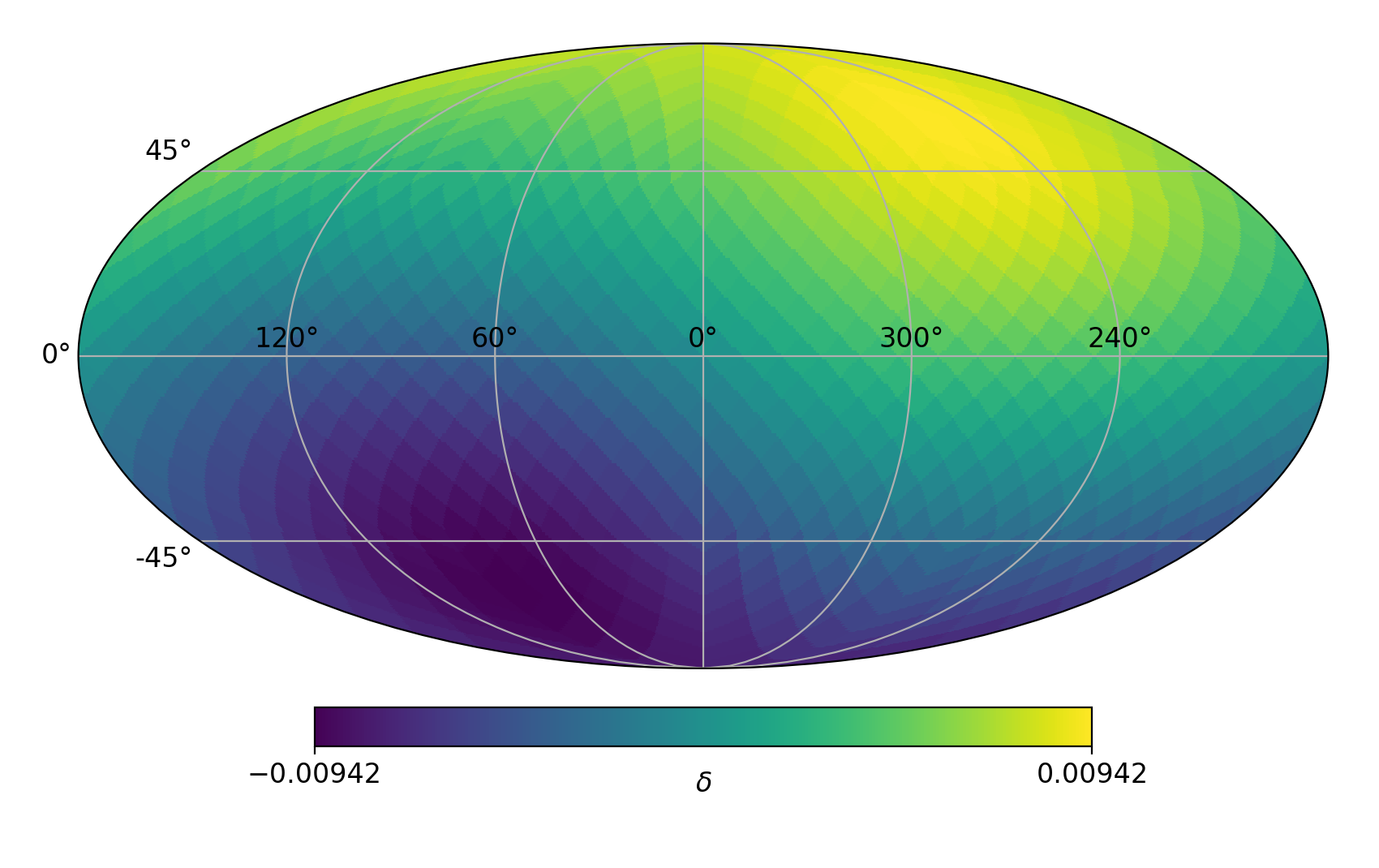}%
}\hspace*{-0.4em}
\subfloat{%
\includegraphics[height=39mm,width=55mm]{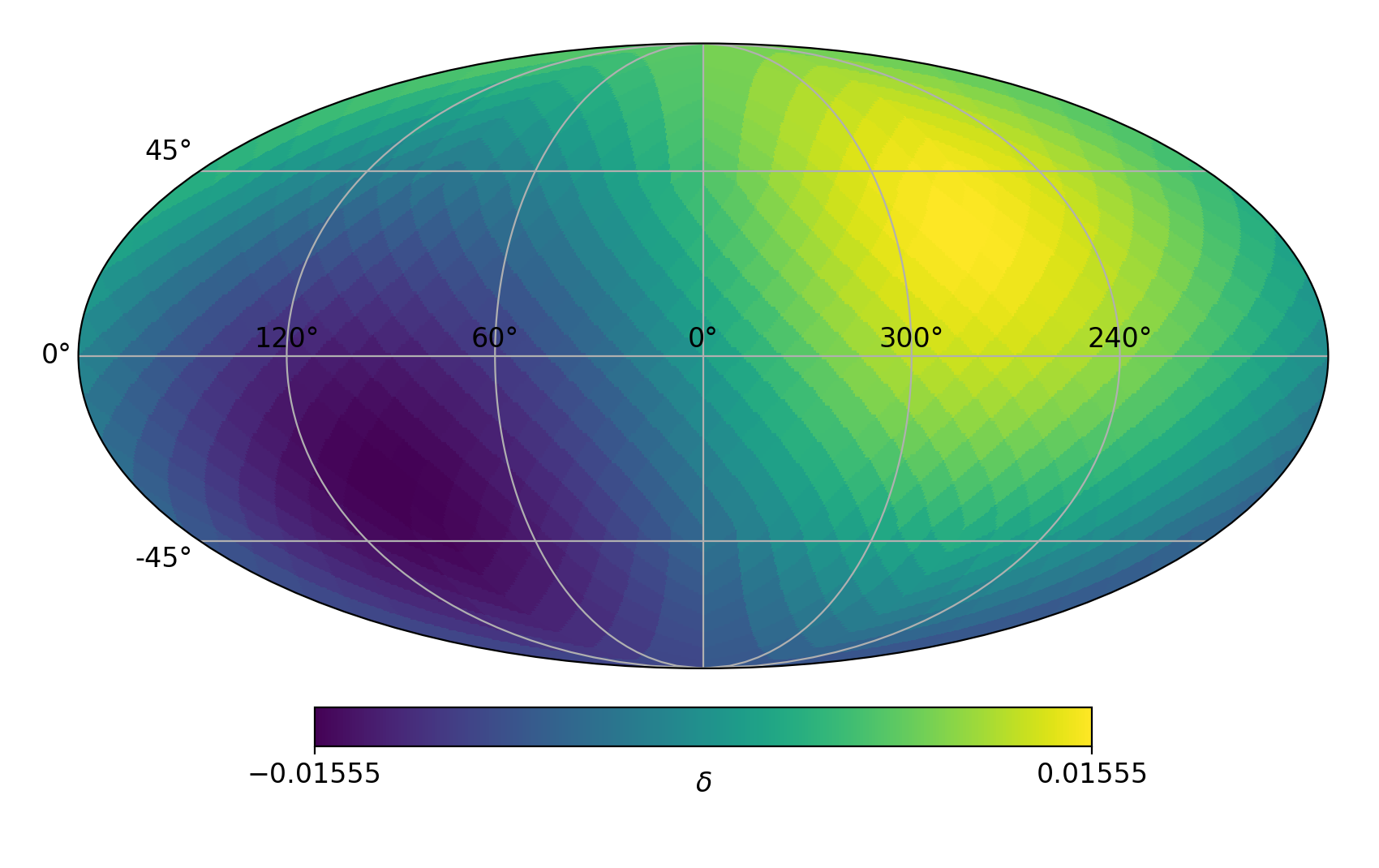}%
}\hspace*{-0.4em}
\subfloat{%
\includegraphics[height=39mm,width=55mm]{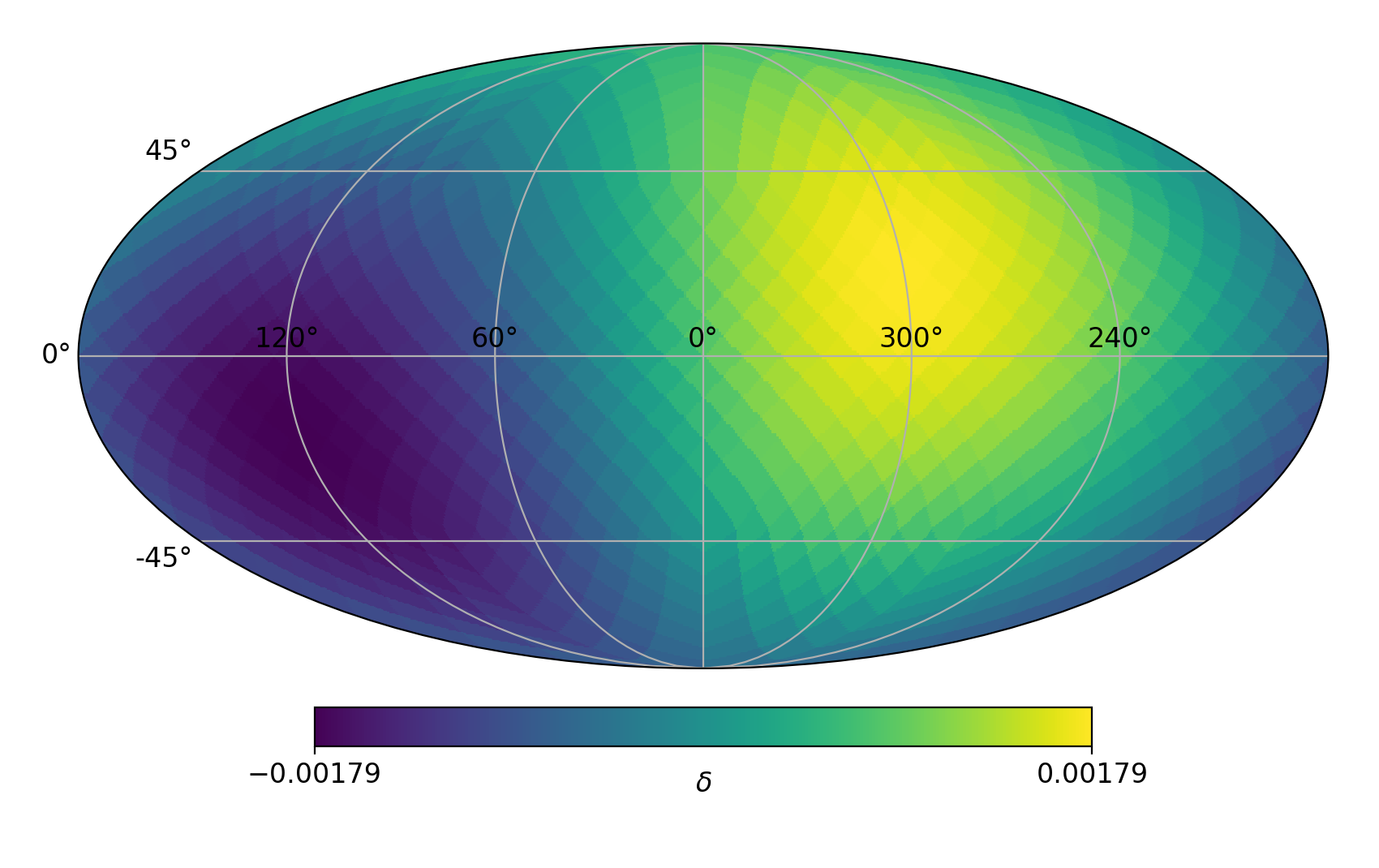}%
}\hfill
\hspace*{-6.2em}
\subfloat{%
\includegraphics[height=39mm,width=55mm]{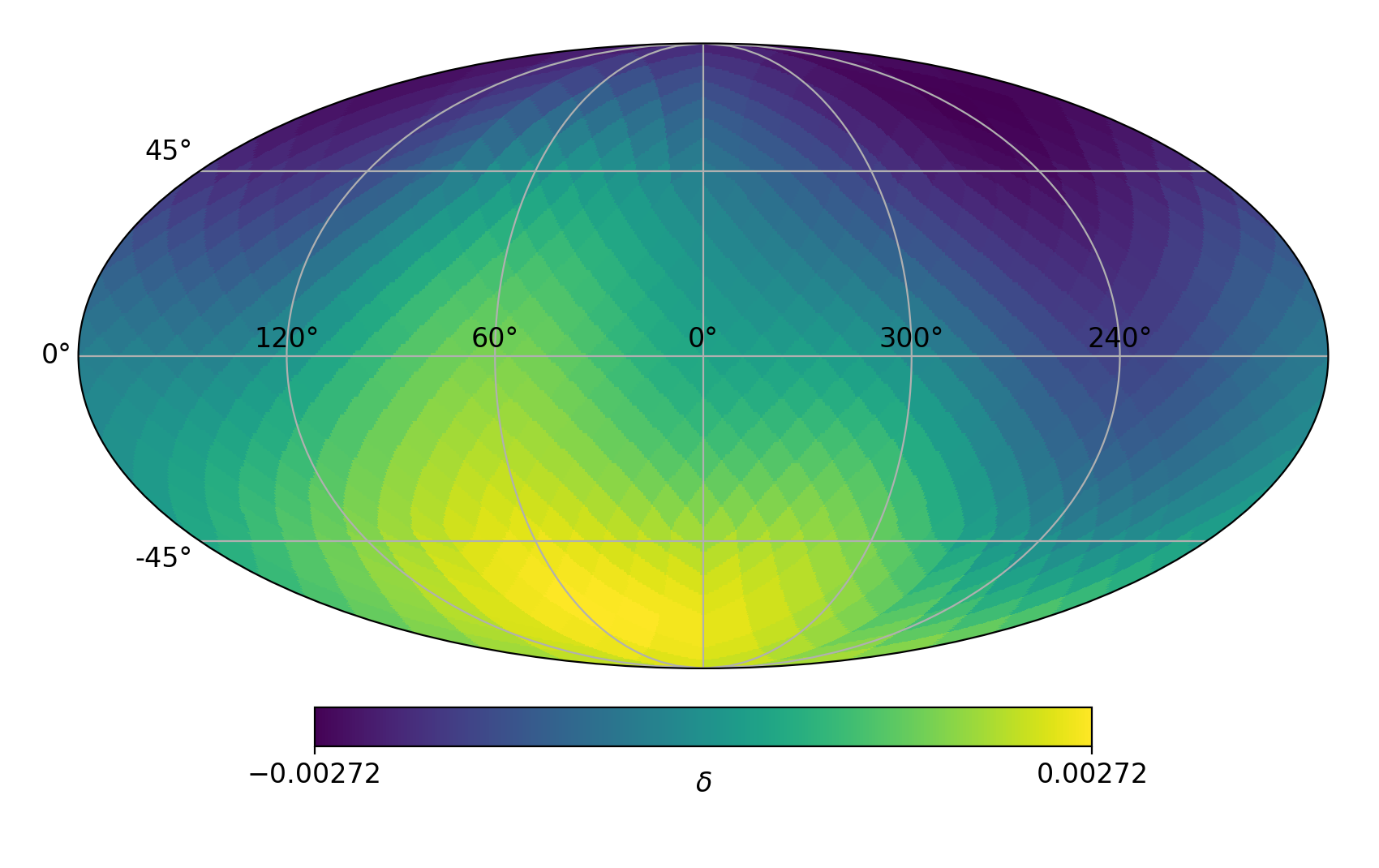}%
}\hspace*{-0.7em}
\subfloat{%
\includegraphics[height=39mm,width=55mm]{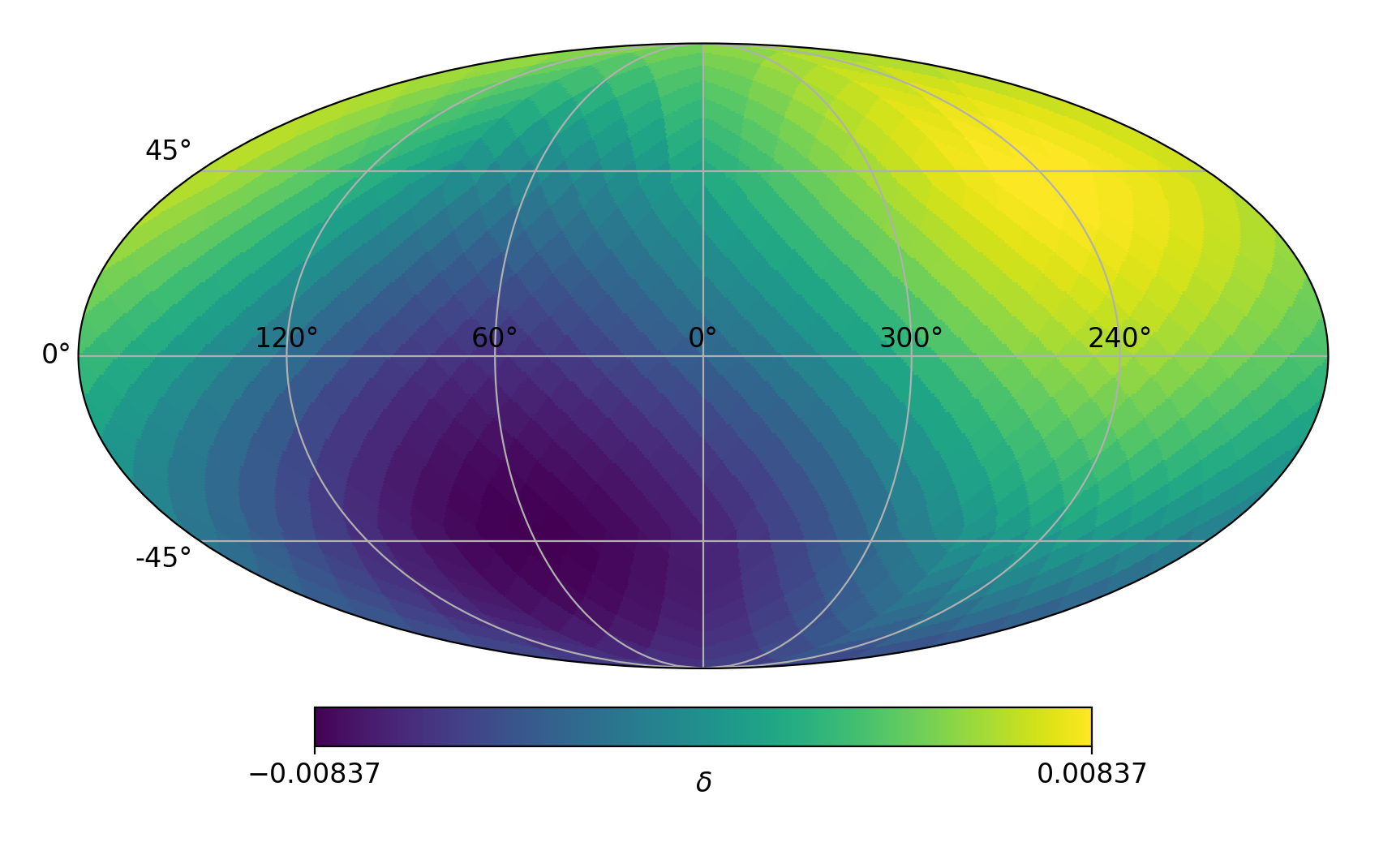}%
}\hspace*{-0.4em}
\subfloat{%
\includegraphics[height=39mm,width=55mm]{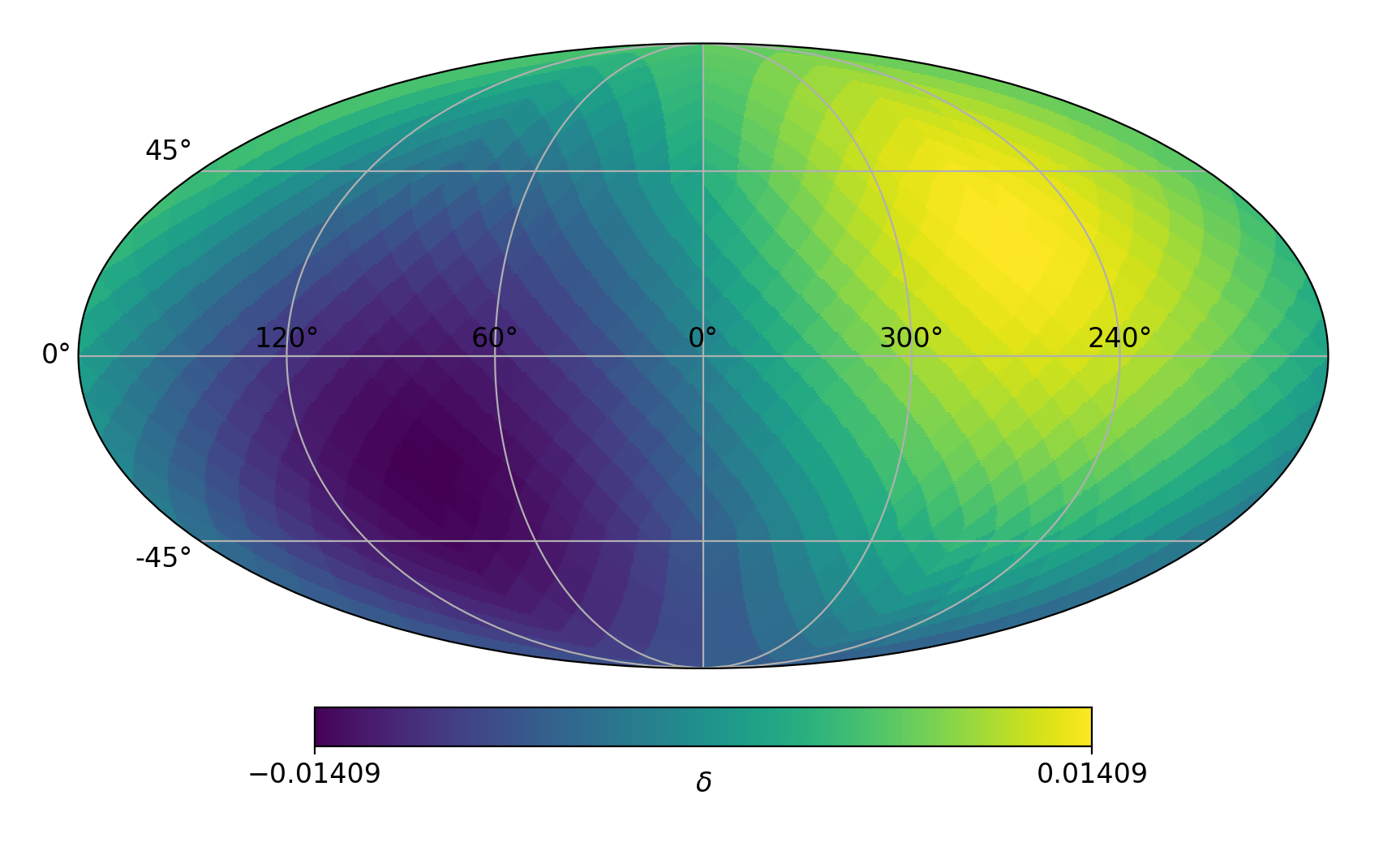}%
}\hspace*{-0.4em}
\subfloat{%
\includegraphics[height=39mm,width=55mm]{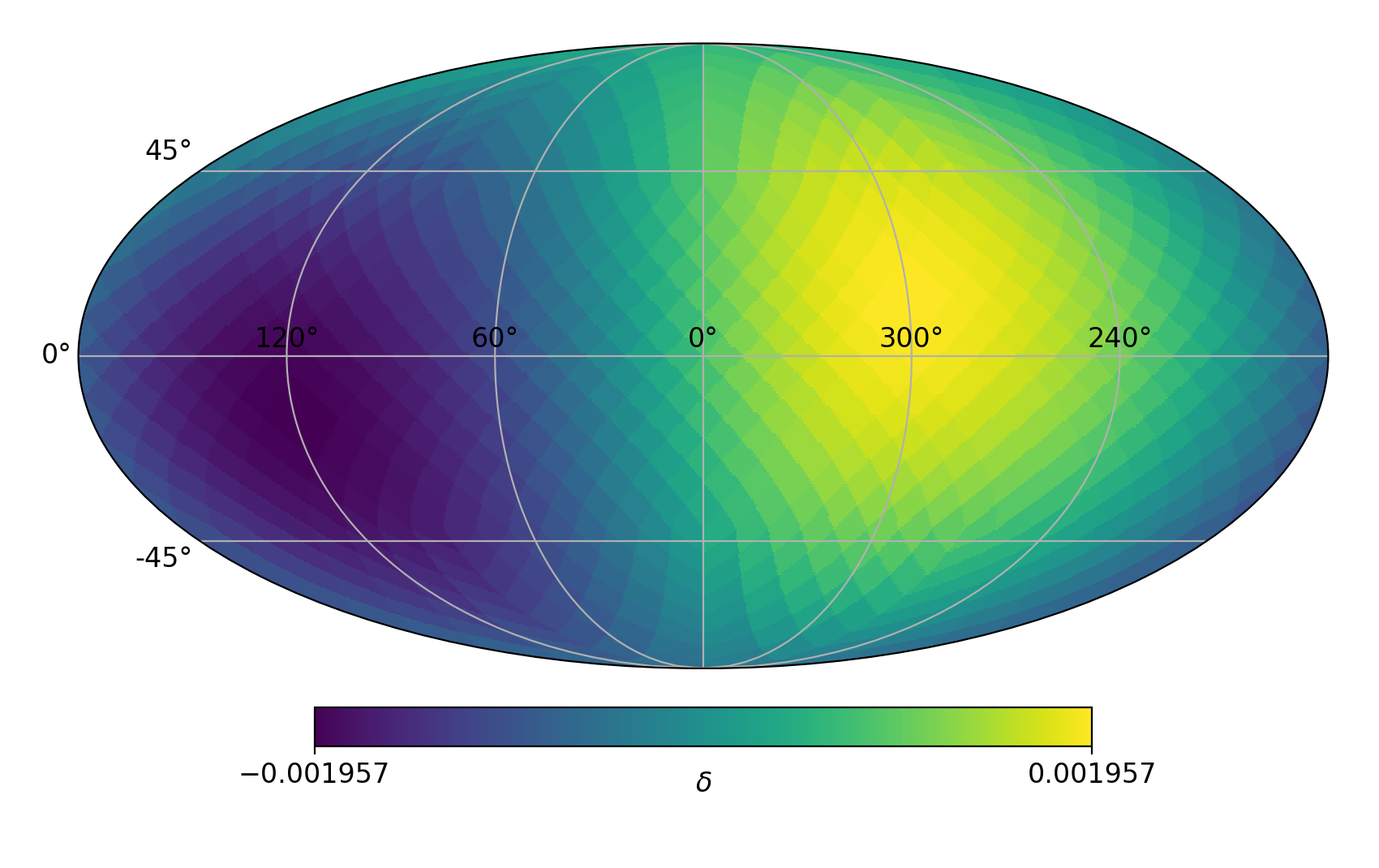}%
}

\caption{ Polarized ITBD event rate relative to the unpolarized rate at $m_{\nu} = 0.001$ eV. The sequence of plots follow FIG.5. 
}
\label{anisotropy3}
\end{figure*}

\begin{figure*}[h!]
\hspace*{-6.2em}
\subfloat{%
\includegraphics[height=39mm,width=55mm]{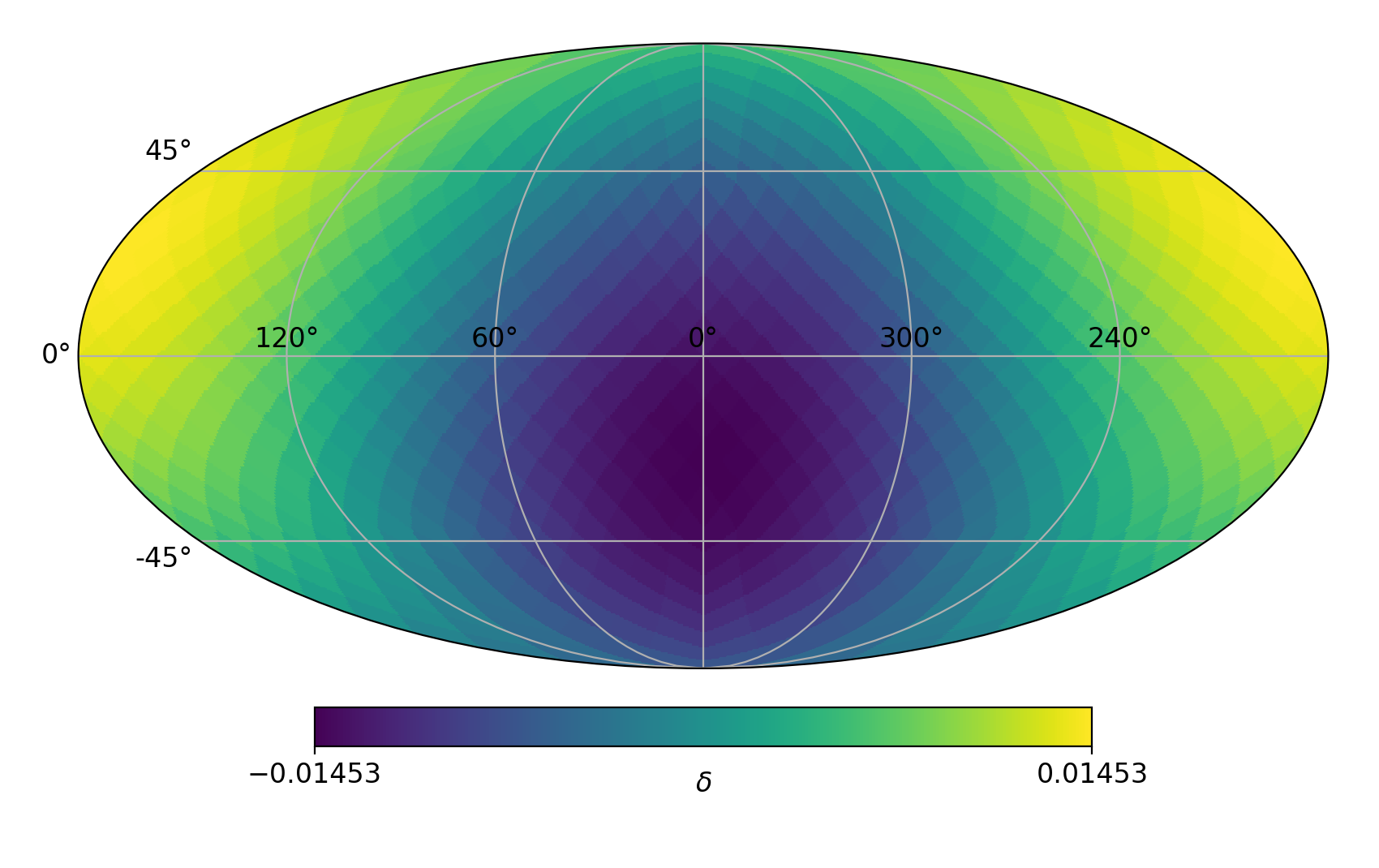}%
}\hspace*{-0.7em}
\subfloat{%
\includegraphics[height=39mm,width=55mm]{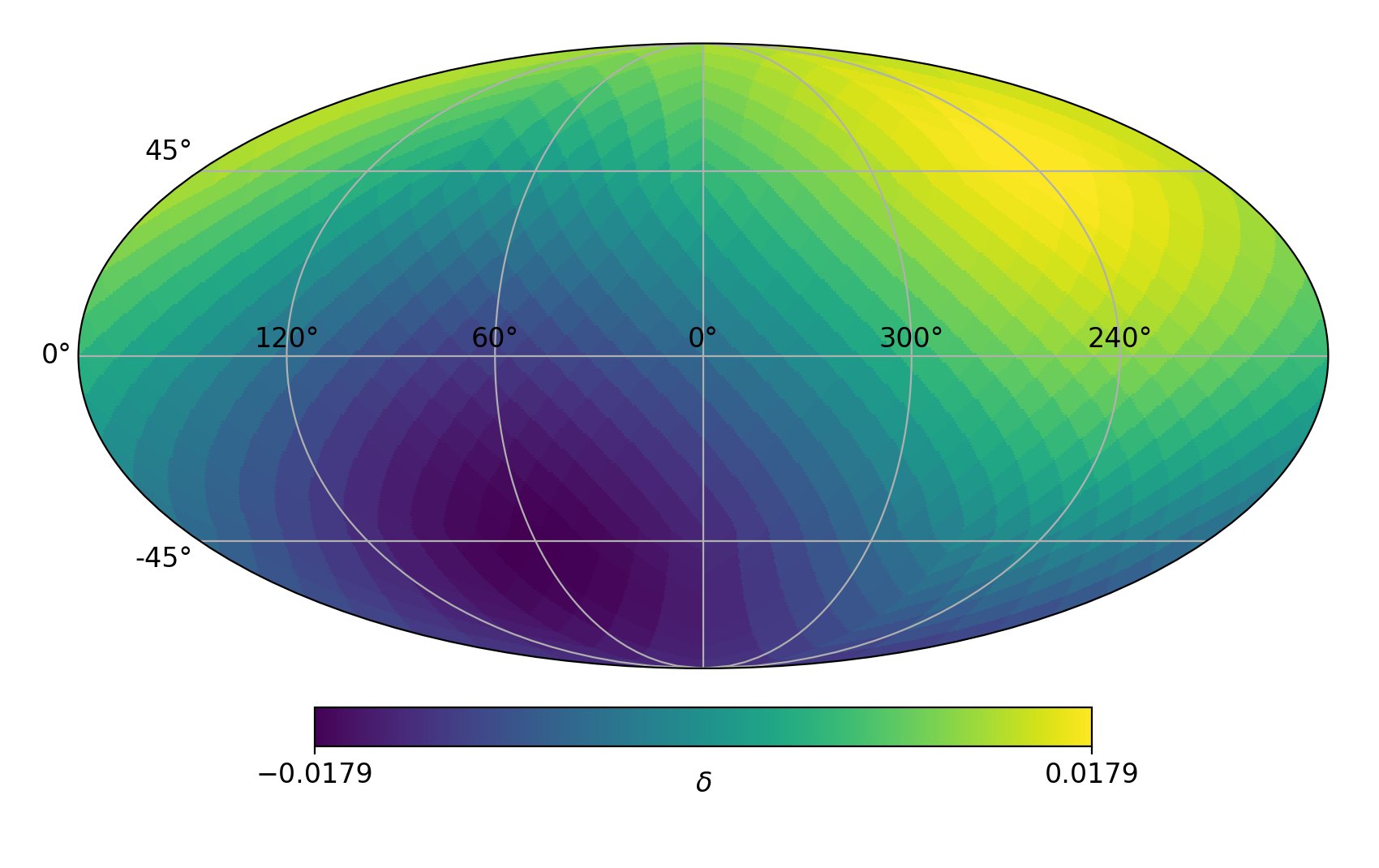}%
}\hspace*{-0.4em}
\subfloat{%
\includegraphics[height=39mm,width=55mm]{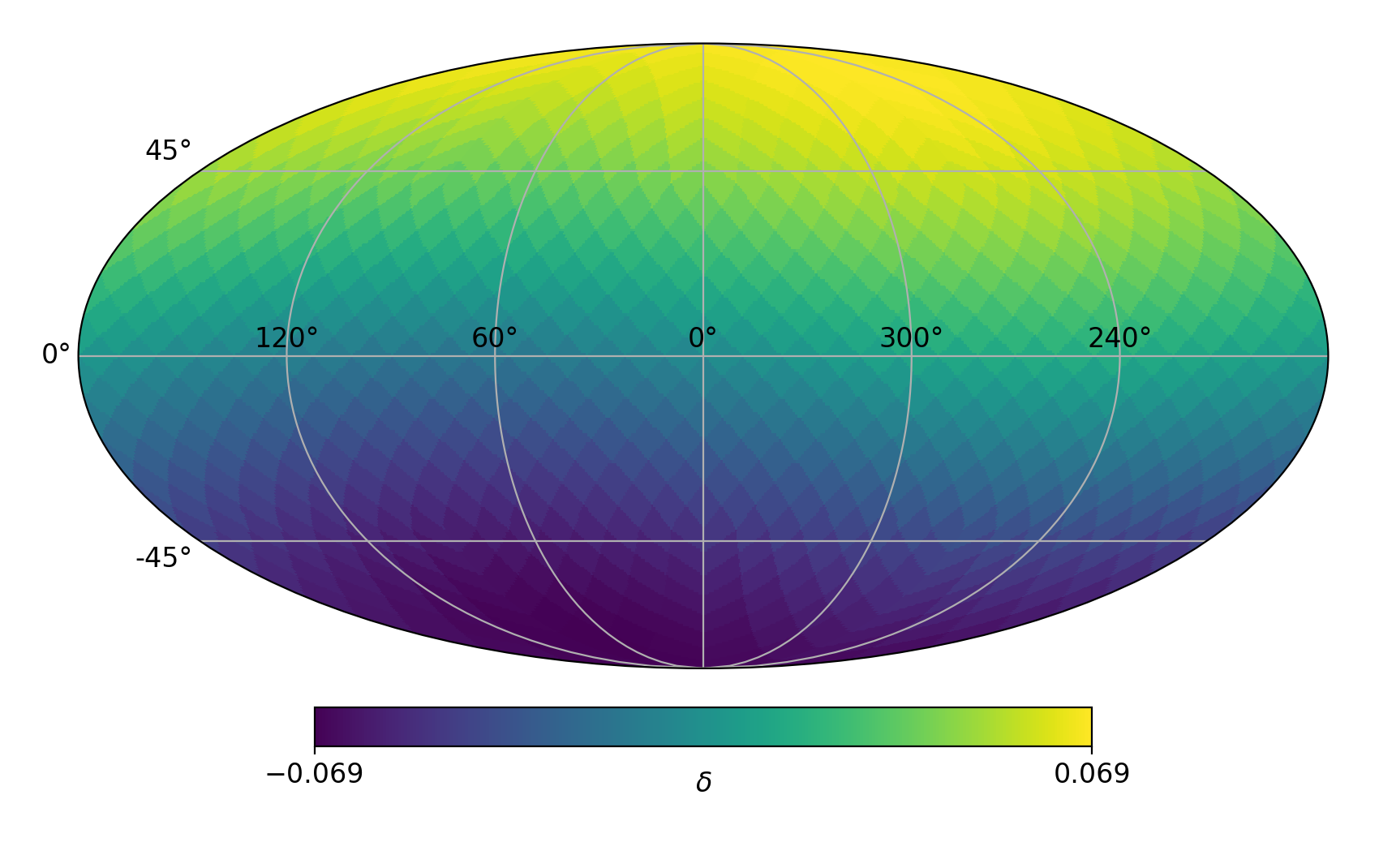}%
}\hspace*{-0.4em}
\subfloat{%
\includegraphics[height=39mm,width=55mm]{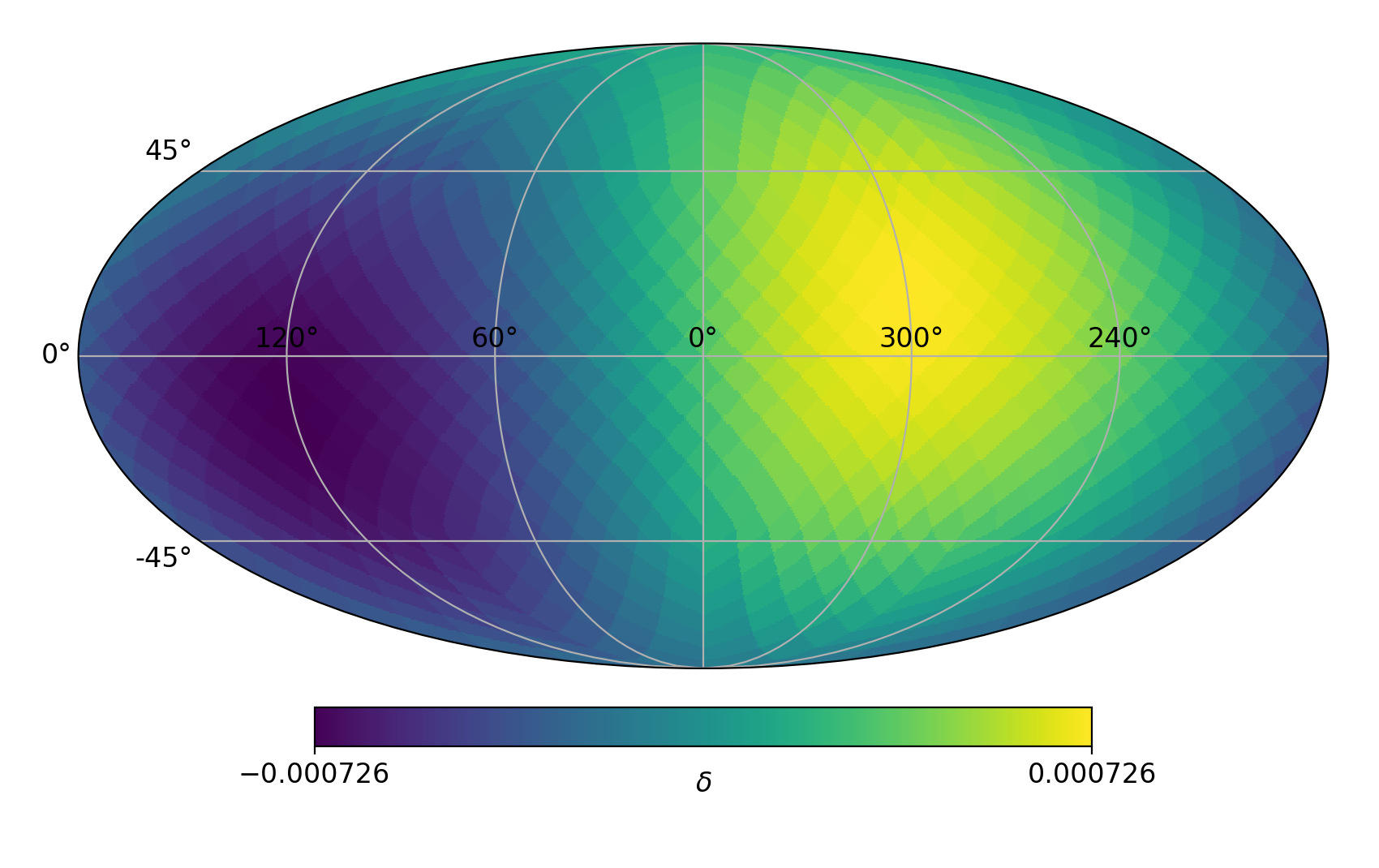}%
}\hfill
\hspace*{-6.2em}
\subfloat{%
\includegraphics[height=39mm,width=55mm]{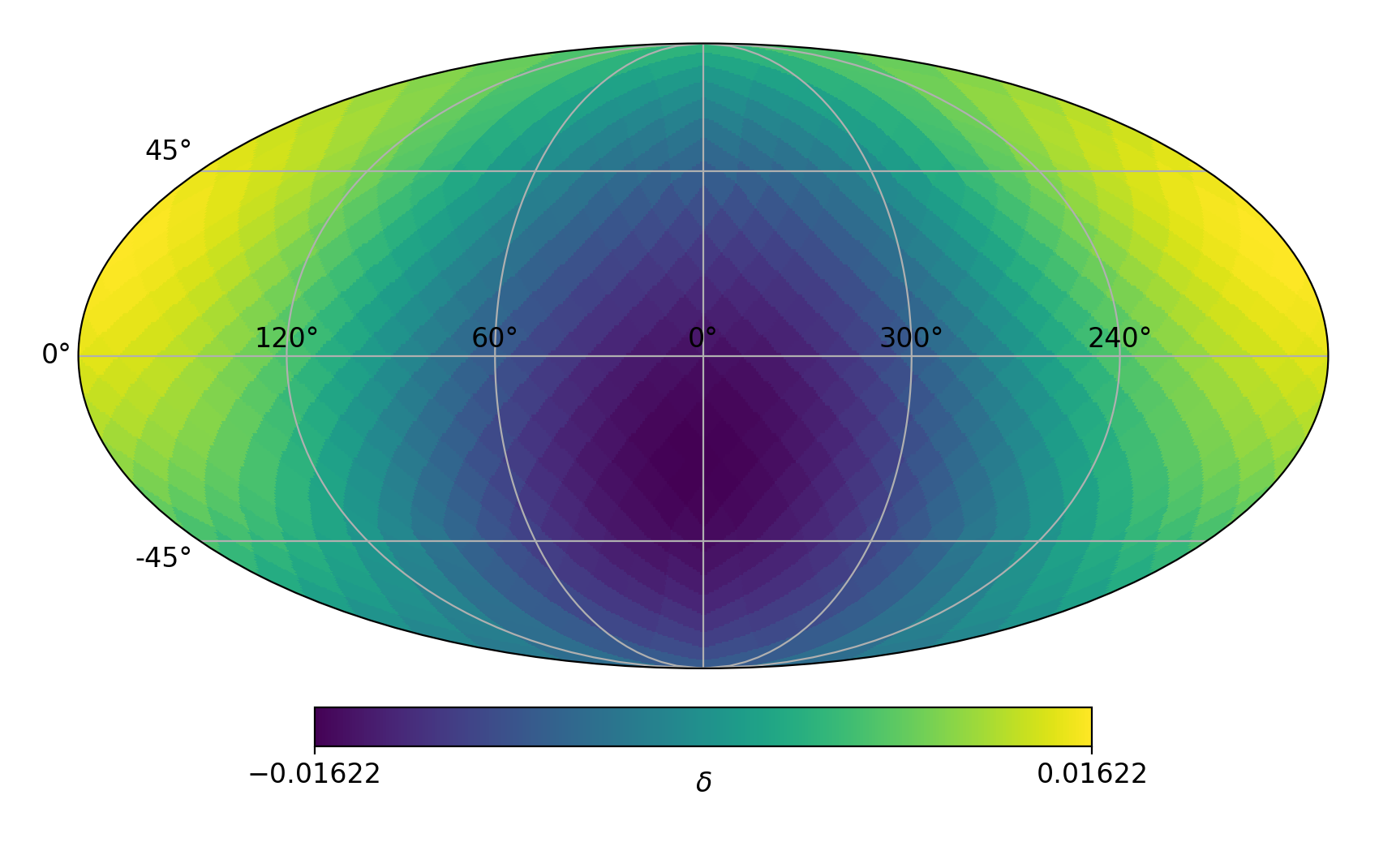}%
}\hspace*{-0.7em}
\subfloat{%
\includegraphics[height=39mm,width=55mm]{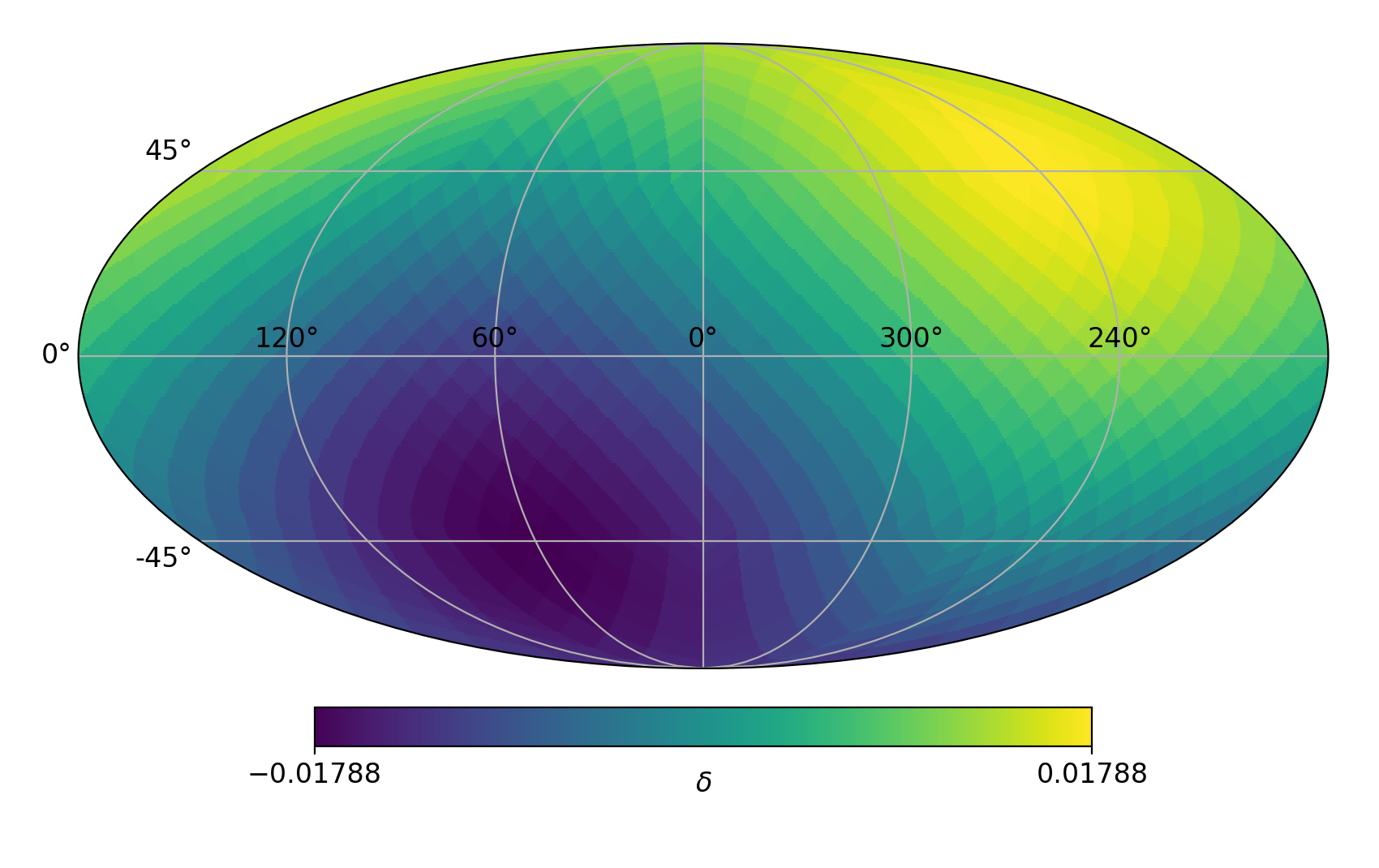}%
}\hspace*{-0.4em}
\subfloat{%
\includegraphics[height=39mm,width=55mm]{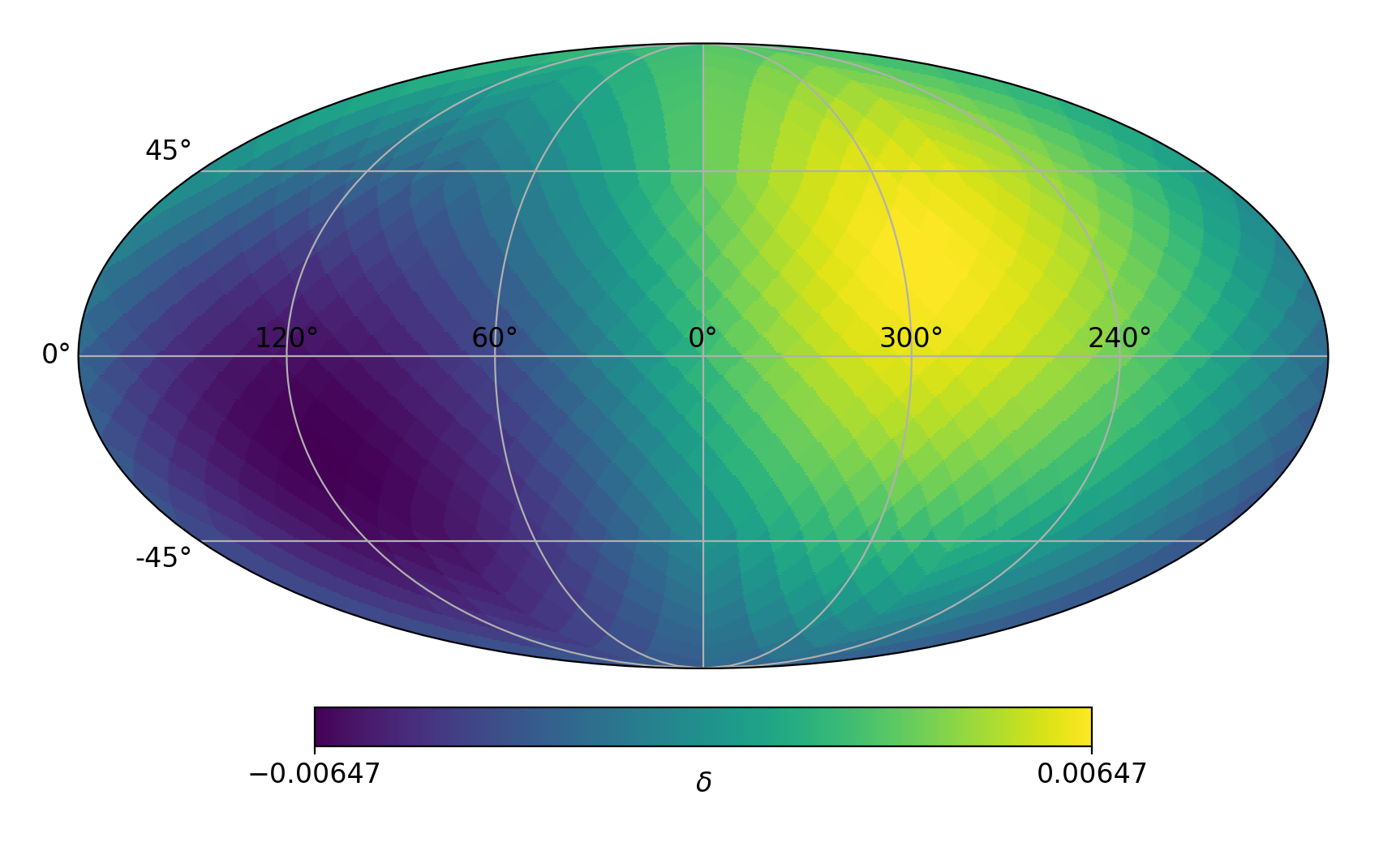}%
}\hspace*{-0.4em}
\subfloat{%
\includegraphics[height=39mm,width=55mm]{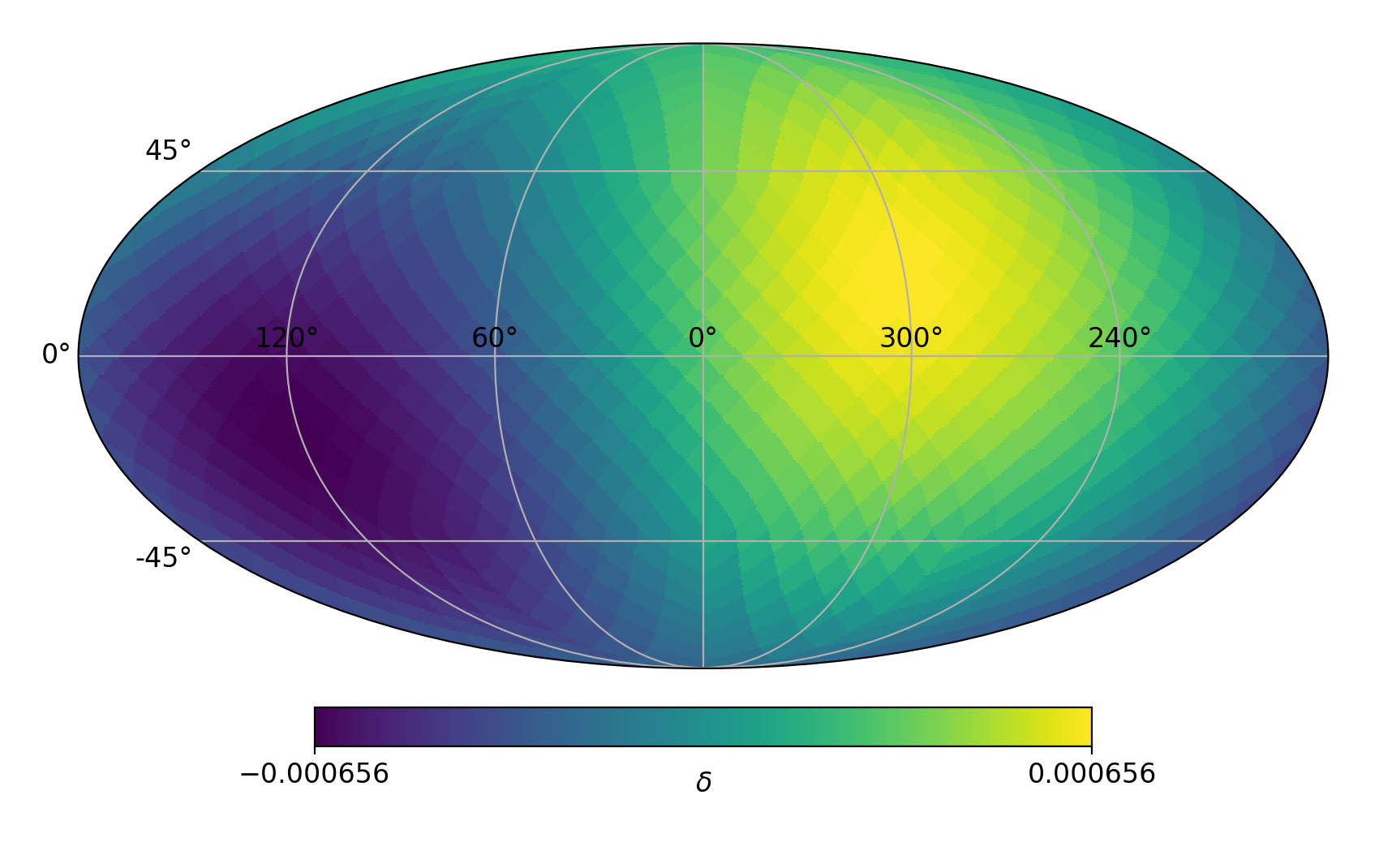}%
}

\caption{ Polarized ITBD event rate relative to the unpolarized rate at $m_{\nu} = 0.0001$ eV. The sequence of plots follow FIG.5. 
}
\label{anisotropy4}
\end{figure*}

\bibliography{NeutrinoMS} %Produces the bibliography via BibTeX.

\end{document}